# The galaxies' energy balance problem solved

Christopher J. Inman [1★], Cristina C. Popescu[1,2★] and David Murphy[1]

[1] *Jeremiah Horrocks Institute, University of Lancashire, Preston PR1 2HE, UK*
[2] *Max Planck Institut für Kernphysik, Saupfercheckweg 1, D-69117 Heidelberg, Germany*

## ABSTRACT

We attempt to resolve the long-standing energy balance problem encountered by radiative transfer (RT) models, particularly in edge-on galaxies by incorporating a treatment for the clumpy structure of the interstellar medium (ISM). A subgrid approach is adopted, treating the quiescent dust clouds as pseudo-dust grains, with equivalent optical and thermal emission properties. Deriving key quantities such as the absorption, scattering, and extinction cross-sections enables the virtualization of the macroscopic clump into a microscopic pseudo-grain that can be included alongside the existing dust model constituents. The addition of the pseudo-grain results in a flatter extinction curve. A library of clump emission spectral energy distributions (SEDs) is constructed for radiation fields of various colours and intensities. The new clumpy model is applied to the edge-on galaxy NGC 891 and, for the first time, is able to achieve a good energy balance, simultaneously fitting both the submillimetre and near-infrared (NIR) data. The clumpy model is further applied to a small sample of seven galaxies of various inclinations, and the results are compared with those from the purely diffuse models. The clumpy models are characterized by a reduction in dust opacity, and therefore attenuation, compared to their purely diffuse counterparts. Thus, the maximum face-on optical depth in the $B$ band, $\max(\tau^{f}_{B})$, derived from the clumpy models is found to be lower by factors ranging from 1.3 to 2.8. Of the seven galaxies, two are found to be optically thick in their centres, two are found to be moderately optically thick, and three are found to be optically thin.

## 1 INTRODUCTION

Within our current cosmological paradigm (the concordance model), the cosmological simulations have successfully accounted for the effects of gravity due to dark and ordinary matter in shaping the evolution of structure and formation of galaxies, as seen throughout the universe. Our biggest challenge in this field remains to understand the detailed baryonic physics equally critical for galaxy evolution. Already the largest cosmological simulations (e.g. MILLENIUMTNG – C. Hernández-Aguayo et al. 2023, FLAMINGO – J. Schaye et al. 2023) include ever more realistic physics, like neutrino, cosmic rays, and interstellar gas and dust physics. However, even if explicit calculations could be performed self-consistently across all the components of the interstellar medium (ISM) in galaxy simulations, the result would still only be a statistical realization of the galaxy populations. Thus, no specific individual galaxy could be simulated, not even the Milky Way (MW).

An alternative approach (applicable to individual galaxies) is to decode the panchromatic imaging information, through modelling, and derive an intrinsic representation of galaxy components: stars, gas, dust, and cosmic rays, and their interplay. This approach accounts for the formation and evolution of any individual galaxy. From a mathematical point of view, this is solving the inverse problem for a galaxy. In turn, this could better inform cosmological simulations, improving input baryonic physics, in addition to directly allowing advances in our knowledge of fundamental processes, such as star formation in galaxies. The most powerful tools available to self-consistently account for all these processes are radiative transfer (RT) models, and with recent increases in computing power their use is now realisable.

From an observational perspective, the *James Webb Space Telescope* (*JWST*) is revealing amazing detail of the stellar structures and ISM in local Universe galaxies (e.g. T. G. Williams et al. 2024, for the Physics at High Angular resolution in Nearby GalaxieS (PHANGS)-JWST sample), and identified the first forming galaxies in the Universe (S. Carniani et al. 2024). Despite advances in modelling and observations, when decoding galaxies with self-consistent models, even in the nearby Universe, a significant problem has remained unsolved until now. Perhaps brushed under the carpet due to the recent developments, there is a leftover, but fundamental energy balance problem between direct stellar light and dust emission for galaxies seen under a specific orientation (M. Baes et al. 2010; C. C. Popescu et al. 2000). The story behind this is as follows.

★ E-mail: CJInman@lancashire.ac.uk(CJI), CPopescu@lancashire.ac.uk(CCP)

When the first RT models of the surface-brightness distribution of star-forming galaxies were developed (N. D. Kylafis & J. N. Bahcall 1987), they were first applied to edge-on galaxies, since images obtained under this orientation reveal the vertical stratification of stars and dust, allowing for additional model constraints, through the determination of the scale heights of the different galaxy components. Usually three advantages were invoked for using this view of galaxies: (i) easy to separate the stellar disc, the bulge, and the dust; (ii) the averaging of details allowing for simpler modelling; and (iii) the dust is prominently seen in the form of dust lanes (E. M. Xilouris et al. 1997).

The RT modelling in the optical bands of the prototype nearby edge-on spiral NGC 891, together with a few other edge-on galaxies, allowed E. M. Xilouris et al. (1997, 1998, 1999) to derive the spatial distribution of stars (mainly older stellar populations) and associated dust. They found that spirals are optically thin in the optical bands, with a central face-on *B*-band optical depth of around 1 ($\tau_f^c(B) \approx 1$), so just becoming optically thick towards shorter ultraviolet (UV) wavelengths. This would have been seen as solving the long debate of whether spiral galaxies are optically thin or thick (M. Disney, J. Davies & S. Phillipps 1989). However, when C. C. Popescu et al. (2000) developed the first panchromatic RT modelling of spiral galaxies, whereby the dust emission was self-consistently calculated with the dust attenuation, they found that the solution obtained in the optical bands for NGC 891 by E. M. Xilouris et al. (1999), was underestimating the far-infrared (FIR)/submillimetre (submm) emission by a factor of around 3. This was later called the energy balance problem (S. Bianchi 2007, 2008; M. Baes et al. 2010). To solve the problem, C. C. Popescu et al. (2000) proposed several scenarios, including the existence of additional large-scale diffuse or clumpy components and of modified dust grain properties, and adopted a solution with an additional disc of diffuse dust, with the dust opacity constrained to fit the submm data. This solution, with $\tau_f^c(B) \sim 4$ provided a good fit to all FIR/submm images available at that time, including the optical images. It was clear that the solution derived for the optical images alone is degenerate. In other words, the appearance of the dust lane in the optical can be accommodated with both a face-on optically thin or thick solution. The degeneracy is lifted in the near-infrared (NIR) *J* and *K* bands, where edge-on lines of sight make the transition between optically thin and thick at these wavelengths. It was shown by K. M. Dasyra et al. (2005) that the model of NGC 891 with $\tau_f^c(B) \sim 4$ produces too prominent dust lanes in the NIR, not observed in the then-new 2MASS (Two Micron All Sky Survey; T. H. Jarrett et al. 2003) images. The energy balance problem in edge-on galaxies was confirmed by all later studies of edge-on galaxies, including S. Bianchi (2008), M. Baes et al. (2010), G. De Geyter et al. (2015), and A. V. Mosenkov et al. (2016, 2018).

One other scenario proposed in C. C. Popescu et al. (2000) was to take into account the clumpy nature of the ISM. Clumps are known to reduce the overall attenuation of the stellar light (e.g. A. N. Witt & K. D. Gordon 1996; F. Városi & E. Dwek 1999). C. C. Popescu et al. (2000) included a clumpy component associated with the star-forming regions (see also C. C. Popescu et al. 2011, hereafter PT11 for further developments), although this was not enough to solve the issue. Models with quiescent clumps have been also used (S. Bianchi, J. I. Davies & P. B. Alton 2000; S. Bianchi 2008), but they could not account for the energy balance problem either.

W. Saftly et al. (2015) argued that using galaxy simulations, whereby the input distribution of stars and dust is known, does not result in an energy balance problem, and that therefore the solution to the problem could be to take into account the complex distribution of stars and dust, as seen in face-on galaxies, departing from the azimuthal symmetry involved in modelling the edge-on galaxies. The focus then shifted towards face-on galaxies. Using the SKIRT code and non-axisymmetric models, I. De Looze et al. (2014), S. Viaene et al. (2014), and S. Verstocken et al. (2020) fitted the panchromatic images of the face-on spirals M51, M31, and M81, and obtained a reasonable agreement between models and observations. However, the axisymmetric models of PT11 were also used to model face-on galaxies (J. J. Thirlwall et al. 2020, hereafter TP20, for M33; M. T. Rushton et al. 2022 for NGC 628; C. J. Inman et al. 2023 for M51; D. Pricopi et al. 2025 for M101 and NGC 3938), and produced good agreement between models and observations. It became obvious that the use of axisymmetric versus non-axisymmetric models was not at the root of the problem. Dust with modified optical properties cannot provide the solution either, since the properties of dust should not vary with the viewing angle. Thus, the energy balance problem remained unsolved: RT models accounted for the panchromatic surface-brightness distributions of face-on spiral galaxies, but not for their edge-on counterparts. In edge-on systems either an optically thin solution is derived from the optical data alone, but this underestimates the FIR/submm data by factors between 2–4 (C. C. Popescu et al. 2000; A. V. Mosenkov et al. 2018), or an optically thick solution is derived from the submm data alone (C. C. Popescu et al. 2000, 2011), but this overestimates the strength of the dust lane in the NIR.

In this paper, we show that the clumpy nature of the ISM on scales of around 1 pc provides the resolution to the energy balance problem. We introduce a subgrid approach, in the form of a novel concept, that of the so-called pseudo-grain. We show that this provides a consistent solution for both face-on and edge-on galaxies, thus solving the long-standing energy balance problem in galaxy models.

## 2 THE PSEUDO-GRAIN CONCEPT

The diffuse dust, pervading the ISM of star-forming galaxies, is a fundamental component of any dusty RT model, including the one used in this work (PT11). A significant fraction of 'diffuse dust' is distributed in cirrus-like clouds; however, since these structures are optically thin at optical wavelengths, they yield RT solutions equivalent to those of a purely diffuse medium. Thus, for all practical purposes, optically thin clumps have not been explicitly incorporated into RT models until now. However, while this assumption holds very well in the optical bands, it will progressively break at shorter UV wavelengths, in particular within the clump cores on scales of approximately 1 pc or smaller, where clouds become optically thick. As a result, a fraction of the dust is more effectively shielded from UV radiation than would be expected in a purely diffuse medium.

Sampling galactic discs on tens-of-kiloparsec scales with parsec-scale resolution exceeds the computational capabilities of existing RT codes. To address this limitation, a subgrid approach is introduced, which we call the pseudo-grain approach. This concept is similar to the mega-grains used by M. P. Hobson & R. Padman (1993), and F. Városi & E. Dwek (1999) to describe a two-phase clumpy medium. In our approach, a macroscopic, parsec-sized dust cloud can be analysed in terms of its attenuation prop-erties of the interstellar radiation fields (RFs), as well as of its dust emission properties (dust emission spectral energy distribution, SED) in the same way a dust particle can be characterized by its optical properties when interacting with a planar wave. By performing dedicated RT calculations of a single dust cloud illuminated externally by stellar RFs, one can determine the amount of absorption and scattering produced by the clump, and thus define an efficiency for absorption and scattering of the cloud, $Q_{\rm abs}$ and $Q_{\rm sca}$. Based on these efficiencies, the corresponding absorption and scattering cross-sections, $C_{\rm abs}$ and $C_{\rm sca}$, can also be calculated. Furthermore, for given incident RFs, the temperature of the dust within the cloud can be calculated using the RT codes, and a dust emission SED derived, analogous to the IR emissivity of an individual dust grain. Because of all these similarities in methodology, the cloud can be replaced in the model with a virtual particle of dust, referred to as a pseudo-grain, that carries the same photon interaction efficiencies and corresponding emissivities defined by the above macroscopic calculation. As such, the resulting effect is as if the dust model would have a new component of dust, supplementing the existing species present in the dust model such as silicates, graphites, and polycyclic aromatic hydrocarbons (PAHs).

### 2.1 Pseudo-grain parametrization

As introduced above, each pseudo-grain is a single macroscopic dust cloud (associated with a gas cloud). The volume density (by mass) of dust in the cloud can be defined by a spherically symmetric I. King (1962) profile with

$$\rho^{\rm c,s}(r) = \begin{cases} \frac{\rho_0^{\rm c,s}}{[1+(r/r_{\rm c})^2]^{3/2}} & r < r_{\rm t} \\ 0 & \text{otherwise,} \end{cases} \tag{1}$$

where the abbreviation 'c,s' stands for a 'single cloud', $\rho_0^{\rm c,s}$ is the volume density of dust in the cloud at $r = 0$, $r_{\rm t}$ is the truncation radius of the clump, and $r_{\rm c}$ is a scale factor (core radius) given by

$$r_{\rm c} = \sqrt{\frac{k_{\rm B} T}{4\pi G \rho_{\rm gas,0}\, \mu\, m_{\rm H}}}\,, \tag{2}$$

where $\rho_{\rm gas,0}$ is the volume density by mass of gas at the cloud centre, $T$ is the gas temperature at the centre, $k_{\rm B}$ is the Boltzmann constant, $G$ is the gravitational constant, $m_{\rm H}$ is the mass of the hydrogen atom, and $\mu$ is the mean mass per gas particle in units of $m_{\rm H}$.

For $r \lesssim r_{\rm c}$, equation (1) is a good representation of an isothermal pressure-supported gas cloud. It should be noted that this is valid only if $\mu$ is not a function of $r$; that is, that the gas does not undergo a phase transition from H I to $H_2$. Moreover, $r_{\rm t}/r_{\rm c}$ is required to be a reasonably small multiple. These assumptions are reasonably satisfied in diffuse H I-dominated clouds observed in the solar neighbourhood with central H densities of a few hundred $\rm cm^{-3}$ or less and $T$ on the order of 50–100 K (T. P. Snow & B. J. McCall 2006). In such clouds, the H I emission is also observed to be tightly linked to the dust column, justifying the further implicit assumption that both dust and gas distributions follow equation (1).

If $\kappa_\nu^{\rm tot}$ is the total mass extinction coefficient of the microscopic grains, in units of area per mass, due to scattering and absorption from all sizes and composition of grains, then the opacity $\tau_\nu$ along any ray $\underline{l}$ is defined by

$$\frac{{\rm d}\tau_\nu^{\rm c,s}}{{\rm d}l}(r) = \kappa_\nu^{\rm tot} \rho^{\rm c,s}(r), \tag{3}$$

with $\rho^{\rm c,s}(r)$ given by equation (1). Then, the parameter $\tau_\nu^{\rm c,s}$ is the total opacity at frequency $\nu$ along a line of sight through the cloud centre:

$$\tau_\nu^{\rm c,s} = 2\kappa_\nu^{\rm tot} \int_0^{r_{\rm t}} \rho^{\rm c,s}(r){\rm d}r \tag{4}$$

$$= 2\kappa_\nu^{\rm tot} \rho_0^{\rm c,s} r_{\rm c} \xi\,, \tag{5}$$

with

$$\xi = \sin\left(\arctan(r_{\rm t}/r_{\rm c})\right). \tag{6}$$

The total mass of dust in each cloud, i.e. each pseudo-grain, is:

$$M^{\rm c,s} = \int_0^{r_{\rm t}} \rho^{\rm c,s}(r) \cdot 4\pi r^2 {\rm d}r \tag{7}$$

$$= 4\pi \rho_0^{\rm c,s} r_{\rm c}^3 \left(\ln\sqrt{\tfrac{1+\xi}{1-\xi}} - \xi\right). \tag{8}$$

Equation (5) can be re-arranged to solve for the density by mass at the centre of the clump, $\rho_0^{\rm c,s}$, and the dust opacity in the $B$ band:

$$\rho_0^{\rm c,s} = \frac{\tau_B^{\rm c,s}}{2 r_{\rm c} \xi \kappa_B} \tag{9}$$

where $\tau_B^{\rm c,s}$ is the opacity in the $B$ band of a single clump and $\kappa_B$ is the mass extinction coefficient in the $B$ band with units [$\rm pc^2\ M_\odot^{-1}$].

By substituting equation (9) into equation (8) the mass of the clump is

$$M^{\rm c,s} = 2\pi \frac{\tau_B^{\rm c,s} r_{\rm c}^2}{\kappa_B} \chi \tag{10}$$

where

$$\chi = \frac{\ln\left(\sqrt{\frac{1+\xi}{1-\xi}}\right)}{\xi} - 1. \tag{11}$$

Thus, the mass of a single dust clump is $M^{\rm c,s}(\tau^{\rm c,s}, r_{\rm c}, r_{\rm t})$. In this work, clumps are considered with $r_{\rm c} = 1$ pc and $r_{\rm t} = 1$ pc ($\xi = 0.7071$ and $\chi = 0.246$) and dust opacity through the clump (from the centre to the truncation radius) of $\tau_B^{\rm c,s} = 1$. The dust model used has $\kappa_B = 8.07\ \rm pc^2\ M_\odot^{-1}$ which leads to a dust clump with mass $M^{\rm c,s} = 0.192\ \rm M_\odot$. The choice of the values for $r_{\rm c}$, $r_{\rm t}$, and $\tau_B^{\rm c,s}$ was motivated by our physical interpretation of the clumps as being pressure-supported, and being optically thin in the optical/NIR wavelengths, and optically thick in the UV. From a conceptual point of view any dust outside the truncation radius of the clump would belong to the diffuse medium. We note here that by considering $r_{\rm t} = r_{\rm c}$ we do not allow for a large variation in the density profile, so the calculations could have been equally done with an uniform distribution. None the less, in order to be self-consistent with our physical interpretation of the clumps, as well as to allow for flexibility in any future application, we decided to use a physically motivated profile.

### 2.2 Extinction efficiency and cross-section of the pseudo-grain

Having prescribed the dust density distribution of the cloud, the next step is to describe the interaction with the interstellar RFs. The RT codes from PT11 are used to determine the efficiencies of absorption, scattering, and extinction of the cloud, denoted by $Q^{\rm c}_{\rm abs}$, $Q^{\rm c}_{\rm sca}$, and $Q^{\rm c}_{\rm ext}$, respectively. The RT codes from PT11 were built for modelling spiral galaxies, but can be easily adapted to describe the interaction of stellar light with dust grains in any environment, including in a quiescent gas cloud, as considered in this work. The model uses a Milky-Way-type dust consisting of a mixture of silicates, graphites, and PAH molecules with optical properties from J. C. Weingartner & B. T. Draine (2001) and B. T. Draine & A. Li (2007).

#### 2.2.1 $Q^{c}_{abs}$

The efficiency of absorption or absorptivity, $Q^{c}_{\rm abs}$, relates the energy absorbed by the dust in the clump to the energy of the photons incident on the clump.

In attempts to analogise the scenario where the clump is being heated by some ambient RF, the clump is placed at the centre of a spherical shell of stellar emissivity. Thus, the spherical dust cloud is uniformly illuminated from outside. The RT code (including scattering) is then applied for this specific configuration. The spatial resolution of the calculation is 0.1 pc.

The $Q^{c}_{\rm abs}$ of the clump is calculated as:

$$Q^{c}_{\rm abs}(\lambda) = \frac{4\pi \int_0^{r_t} e_{\rm abs}(\lambda, r) r^2\, {\rm d}r}{L_{\rm inc}(\lambda)}, \tag{12}$$

where $e_{\rm abs}$ is the volume density of energy absorbed by dust in the clump, and $L_{\rm inc}(\lambda)$ is the luminosity of the incident radiation at the surface of the clump, at $r = r_{\rm t}$. This specific geometric configuration is also ideal for calculating the heating of the grains in the clump, and therefore it is used to derive the IR emissivities of the clump, as detailed in Section 2.3.

#### 2.2.2 $Q^{c}_{sca}$ and $Q^{c}_{\rm ext}$

The scattering efficiency, $Q_{\rm sca}$ relates the energy scattered by the clump to the energy of the photons incident on the clump, while the extinction efficiency $Q^{c}_{\rm ext}$ relates both the energy absorbed and scattered by the clump to that of the incident photons. To calculate these two quantities, the clump is placed in front of an illuminating disc whose cross-sectional area is equal to that of the clump. RT calculations are performed to obtain the surface-brightness maps of the illuminating disc, with and without the attenuating clump. The extinction efficiency obtained from the RT calculations is then:

$$Q^{c}_{\rm ext}(\lambda) = \frac{\left(L_0^{\rm map}(\lambda) - L_{\rm att}^{\rm map}(\lambda)\right)}{L_0^{\rm map}(\lambda)}, \tag{13}$$

where the luminosities of the clump-attenuated and clump-free maps are $L_{\rm att}^{\rm map}$ and $L_0^{\rm map}$, respectively. The difference between $L_0^{\rm map}$ and $L_{\rm att}^{\rm map}$ is the attenuation map (due to the combined effect of scattering and absorption).

The scattering efficiency is then

$$Q^{c}_{\rm sca} = Q^{c}_{\rm ext} - Q^{c}_{\rm abs} \tag{14}$$

#### 2.2.3 Cross-sections of interaction

Relating $Q^{c}_{\rm abs}$, $Q^{c}_{\rm sca}$, and $Q^{c}_{\rm ext}$ to the geometric cross-section of the clump $\pi r_{\rm t}^2$, and the mass of the dust within the cloud, $M^{\rm c,s}$, gives the cross-sections of absorption $C^{c}_{\rm abs}$, scattering $C^{c}_{\rm sca}$, and extinction $C^{c}_{\rm ext}$ per unit mass of dust:

$$C^{c}_{\rm abs} = \frac{\pi\, r_{\rm t}^2\, Q^{c}_{\rm abs}}{M^{\rm c,s}}, \tag{15}$$

$$C^{c}_{\rm sca} = \frac{\pi\, r_{\rm t}^2\, Q^{c}_{\rm sca}}{M^{\rm c,s}}, \tag{16}$$

$$C^{c}_{\rm ext} = \frac{\pi\, r_{\rm t}^2\, Q^{c}_{\rm ext}}{M^{\rm c,s}}, \tag{17}$$

where $M^{\rm c,s}$ is defined in the previous subsection, by equation (10). $C^{c}_{\rm abs}$, $C^{c}_{\rm sca}$, and $C^{c}_{\rm ext}$ have units of [$\rm pc^2\, M_\odot^{-1}$].

Fig. 1 shows the optical properties of the clump, as derived from the methods described above, plotted together with the extinction curve for the diffuse dust, which, as mentioned before (Section 2.2), is a Milky-Way-type dust with the optical properties from J. C. Weingartner & B. T. Draine (2001) and B. T. Draine & A. Li (2007). It can be seen that the two extinction curves differ significantly at shorter wavelengths ($\lambda \lesssim 1\,\mu$m), then begin to converge at longer wavelengths ($\lambda \gtrsim 1\,\mu$m), as is expected for $\tau_{\rm B} = 1$ clumps.

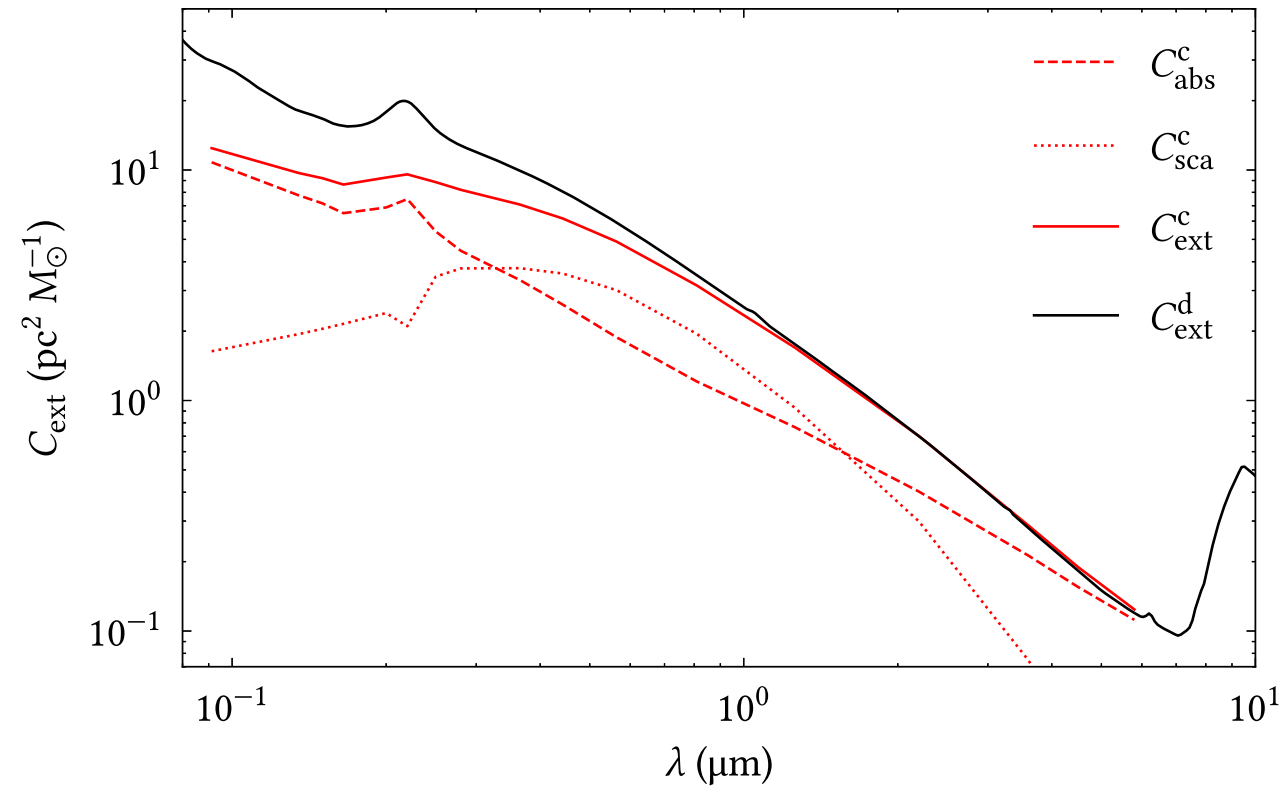

**Figure 1.** Absorption, scattering, and extinction cross-sections of a single clump, $C^{c}_{\rm abs}$, $C^{c}_{\rm sca}$, and $C^{c}_{\rm ext}$, plotted in red, with the dashed, dotted, and solid lines, respectively. Also plotted with a solid black line is the extinction cross-section $C^{d}_{\rm ext}$ of the diffuse dust used in these models, which is a Milky-Way-type dust, with the optical properties from J. C. Weingartner & B. T. Draine (2001) and B. T. Draine & A. Li (2007).

### 2.3 The infrared emissivities of the pseudo-grain

In the previous Section 2.2, the extinction properties of the pseudo-grain were derived. Here, the emission properties of the clump are calculated, to obtain a complete characterization of the pseudo-grain.

The IR emissivity of the pseudo-grain (the clump SED) is calculated for the geometric configuration from Section 2.2.1 used for the calculation of $Q^{c}_{\rm abs}$, with the clump being uniformly illuminated by a spherical shell of stellar emissivity that is concentric with the centre of the clump. As with $Q^{c}_{\rm abs}$, the IR emissivities are calculated with a 0.1 pc resolution. However, unlike the derivation of the absorption efficiency, which only depends on the geometry and optical depth of the dust cloud (for dust grains with fixed optical properties), the calculation of the dust emission SED also depends on the intensity and colour of the RFs heating the dust in which the clump is embedded. This is because unlike the energy absorbed, the heating of the grains does not linearly scales with the total luminosity of the heating sources, even for grains heated at equilibrium temperature, and has an even more complex dependence for grains stochastically heated. Therefore for each different ambient RF, an explicit calculation of the dust temperature needs to be performed at each position in the clump (and for each grain specie and size). For a comprehensive characterization, the dust emission SEDs are calculated for clumps placed in RFs with different strengths and colours to produce a corresponding library of dust emission SEDs for the clump. For ease of use in external applications (e.g. for the incorporation of the clumps into the large-scale modelling of the whole galaxy), two parameters – the intensity of the RFs and their UV-to-optical colour were found to be sufficient in taking into account the most relevant features of the RFs affecting the dust emission SEDs (see G. Natale et al. (2015)). In our current application, we use an equivalent parametrization, which is the 'UV strength' and an 'optical strength'.

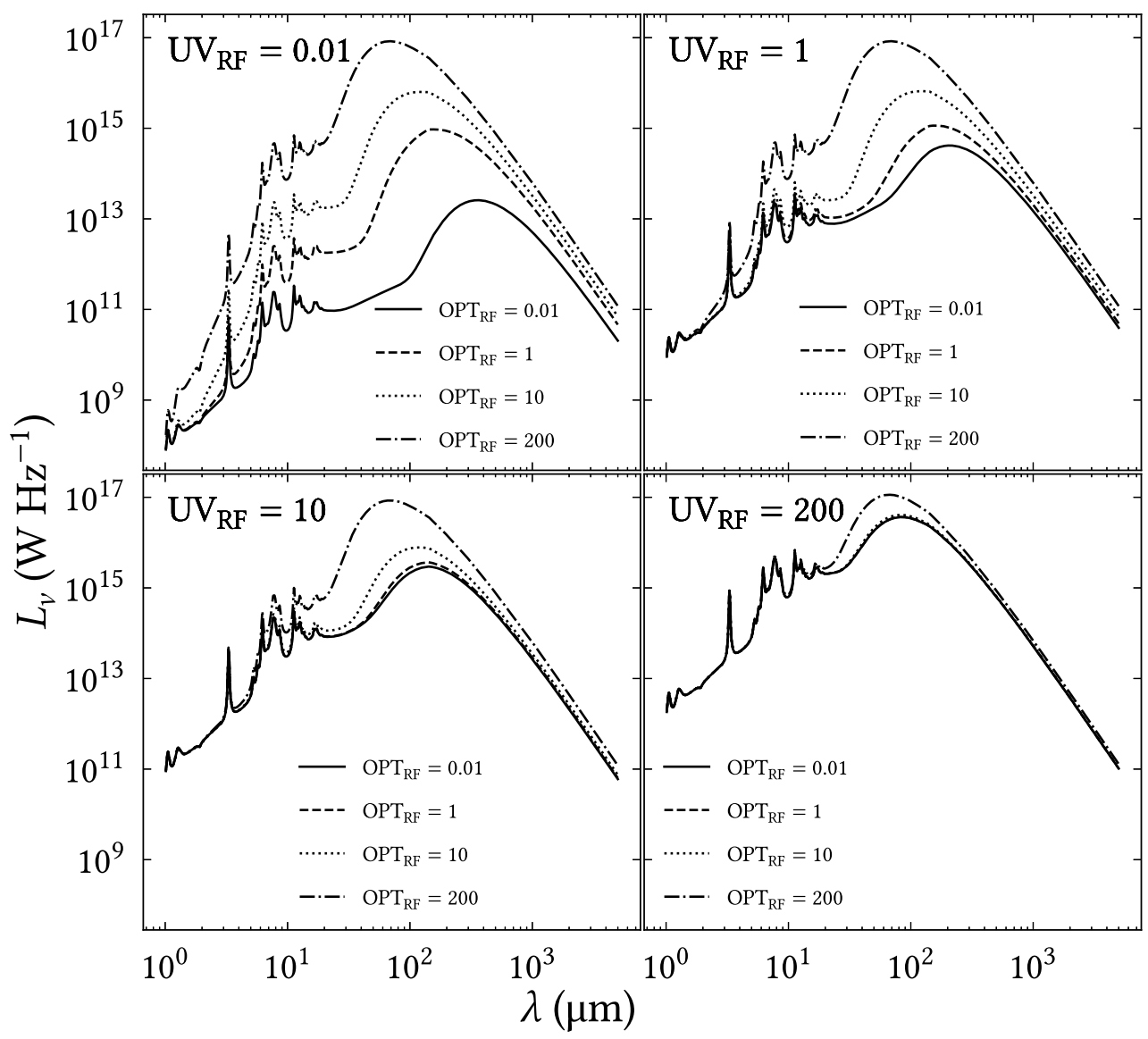

**Figure 2.** Clump dust emission SEDs, for clumps embedded in RFs with various values of the parameters $\mathrm{UV_{RF}}$ and $\mathrm{OPT_{RF}}$.

The detailed procedure is as follows: we use the RFs of the MW derived at the solar position in C. C. Popescu et al. (2017), which are referred to as 'unit' RFs. Two quantities are defined as

$$u_{\mathrm{UV}}^{\mathrm{unit}} = \int_{0.0912\,\mu\mathrm{m}}^{0.365\,\mu\mathrm{m}} u_\lambda^{\mathrm{unit}}\,\mathrm{d}\lambda \tag{18}$$

$$u_{\mathrm{opt}}^{\mathrm{unit}} = \int_{0.443\,\mu\mathrm{m}}^{5.800\,\mu\mathrm{m}} u_\lambda^{\mathrm{unit}}\,\mathrm{d}\lambda \tag{19}$$

where $u_\lambda^{\mathrm{unit}}$ is the RF energy density at the Solar position (C. C. Popescu et al. 2017) – 'the unit RF energy density', $u_{\mathrm{UV}}^{\mathrm{unit}}$ is the spectral integration over the UV of the unit RF – 'the unit UV RF energy', and $u_{\mathrm{opt}}^{\mathrm{unit}}$ is the spectral integration over the optical/NIR range of the unit RF – 'the unit optical RF energy'. The strengths of the UV and optical/NIR RFs are then expressed in units of $u_{\mathrm{UV}}^{\mathrm{unit}}$ and $u_{\mathrm{opt}}^{\mathrm{unit}}$,

$$\mathrm{UV_{RF}} = \frac{\int_{0.0912\,\mu\mathrm{m}}^{0.365\,\mu\mathrm{m}} u_\lambda\,\mathrm{d}\lambda}{u_{\mathrm{UV}}^{\mathrm{unit}}} \tag{20}$$

$$\mathrm{OPT_{RF}} = \frac{\int_{0.443\,\mu\mathrm{m}}^{5.800\,\mu\mathrm{m}} u_\lambda\,\mathrm{d}\lambda}{u_{\mathrm{opt}}^{\mathrm{unit}}}. \tag{21}$$

These are the two parameters of the library of dust emission clump SEDs. The following values were used for $\mathrm{UV_{RF}}$ and $\mathrm{OPT_{RF}}$: 0.001, 0.01, 0.1, 0.2, 0.6, 1, 2, 6, 10, 20, 60, 100, 160, 500, 800, and 1000. The minimum and maximum values were chosen to encompass the range of RF energy densities seen in local spiral galaxies, while the intervals were chosen to balance sampling with computational efficiency. The clump SEDs were calculated for RFs corresponding to all the combinations of $16 \times 16 = 256$ values for $\mathrm{UV_{RF}}$ and $\mathrm{OPT_{RF}}$.

Fig. 2 shows the clump dust emission SED for various combinations of $\mathrm{UV_{RF}}$ and $\mathrm{OPT_{RF}}$. One can see that if the RFs have a small UV component (e.g. top-left panel of Fig. 2), then an increase in $\mathrm{OPT_{RF}}$ will have a large effect in the dust emission SED: the peak of the dust emission will shift to shorter wavelengths, and the amplitude of the emission will strongly increase. On the contrary, if the RFs are UV dominated (bottom-right panel of Fig. 2, then the same increase in $\mathrm{OPT_{RF}}$ will produce very small effects.

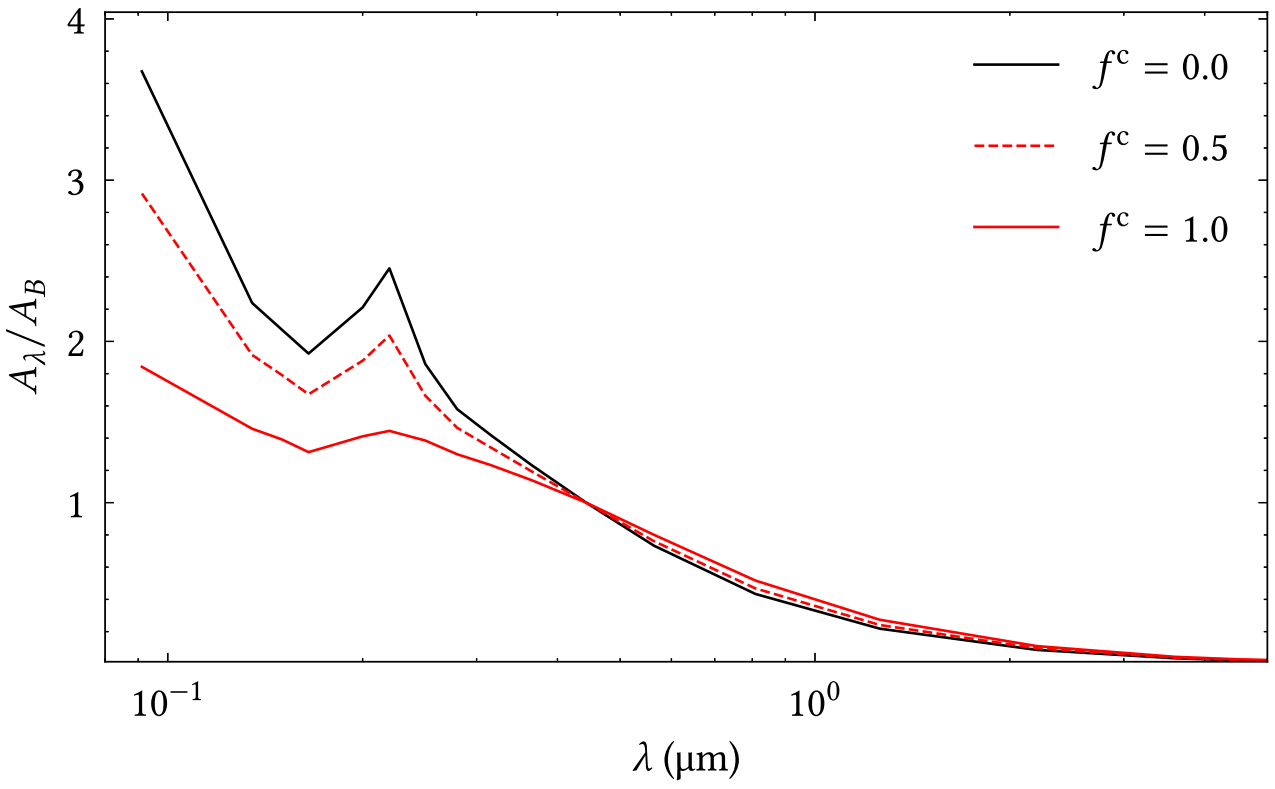

**Figure 3.** Extinction curves normalized in the $B$ band for three values of the fraction of mass in clumps: $f^{\mathrm{c}} = 0$ in solid black, $f^{\mathrm{c}} = 0.5$ in dashed red, and $f^{\mathrm{c}} = 1.0$ in solid red.

## 2.4 Incorporation of the pseudo-grain component into the large-scale, galactic calculation

The calculation of the absorption and scattering efficiencies of the clump, as well as of the IR emissivities enables the replacement of the macroscopic clump with a pseudo-grain with equivalent extinction and emission characteristics. Accordingly, the dust grain mixture, originally composed of Si, Gra, $\mathrm{PAH^0}$ and $\mathrm{PAH^+}$, was supplemented with a fifth component: the pseudo-grain.

### *2.4.1 The extinction law of the new dust model*

The RT code and the overall formalism remain unchanged from the previous (diffuse) version of the model. However, because the input albedo and extinction coefficients will be different, a different extinction law will be considered in the calculation, depending on the assumed fraction of mass in clumps, $f^{\mathrm{c}}$, defined as $f^{\mathrm{c}} = \frac{M_{\mathrm{c}}}{M_{\mathrm{tot}}}$, where $M_{\mathrm{c}}$ is the mass of dust in clumps and $M_{\mathrm{tot}}$ is the total dust mass. Examples of extinction laws for a few values $f^{\mathrm{c}}$ are given in Fig. 3. One can see that the new dust model will produce less extinction in the UV with respect to a pure diffuse model.

### *2.4.2 The dust emission SED*

The RT model of PT11, used in this work, explicitly calculates the heating of the dust grains, including the stochastic heating. Thus, for each grain size and composition, the probability distribution of dust temperature, and therefore the IR emissivity, is calculated based on the RFs heating the dust at each individual location in the galaxy. However, a similar calculation applied to the pseudo-grain would entail doing a separate calculation for the macroscopic clump heated by the RFs at each sampled position in the model galaxy. Taking into account that we usually sample around several hundred positions in the modelling of an individual galaxy, and that the modelling is an iterative process, whereby the RFs are recalculated for each iteration, and that each galaxy has different RFs, the explicit calculation for the pseudo-grain would be very time consuming. We remind the reader that the calculation of the dust emission SED of the pseudo-grain in itself is time consuming, since we calculate the heating of the grains inside the clump, including the stochastic heating, at resolutions of 0.1 pc in the clump. Because of all these reasons, we use the pre-computed SEDs from the library (derived in Section 2.3), that correspond to an RF that has the same strength in the UV and optical as the RF calculated in the simulation. Since the library of dust emission SEDs for the clump has a finite number of entries (256 SEDs), we take the two parameters of the library and perform a double interpolation in $UV_{RF}$ and $OPT_{RF}$ to derive the SED that most closely matches the RF in the simulation.

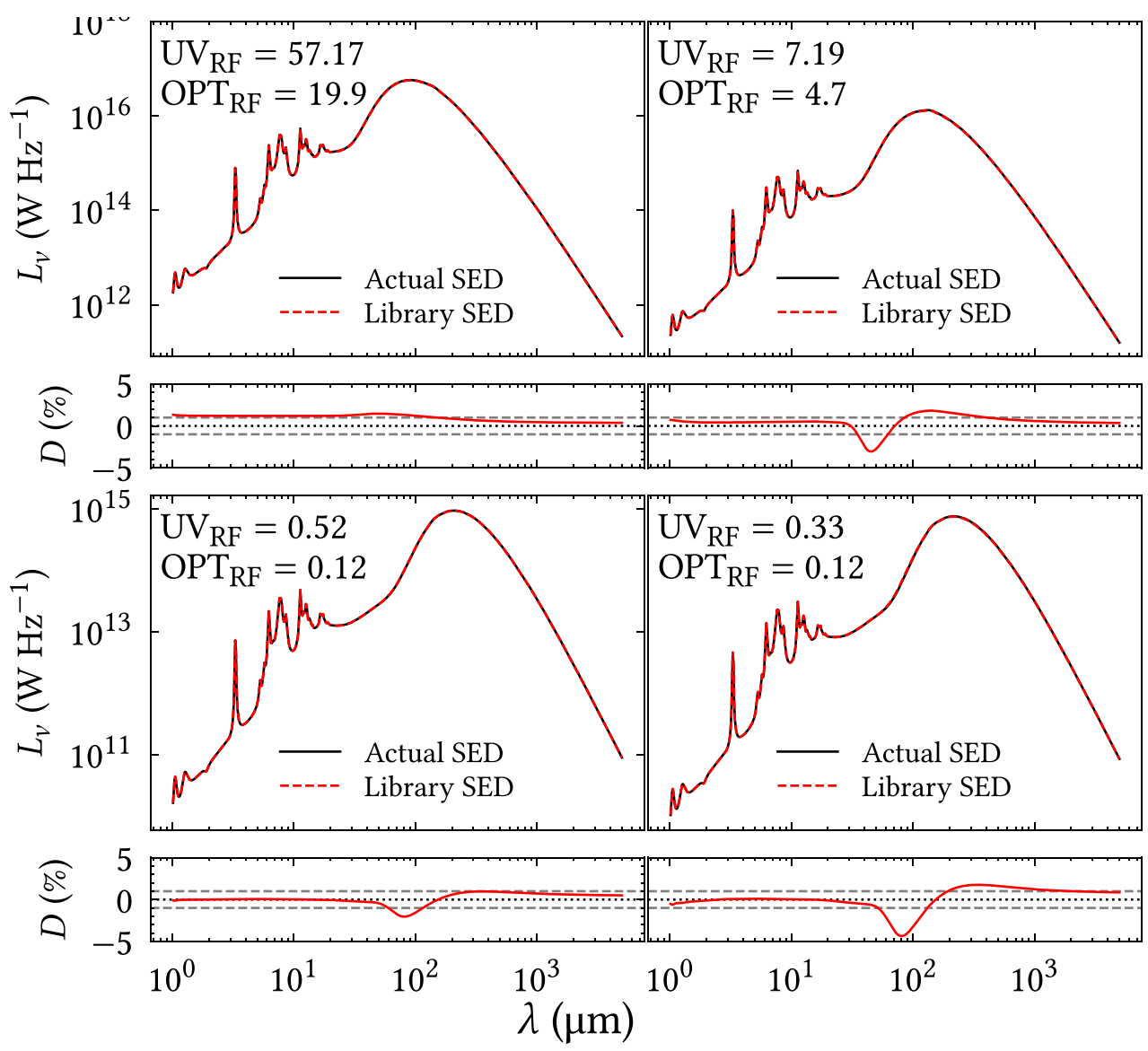

**Figure 4.** Comparison between explicitly calculated dust emission clump SEDs (solid black line) heated by the model RFs of M33, and the corresponding SEDs (dashed red line) derived using the model library approach. The examples correspond to four positions $(R, z)$ in the M33 galaxy: (from left to right, then top to bottom) $(R, z) = (0, 0)$ pc, $(R, z) = (0, 300)$ pc, $(R, z) = (7000, 0)$ pc, and $(R, z) = (7000, 300)$ pc. The lower panels plot the percent differences, with the dashed horizontal lines indicating ±1 per cent deviation. Each panel indicates the strength of the UV and optical part of the radiation fields, $UV_{RF}$ and $OPT_{RF}$, corresponding to the chosen spatial location in M33.

To demonstrate the efficacy of this approach we show in Fig. 4 examples of a comparison between explicitly calculated dust emission clump SEDs and those calculated via the pre-computed library approach described above. This comparison was done for four regions in M33; $(R, z) = (0, 0)$ pc (top left), $(R, z) = (0, 300)$ pc (top right), $(R, z) = (7000, 0)$ pc (bottom left), and $(R, z) = (7000, 300)$ pc (bottom right). In each case, we find that the difference is typically within one per cent. The interpolated SEDs are then scaled according to the dust density (of the clumpy component) at each position in the modelled galaxy.

## 3 THE LARGE-SCALE MODEL OF GALAXIES

The large-scale model used to simulate a galaxy as a whole is the axisymmetric RT model of PT11, which describes the dust opacity and stellar emissivity for spiral galaxies from the UV to submm wavelengths using parametrized analytic functions. Full details of the model and the physical justification for the choice of parameters and methodology can be found in PT11, while a detailed description of the fitting procedure used for modelling face-on and edge-on galaxies is given in the subsequent application papers, TP20; M. T. Rushton et al. (2022), G. Natale et al. (2022), C. J. Inman et al. (2023), and D. Pricopi et al. (2025). Here, we only give a very brief description of the main elements of the model and fitting methodology.

As mentioned before, the model considers the absorption and anisotropic scattering of stellar photons with dust grains of various size and chemical composition. Following PT11, all galaxies are modelled using a stratified 'main disc' made of a disc of old stars (the stellar disc), a thinner disc of younger stars (the thin stellar disc), a disc of dust (the dust disc), and a thin disc of dust associated to young stars (the thin dust disc). Some galaxies are modelled with additional nuclear, inner and outer discs of both stars and dust, with, in most cases, the same stratification as the main disc (see TP20). Finally, most models have a bulge of stars, for which there is no stratification. The stellar volume emissivity and the dust density distribution for each disc component are described by the following generic formula:

$$w_j(R, z) = \begin{cases} 0, & \text{if } R < R_{\text{tin,j}} \\ A_j \left[ \frac{R}{R_{\text{in,j}}} (1 - \chi_j) + \chi_j \right] \\ \quad \times \exp\left(-\frac{R_{\text{in,j}}}{h_j}\right) \exp\left(-\frac{z}{z_j}\right), & \text{if } R_{\text{tin,j}} \le R < R_{\text{in,j}} \\ A_j \exp\left(-\frac{R}{h_j}\right) \exp\left(-\frac{z}{z_j}\right), & \text{if } R_{\text{in,j}} \le R \le R_{\text{t,j}} \end{cases} \tag{22}$$

with:

$$\chi_j = \frac{w_j(0, z)}{w_j(R_{\text{in,j}}, z)} \tag{23}$$

where j is 's' (stellar) or 'd' (dust), $R$ and $z$ are the radial and vertical coordinates, $h_j$ and $z_j$ are the scale length and scale height, respectively, $A_j$ is a constant (amplitude), and $\chi_j$ describes a linear slope of the radial distributions from the inner radius, $R_{\text{in}}$, to the centre of the galaxy. Note that for the case where $R_{\text{tin,j}} > 0$, $\chi_j$ is defined using the extrapolation of the emissivity to the centre rather than the truncated profile. $R_{\text{tin}}$ and $R_{\text{t}}$ are the inner and outer truncation radii, respectively. Equations (22) and (23) are wavelength dependent.

The bulge is described by the 2D Sérsic distribution (J. L. Sérsic 1963, 1968). Since there is no exact analytic solution for the deprojection of this distribution into 3D, several approximations have been proposed in the literature. In these calculations, the volume emissivity as derived in G. Natale et al. (2022) is used:

$$w_{\text{bulge}}(R, z) = w(0, 0) \sqrt{\frac{b_s}{2\pi}} \frac{(a/b)}{R_e} \eta^{(1/2n_s)-1} \exp\left(-b_s \eta^{1/n_s}\right) \tag{24}$$

with:

$$\eta(R, z) = \frac{\sqrt{R^2 + z^2(a/b)^2}}{R_e} \tag{25}$$

where $w(0, 0)$ is the stellar emissivity at the centre of the bulge, $R_e$ is the effective radius of the bulge, $a$ and $b$ are the semimajor and semiminor axes of the bulge, respectively, and $b_s$ is a constant which depends on the value of the Sérsic index $n_s$, given by the approximation (L. Ciotti & G. Bertin 1999):

$$b_s \approx 2n_s - \frac{1}{3} + \frac{4}{405n_s} - \frac{46}{25515n_s^2}. \tag{26}$$

### 3.1 Stellar components

#### 3.1.1 The stellar disc

The stellar disc is made up of the old stellar populations and primarily emits in the optical and NIR wavelengths. It is described by the four geometrical parameters: $h_s^{disc}$, $z_s^{disc}$, $R_{in,s}^{disc}$, and $\chi_s^{disc}$, and one amplitude parameter, $A_s$, or equivalent the total spatially integrated luminosity $L^{disc}$. $L^{disc}$ is allowed to be a free parameter for each wavelength for which data are available, usually corresponding to $U, B, V, I$, or to the $u, g, r, i, z$ SDSS (Sloan Digital Sky Survey) bands, and to the NIR 2MASS $J$, $K$ bands, and *Spitzer*$I1$, $I2$, $I3$ bands (see e.g. Section 4.1 for the description of the data used for NGC 891). $h_s$ is also wavelength dependent and treated as a free parameter at each modelled wavelength. $\chi_s^{disc}$ and $R_{in,s}^{disc}$ are usually wavelength independent, and fitted (former) or fixed (latter) from data. If the galaxy being modelled is edge-on and resolved in vertical direction, the vertical distribution can be fit, thus $z_s$ is allowed to be wavelength dependent and fitted for the previously mentioned optical and NIR bands. In face-on galaxies $z_s$ is fixed from general trends found for edge-on galaxies.

#### *3.1.2 The bulge*

If present, an ellipsoidal bulge containing old stellar populations is also included in the model. Similarly to the stellar disc, the bulge emits predominantly in the optical/NIR range. The model geometric parameters $R_e$, $\frac{a}{b}$, and $n_s$ are usually kept wavelength independent. As with the stellar disc, the spatially integrated luminosity of the bulge, $L^{bulge}$, is a free parameter derived at each optical/NIR wavelength considered in this work.

#### *3.1.3 The thin stellar disc*

The thin stellar disc hosts the young stellar population. Primarily emitting in the UV regime, the young stellar disc contributes the majority of the UV output of the galaxy and has a smaller scale height than that of the stellar disc. The thin stellar disc is described by the four geometrical parameters: $h_s^{tdisc}$, $z_s^{tdisc}$, $R_{in,s}^{tdisc}$, and $\chi_s^{tdisc}$, and one amplitude parameter, $A_s^{tdisc}$, or equivalently the spatially integrated luminosity $L^{tdisc}$. In face-on galaxies, all these parameters are usually fitted at the wavelengths for which data are available, in particular in the *GALEX* far-UV (FUV) and near-UV (NUV) bands, where the thin disc is the only contributor (in the model the stellar disc and bulge do not contribute in the UV bands). In edge-on galaxies, the thin stellar disc is completely obscured by dust, and its UV luminosity and main geometric parameters are constrained from the FIR imaging data, while its UV spectral distribution is fixed to the spectral template from PT11.

$L^{tdisc}$ is related to the star formation rate (SFR) using eqs (16)–(18) from PT11. These are valid for application to edge-on galaxies, where the spectral templates from PT11 are used for the UV emissivity. For face-on galaxies, the equations are altered to take into account the different intrinsic SEDs derived from fitting the spectral luminosity density at each observed wavelength.

### 3.2 Dust components

#### *3.2.1 The dust disc*

The dust disc describes the large scale diffuse distribution of dust associated with the stellar population in the stellar disc and the H I gas. The dust disc is usually more radially extended than the old stellar disc (E. M. Xilouris et al. 1999), while having a scale height $z_d^{disc}$ larger than that of the thin stellar disc $z_s^{tdisc}$, and smaller than that of the stellar disc $z_s^{disc}$. The dust disc is described by the geometrical parameters $h_d^{disc}$, $z_d^{disc}$, $R_{in,d}^{disc}$, and $\chi_d^{disc}$ and the amplitude parameter, the face-on optical depth of the $B$ band at the inner radius $\tau_B^{f,disc}\,(R_{in,d}^{disc})$.

#### *3.2.2 The thin dust disc*

The thin dust disc represents the diffuse dust distribution associated with the young stellar population and is described by the geometrical parameters $h_d^{tdisc}$, $z_d^{tdisc}$, $R_{in,d}^{tdisc}$, and $\chi_d^{tdisc}$, and the amplitude parameter, the face-on optical depth of the $B$ band at the inner radius $\tau_B^{f,tdisc}\,(R_{in,d}^{tdisc})$. Following PT11, the geometric parameters are usually fixed to that of the young stellar disc.

### 3.3 Fitting procedures

#### *3.3.1 Face-on fitting procedure*

For face-on galaxies, we follow the fitting procedure described in TP20. We use azimuthally averaged surface-brightness profiles obtained from model images to fit the corresponding observed profiles. For the very nearby galaxies used in this work several morphological components were needed to account for the complex structure. This was identified in the observed profiles, which show several well-defined breaks and distinct slopes, hinting at several concentric exponential discs with different scale lengths. Many parameters, like inner and outer truncation radii, $R_{tin}$ and $R_t$, inner radii, $R_{in}$, were identified from the appearance of the profiles and fixed from on-set without further iterations. Occasionally minor adjustments were needed during the development of the solution, but usually they were not subject to iterations. When this is the case they are identified as fixed parameters (from data) in the tables presented in this work (see Appendix B).

In the optical and NIR wavelengths, initial guess parameters for the stellar discs and bulge can be obtained at each modelled wavelength by assuming a dust-free case. This then allows the main two parameters, total luminosity ($L_s^{disc}$ and $L_s^{bulge}$) and length ($h_s^{disc}$ and $R_e$) to be identified for each morphological component separately at each wavelength, and kept fixed during the main iteration process. In the NIR, these initial guess parameters do not need further amendment when dust is included, due to the very optically thin nature of the solution in these bands.

The main procedure involves the fitting of the pair of two main parameters, amplitude and length, for the thin stellar disc ($L_s^{tdisc}$, $h_s^{tdisc}$) and for the dust disc ($\tau_B^{f,disc}$, $h_d^{disc}$). When needed, a third parameter, the inner linear slope $\chi^{tdisc}$ is also fitted in this process. This parameter can mimic central depressions in the dust distribution and/or UV emissivity, or, for outer discs, provide a smooth overlap between different morphological components. We use the *GALEX* NUV and the *Herschel*500 μm bands to constrain these parameters. The fitting involves running the RT code several times until the NUV and 500 μm profiles are well fitted. In these iterations, the UV spectral template from PT11 is used and a fixed scale length for the UV stellar emissivity, with the optical/NIR parameters fixed to the initial guess. Once a good energy balance is achieved, final adjustments are made to also fit the detail of the FUV and optical profiles. We note that, apart from the 500 μm band, the rest of the dust emission SED is predicted rather than fitted.

The goodness of fit was evaluated using the residuals $D$ between the surface-brightness profiles of the observed data and the model, as well as a reduced chi-squared which becomes more relevant when finalizing the model solution (see details in TP20).

**Table 1.** Summary of observational data used to model NGC 891.

| Telescope | Filter/ Instrument | Wavelength (μm) | Flux (Jy) | pixel scale (arcsec) | (pc) | Calibration error (per cent) |
|---|---|---|---|---|---|---|
| Skinakas | $B$ | 0.443 | $0.17 \pm 0.02$ | 0.75 | 35 | |
| | $V$ | 0.564 | $0.29 \pm 0.03$ | 0.75 | 35 | |
| | $I$ | 0.809 | $0.84 \pm 0.08$ | 0.75 | 35 | |
| 2MASS | $J$ | 1.2 | $1.79 \pm 0.05$ | 1 | 46 | 3 |
| | $K_s$ | 2.2 | $2.62 \pm 0.08$ | 1 | 46 | 3 |
| *Spitzer* | IRAC | 3.6 | $1.84 \pm 0.06$ | 0.6 | 28 | 3 |
| | IRAC | 4.5 | $1.27 \pm 0.04$ | 0.6 | 28 | 3 |
| | IRAC | 5.8 | $3.21 \pm 0.10$ | 0.6 | 28 | 3 |
| | MIPS | 24 | $5.32 \pm 0.06$ | 1.5 | 69 | 1 |
| *Herschel* | PACS | 70 | $106.7 \pm 14.4$ | 1.4 | 64 | 5 |
| | PACS | 100 | $227.7 \pm 28.0$ | 1.7 | 78 | 5 |
| | PACS | 160 | $294.1 \pm 34.4$ | 2.85 | 131 | 5 |
| | SPIRE | 250 | $158.7 \pm 22.5$ | 6 | 276 | 6 |
| | SPIRE | 350 | $66.6 \pm 9.4$ | 8 | 368 | 6 |
| | SPIRE | 500 | $23.5 \pm 3.2$ | 12 | 553 | 6 |

#### 3.3.2 *Edge-on fitting procedure*

The fitting procedure follows the formalism from G. Natale et al. (2022), used to model the MW, and is similar to that used to fit face-on galaxies, as described in Section 3.3.1, but has a few differences from either the MW model or the face-on case.

First, unlike in face-on galaxies, in the edge-on view the vertical profiles can be used to derive the scale heights of stars and dust. In addition, because of the large line-of-sight optical depth, the UV emission from the thin stellar disc is completely obscured by dust, and, as in the MW, the UV emission is derived to best fit the FIR emission, in particular at the peak of the dust emission SED (G. Natale et al. 2022). The lack of direct constraints in the UV and the use of FIR data is well documented in the previous modelling of NGC 891 in PT11.

The fitting procedure also differs from that applied to the MW, because, unlike for our Galaxy, the FIR/submm vertical profiles are not resolved. In the MW, the *Planck* profiles were used to constrain the vertical distribution of the dust. For edge-on galaxies, the dust scale heights are derived from fitting the optical images instead. This means that there is the need to scan a larger parameter space, involving the additional parameters $z_s^{disc}$ and $z_d^{disc}$.

In edge-on galaxies, horizontal- and vertical-surface-brightness profiles are used for fitting model to observations.

## 4 APPLICATION OF THE CLUMPY MODEL TO THE PROTOTYPE EDGE-ON SPIRAL NGC 891

As discussed in Introduction, the solution obtained for NGC 891 by C. C. Popescu et al. (2000, 2011) matched the FIR data available at the time, but produced an overly prominent dust lane in the NIR $J$ and $K_s$ bands. In principle, this earlier model could be used for comparison with the new clumpy model. However, the previous model was fitted using the *Infrared Space Observatory*'s (*ISO*; M. F. Kessler et al. 1996) Photopolarimeter (ISOPHOT) data (D. Lemke et al. 1996), which, at that time, were the only ones available to cover the peak of the dust emission SED in the FIR. With the advent of the *Herschel Space Observatory* (G. L. Pilbratt et al. 2010) data, it became apparent that there was a discrepancy between the fluxes measured by the two missions, with the *Herschel* flux densities exceeding the *ISO* measurements by a factor of approximately 1.5 (see S. Bianchi & E. M. Xilouris 2011). This discrepancy does not seem to be confined to NGC 891 only, but also to other resolved galaxies. Given that most of the recent work in the field has been calibrated on *Herschel* rather than on *ISO* data, including this work on modelling NGC 891 with clumps, it was decided to refit the previous model as well, using the updated *Herschel* data. This section presents both the updated diffuse and the clumpy model of NGC 891.

To be consistent with the previous papers on NGC 891 (C. C. Popescu et al. 2000, 2011), a distance of 9.5 Mpc and an inclination of $i = 89.7^{\circ}$ (E. M. Xilouris et al. 1998, 1999) is adopted in this work.

### 4.1 Observational data

Following from previous work on NGC 891, all available panchromatic imaging data spanning from the optical to submm range were used, while UV imaging data were not used. This is because, in an almost perfectly edge-on galaxy like NGC 891, most UV emission from the disc is completely obscured by dust, thus making it impossible to rely on direct constraints from this wavelength range. In Table 1 (columns 1–3), the instruments and corresponding wavelengths of the data used in this work for modelling NGC 891 are listed.

#### 4.1.1 *Skinakas observatory*

The optical images in the $B$, $V$, and $I$ bands obtained using the 1.3 m telescope at the Skinakas Observatory by E. M. Xilouris et al. (1998) were utilized in this work. The pixel scale of these data is 0.75 arcsec corresponding to 35 pc at the assumed distance.

#### 4.1.2 *2MASS*

NGC 891 was imaged in the NIR bands $J$, $H$, and $K_s$ as part of the 2MASS (M. F. Skrutskie et al. 2006) Large Galaxy Atlas (T. H. Jarrett et al. 2003). These data have a pixel scale of 1.0 arcsec corresponding to 46 pc at the assumed distance. Only the $J$ and $K_s$ bands were utilized in this study as the effective wavelength of the $H$ band is not included in this model.

#### 4.1.3 Spitzer

The *Spitzer Space Telescope* conducted imaging in the IR with two instruments, the Infrared Array Camera (IRAC; G. G. Fazio et al. 2004) and the Multiband Imaging Photometer for Spitzer (MIPS; G. H. Rieke et al. 2004). IRAC was a four channel camera that simultaneously imaged the 3.6, 4.5, 5.8, and 8 μm bands. The pixel scale of the IRAC images are 0.6 arcsec corresponding to 28 pc. MIPS imaged from the MIR (mid-infrared) to FIR (24, 70, and 160 μm). In this study, only the 24 μm images were utilized as higher quality images for the 70 and 160 μm bands are used instead. The pixel scale of the 24 μm data are 1.5 arcsec (69 pc).

#### 4.1.4 Herschel

The *Herschel Space Observatory* imaged NGC 891 in the FIR and submm. The Photodetecting Array Camera and Spectrometer (PACS; A. Poglitsch et al. 2010) provided image data at 70, 100, and 160 μm with pixel scales of 1.40, 1.70, and 2.85 arcsec, corresponding to 64, 78, and 131 pc, respectively. The Spectral and Photometric Imaging Receiver (SPIRE; M. J. Griffin et al. 2010) observed NGC 891 at the wavelengths 250, 350, and 500 μm with pixel scales of 6, 8, and 12 arcsec, corresponding to 276, 368, and 553 pc, respectively. All these data are used in modelling the radial profiles of NGC 891, but only the 70 μm data (the only resolved one in minor axis), to model the vertical surface brightness.

#### 4.1.5 Data post-processing

Data from 2MASS, MIPS, and *Herschel* were found on NASA/IPAC Extragalactic Database.[1] The optical data were obtained by E. M. Xilouris et al. (1998). The IRAC data were downloaded from the Spitzer Heritage Archive.[2] While these data are calibrated from their raw form, further processing is required to make them useful for use in these models.

#### 4.1.6 Background removal

The cumulative sum of the averaged surface brightness of the data in the vertical direction was examined to determine the height above the disc at which the emission of the galaxy falls below the background level. A visual inspection of each map was done to ensure that the emission associated with NGC 891 does not extend beyond this point, which was found to be $|z| = 6$ kpc. The region $6 \leq |z| \leq 9$ was chosen to be the background sample. This ensured that a large, representative region of the sky was being sampled, while not extending beyond the dimensions of the image. The average background flux per pixel is then calculated by taking the mean surface brightness through six equally spaced strips in the region $6 \leq |z| \leq 9$. Finally, the average background flux is subtracted from each pixel in the image.

#### 4.1.7 Foreground star removal

Foreground stars are visible in the UV/optical/NIR wavelengths and can influence the measured values of emission within the galaxy. It is therefore important to mask these stars. An automatic masking routine was used which operates on a median-subtracted image, where each pixel of the median map is computed using a $3 \times 3$ pixel filter. Although this filter width is intentionally narrow, set to be no larger than the PSF (point spread function) width, detection sensitivity to faint point sources is instead governed by a minimum flux threshold, which is set for each band. A pixel that is a candidate for masking must exceed this threshold on the median-subtracted image, and must also be brighter than all surrounding pixels within a 1-pixel radius, ensuring that it is a local peak. A separate bias map is produced with $21 \times 21$ pixels, that provides an estimate of the local background surface brightness. A candidate is only retained if its flux exceeds 1.5 times this local background level, which mitigates against false detections due to random surface-brightness fluctuations in the galaxy. A final contrast criteria, measuring the ratio of peak flux to the mean flux of the immediate surroundings on the median-subtracted image, ensures that resolved sources such as H II regions are not erroneously masked. Sources passing all criteria are masked over a circular aperture whose radius scales logarithmically with flux, ranging between a minimum and maximum 150 and 400 pc, respectively, set in physical units to ensure a consistent masking scale across all bands.

After this, each mask is manually inspected for regions that may have been missed, or erroneously masked, such as H II regions. H II regions can appear as point sources (similarly to foreground stars) and thus can be falsely masked by this automated process. To help with identification of these H II regions additional colour information was used. As H II regions comprise localized hot dust they appear brightly in the MIR, thus the 24 μm data were examined to help identify the H II regions and unmask them in the optical/NIR data.

#### 4.1.8 Extraplanar emission

As shown in multiple studies (R. A. Pildis, J. N. Bregman & J. M. Schombert 1994; R. J. Rand 1996; C. G. Hoopes, R. A. M. Walterbos & R. J. Rand 1999; Y.-S. Jo et al. 2018), NGC 891 exhibits extraplanar stellar emission, which is not included in this model. Therefore, in order to compare model with observations, this emission needs to be subtracted from the available optical/NIR images used in this study. The level of the extraplanar emission was estimated from a region beyond the main vertical extent of the disc and bulge. Since the model stellar disc has a maximum scale height of $z_s = 500$ pc, and the model bulge a projected effective radius (on the vertical direction) of $R_e \times \frac{b}{a} = 1600 \times 0.47 = 752$ pc, it is reasonable to assume that beyond 2 kpc the emission from the main morphological components of NGC 891 becomes negligible. This assumption is also validated by a visual inspection of the observed vertical profiles of NGC 891. Thus, the emission in the region $2 < |z| < 3$ kpc was taken to be dominated by extraplanar emission, and subtracted from images using a method similar to that used for the MW in C. C. Popescu et al. (2017) and G. Natale et al. (2022). Since the surface brightness decreases with horizontal distance, the subtraction of the extraplanar emission was done separately for each bin in horizontal distance. Using the same horizontal binning as that used to produce average profiles, the extraplanar emission was calculated to be the average of all the pixels with $2 < z < 3$ kpc within each horizontal bin, and subtracted correspondingly from the image.

[1] https://ned.ipac.caltech.edu
[2] https://irsa.ipac.caltech.edu/data/SPITZER/docs/spitzerdataarchives/

For wavelengths $\lambda \geq 70\,\mu$m, no extraplanar emission was detected in available dust emission images (S. Bianchi & E. M. Xilouris 2011). In any case, the PSF of the observing instrument is coarse enough that true emission from the plane of the galaxy would be seen contributing to any emission at high vertical distances. Because of this no attempt was made to extract extraplanar emission at these wavelengths.

### 4.2 An updated diffuse model of NGC 891

As discussed at the beginning of this section, the main reason to refit the diffuse model of NGC 891 was justified by the need to calibrate it on the same (new) data as the clumpy model. In addition, recent work on modelling face-on galaxies (J. J. Thirlwall et al. 2020; M. T. Rushton et al. 2022; C. J. Inman et al. 2023; D. Pricopi et al. 2025) and the MW (G. Natale et al. 2022) has highlighted the need to incorporate a more detailed geometrical and morphological description. Because of all these reasons it was imperative to refit the diffuse model of NGC 891, to provide a more meaningful comparison with the clumpy model.

Inspection of the averaged horizontal profiles of dust emission (Fig. 5 and Figs A3 and A4) shows that there is a local maximum at $R = 3$ kpc, suggesting a possible inner disc in the dust distribution, as well as an inner disc of young stars that could contribute to the dust heating in the inner regions. This was also investigated by S. Katsioli et al. (2023) who found similar emission enhancements at $\pm 3$ kpc. Using data spanning 24 μm to 1.15 mm, they propose a scenario where the spiral arms lie on either side of the central bulge region when viewed through the line of sight. These locations at $\pm 3$ kpc mark the interface between where the line-of-sight emission is dominated by the bulge region, and the spiral arms. We thus considered a diffuse model of NGC 891 consists of three morphological components: a bulge, a main disc, and an inner disc. As we will see later, the inner disc does not produce the right effect in the fits of the diffuse model, however it improves the fits when clumps are introduced (see Section 4.3). In order to keep a consistent approach for both the diffuse and clumpy model, we retain the inner disc component in both cases.

The main disc morphological component has the typical vertical stratification of the PT11 model, with a stellar disc, a thin stellar disc, a dust disc, and a thin dust disc. The inner disc morphological component has only a thin stellar disc, a dust disc, and a thin dust disc. The fits of this updated diffuse model to the surface-brightness profiles (from the best-fitting solution) at selected wavelengths are shown in Fig. 5. The corresponding surface-brightness maps of the data, model, and their residuals are shown in Fig. 6. The fits and maps at all wavelengths used in this study can be found in Figs A1–A5. The spatially integrated SED of the model is shown in Fig. 7. The geometric parameters of this fit are given in Table 2 (column 'd'), and Tables 3 and Table 4 (upper section 'd').

In comparison with the previous diffuse model of NGC 891, the more complex morphology allows for a somewhat better fit to the surface-brightness profiles, although it still does not resolve the problem with the prominence of the dust lanes. Inspection of Fig. 5 shows that in the *J* band there is a very strong trough in the model vertical profiles, around the mid-plane. By contrast, the observed profiles show only a mild trough. In the $K_s$ band, the observed vertical profiles, in particular in the $2 < |R| < 5$ kpc region, do not show any trough, yet the model predicts one.

Similarly, the *J*-band model images from Fig. 6 show a more pronounced dust lane than observed, with the residual images exhibiting a strong residual dust lane. In the $K_s$, the observed images do not exhibit a dust lane, except for a few dark dots around the mid-plane. Yet, the model predicts a distinct dust lane, also strongly visible in the residual maps. In fact at all wavelengths shorter than 3.6 μm, the model fits show an overestimation of the dust lane (see Fig. A1), and there are some mid-plane residuals at even longer wavelengths. It is clear that the updated diffuse model does not solve the energy balance problem.

It is also apparent from the SB profiles in Fig. A3 that in the 24–100 μm range the model overestimates the emission in the central regions. On the other hand, the overall spatially integrated emission SED (Fig. 7) shows a good match between the data and model, with a slight overestimation in the MIR. None the less, the dust-deattenuated SED predicts that even in $K_s$ and IRAC 3.6 μm bands the line-of sight attenuation is severe, possibly indicating an overestimation of dust attenuation in the diffuse model.

We note that the best-fitting solution for the diffuse model has 2/3 opacity in the thin dust disc and 1/3 in the dust disc. This is the original ratio used in C. C. Popescu et al. (2000, 2011). Solutions with different values of this ratio were also explored, but these resulted in poorer fits. In particular, solutions with 2/3 opacity in the dust disc and 1/3 in the thin dust disc cannot reproduce the shape of the vertical profiles, with the maximum SB being at too great a vertical distance from the mid-plane.

The *B*-band central face-on dust opacity $\tau^f_B$ for the best-fitting model is 4.4, which is even higher than the previous value of 3.5 found in PT11. This is mainly attributed to the difference between the observational data used to fit the model: ISOPHOT versus *Herschel* data. As remarked in the preamble to Section 4, the *Herschel* fluxes are systematically larger than the ISOPHOT ones, requiring a boost not only in the luminosity of the heating sources, but also in the dust opacity. The optically thick solution contrasts to the optically thin value of $\tau^f_B \sim 1$ found by E. M. Xilouris et al. (1998, 1999) in the optical fits of NGC 891, suggesting again an energy balance problem.

### 4.3 The clumpy model of NGC 891

Using the same morphological components and the same number of geometric parameters as in the diffuse model case, the new clumpy model was fitted to the data. This time an additional free parameter was needed, the fraction of mass in clumps $f^c$. The best model fit was found for $f^c = 0.5$. It was also found that most of the geometric and luminosity parameters changed in the new model, with varying degrees of change, as discussed in Section 4.4.

The fits to the surface-brightness profiles from the best-fitting solution for the clumpy model are shown for a selection of wavelengths in Fig. 8, and for all wavelengths in Figs A6–A9. The surface-brightness maps of the data, model and their residuals are shown at a selection of wavelengths in Fig. 9, and for all optical/NIR wavelengths in Fig. A10. The spatially integrated SED of the model is shown in Fig. 10. The geometric parameters of this fit are given in right-hand column of Table 2 and lower half of Table 4.

Inspection of the fits to the surface-brightness profiles (Fig. 8 and Figs A6–A9) shows very good agreement with the data. In particular, the *J*- and $K_s$ -band vertical profiles are well fitted. This is also seen in the model images from Fig. 9, with the residual maps showing no prominent residual dust lanes. Overall, the fits to all wavelengths are improved and lie within 20 per cent errors. This is the first model of NGC 891 that provides consistency with the panchromatic imaging data and solves the energy balance problem.

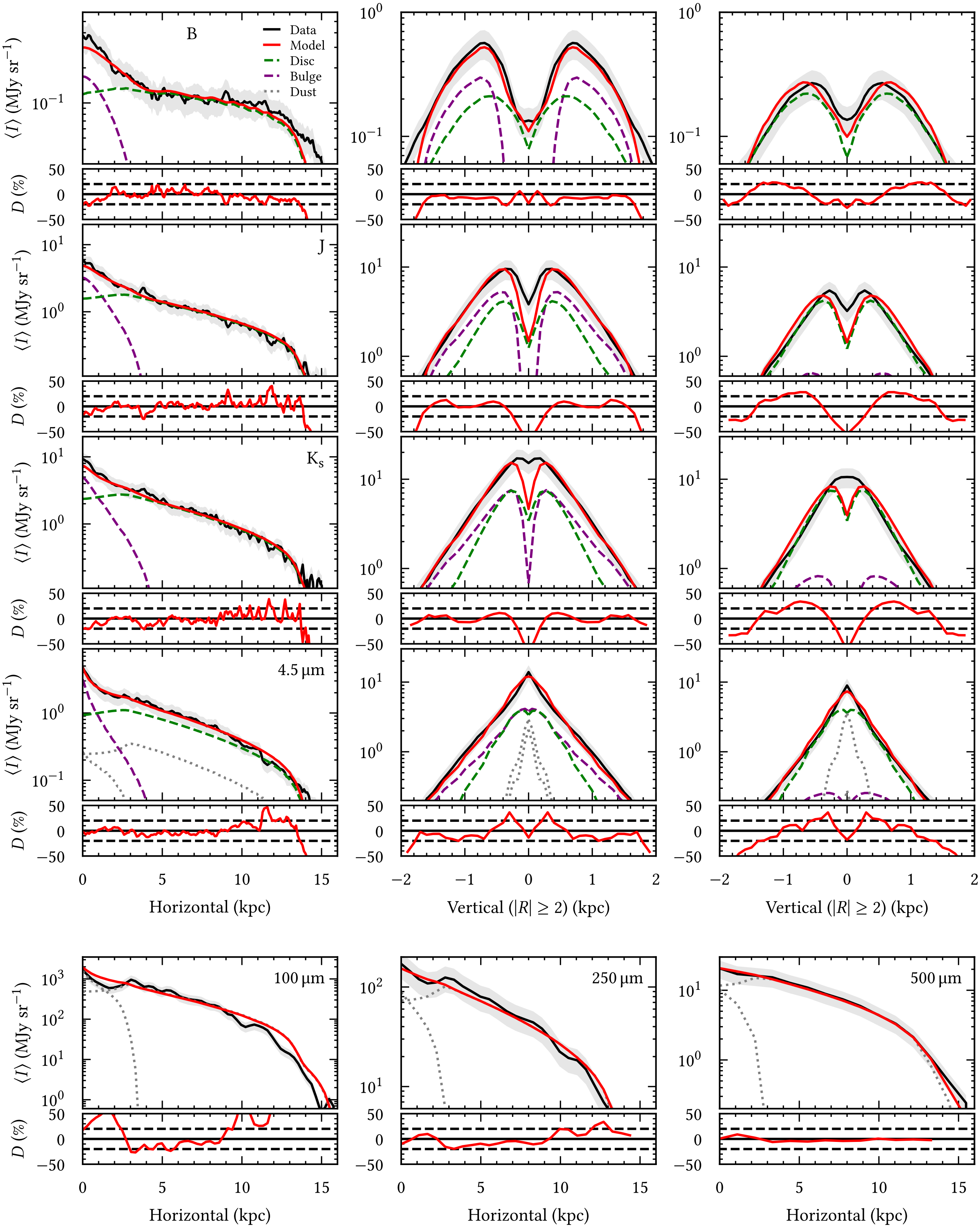

**Figure 5.** First four rows: averaged horizontal SB profiles (mirrored about the central vertical axis, LH panels) and averaged vertical SB profiles (middle and RH panels) of the pure diffuse model of NGC 891 at selected wavelengths in the optical/NIR. Bottom row: averaged horizontal SB profiles (mirrored about a central vertical axis) of the pure diffuse model at selected wavelengths in the FIR/submm. The observed SB profiles are plotted with a solid black line with the shaded banding indicating the uncertainty. The solid red line indicates the model total, with the individual component contributions plotted with the dashed and dotted lines. The percent differences between the model total and the observed profiles, $D$ [ per cent], are plotted in the panels below each profile, with the dashed horizontal lines indicating $\pm$ 20 per cent deviation.

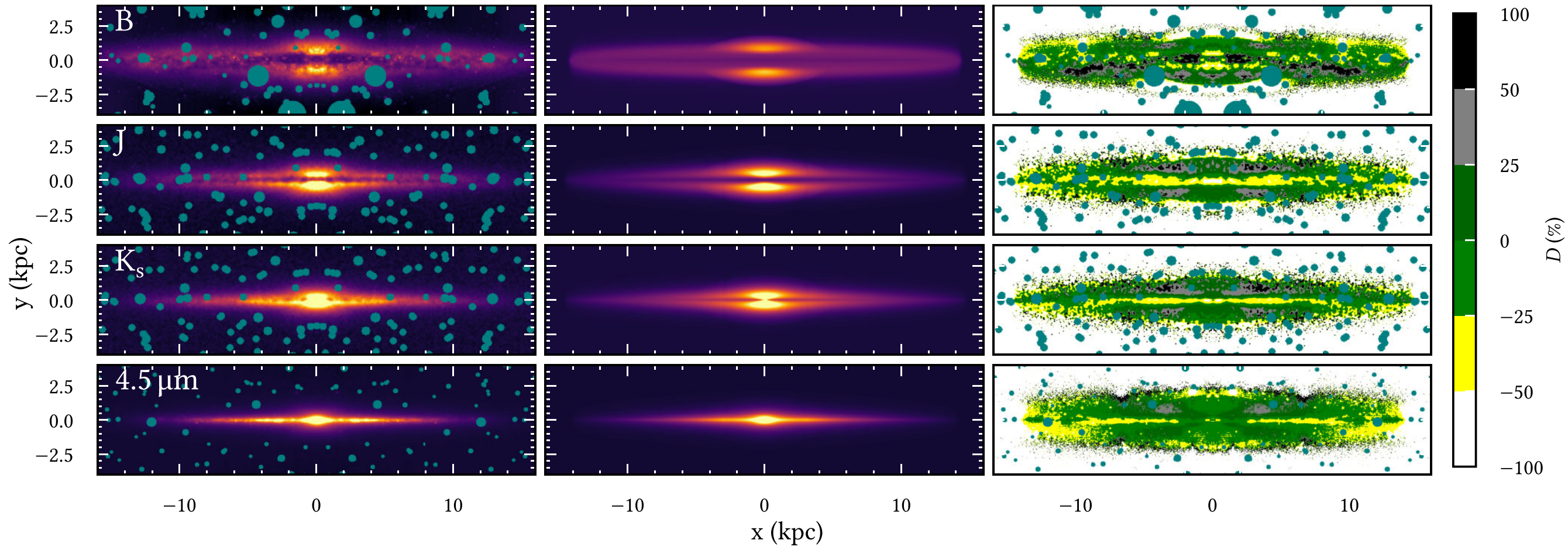

**Figure 6.** Left: surface-brightness maps of NGC 891 at selected optical/NIR wavelengths, mirrored about their vertical central axis. Middle: surface-brightness maps of the corresponding diffuse model images. Right: residuals between the data and model images calculated as $D = (M - O)/O$. Masked foreground stars are marked in teal.

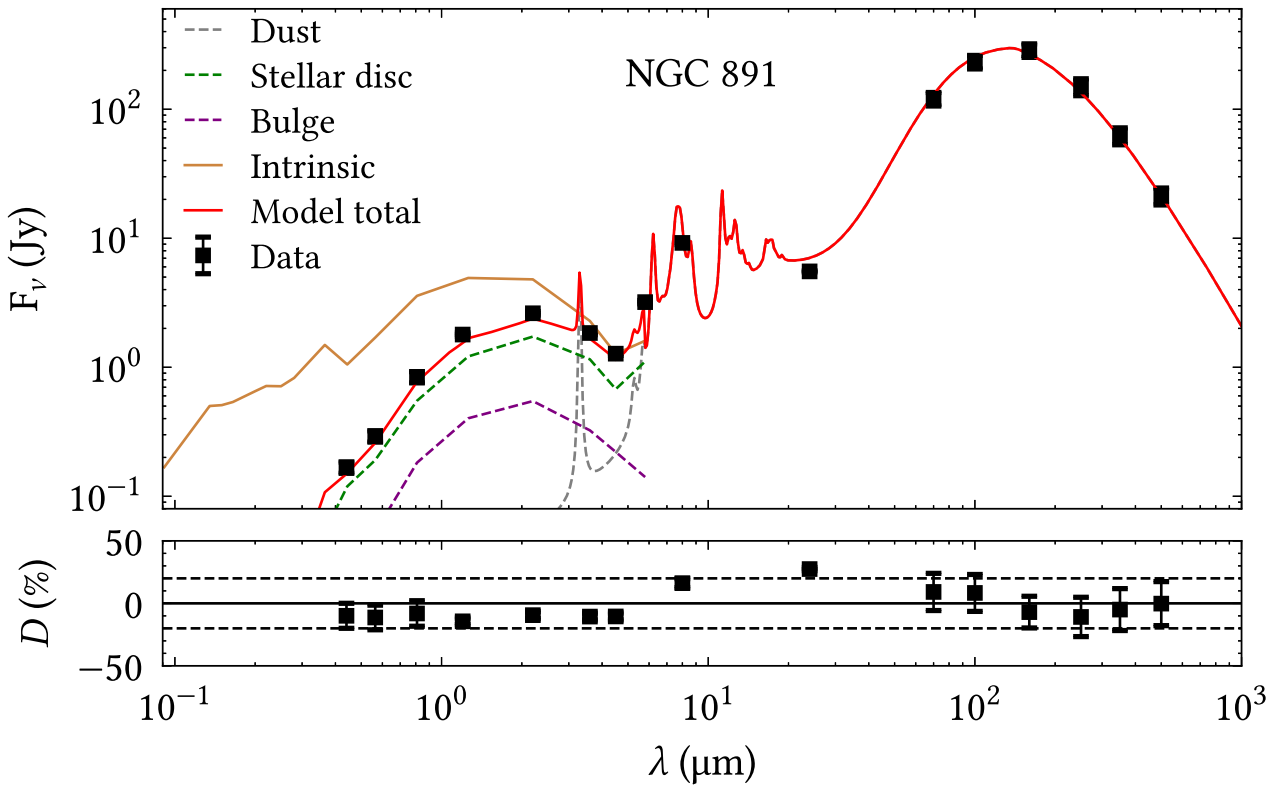

**Figure 7.** The spatially integrated model SED of the pure diffuse solution of NGC 891 plotted with the observed data (black squares) and the associated uncertainties (which are mainly contained in the black squares). Contributions from each of the galaxy components (dashed lines) and the intrinsic stellar SED (solid brown line) are also shown. The lower panel shows the per cent difference between the model SED and the observed data. To guide the eye dashed lines are plotted to show ±20 per cent deviation.

**Table 2.** The free geometrical parameters of NGC 891, for the best-fitting diffuse (middle column), and clumpy models (right column), denoted with 'd' and 'c', respectively. The superscripts 'i' and 'm' found in the notation for the different parameters denote the inner and main discs, respectively.

| Parameter | d | c |
|---|---|---|
| $h_s^{disc-m}(B)$ | $10 \pm 1$ | $10 \pm 1$ |
| $h_s^{tdisc-(i, m)}$ | $(3.0 \pm 0.4, 5.0 \pm 0.3)$ | $(3.0 \pm 0.4, 5.0 \pm 0.3)$ |
| $h_d^{disc-(i, m)}$ | $(3.0 \pm 0.3, 10 \pm 1)$ | $(3 \pm 0.3, 10 \pm 1)$ |
| $\chi_s^{disc-m}$ | $0.0 \pm 0.1$ | $0.0 \pm 0.1$ |
| $\chi_s^{tdisc-(i, m)}$ | $(0, 1.0 \pm 0.2)$ | $(0, 1.0 \pm 0.2)$ |
| $\chi_d^{disc-(i, m)}$ | $(0, -0.9 \pm 0.2)$ | $(0, -0.9 \pm 0.2)$ |
| $z_s^{disc-m}(B)$ | $0.42 \pm 0.02$ | $0.50 \pm 0.02$ |
| $z_d^{disc-(i, m)}$ | $(0.28 \pm 0.03, 0.28 \pm 0.03)$ | $(0.28 \pm 0.03, 0.28 \pm 0.03)$ |
| $R_e(B)$ | $1.4 \pm 0.2$ | $1.4 \pm 0.2$ |
| $\frac{b}{a}(B)$ | $0.43 \pm 0.01$ | $0.47 \pm 0.01$ |

**Table 3.** The geometrical parameters of NGC 891 fixed from data and theoretical considerations. The values are the same between the diffuse and clumpy models. The Sérsic index was fixed to the value considered in E. M. Xilouris et al. (1999). The superscripts 'i' and 'm' found in the notation for the different parameters denote the inner and main discs, respectively. Where applicable, all values are given in kpc.

| Parameter | Value |
|---|---|
| $R_{in,s}^{tdisc-(i, m)}$ | (0, 0) |
| $R_{in,s}^{disc-m}$ | 3 |
| $R_{in,d}^{disc-(i, m)}$ | (0, 3) |
| $R_{tin,s}^{tdisc-(i, m)}$ | (0, 0) |
| $R_{tin,s}^{disc-m}$ | 0 |
| $R_{tin,d}^{disc-(i, m)}$ | (0, 0) |
| $R_{t,s}^{tdisc-(i, m)}$ | (3, 14) |
| $R_{t,s}^{disc-m}$ | 14 |
| $R_{t,d}^{disc-(i, m)}$ | (3, 14) |
| $z_s^{tdisc-(i, m)}$ | (0.09, 0.09) |
| $n_s$ | 4 |

The spatially integrated model SED from Fig. 10 is well matched to the data. This is also the case for the diffuse model. However, unlike the diffuse model, the intrinsic stellar SED is more realistic, because it shows a smaller difference between the intrinsic and dust attenuated stellar SED, predicting thus less attenuation. We otherwise note the MIR SED is very similar to that of the diffuse model. This is because the clump dust emission SEDs are very similar to that of the corresponding diffuse dust, presenting only a mild reduction in the MIR fluxes, as expected for optically thin (in the optical) clumps. The similarity between the dust emission SEDs of the clumpy and diffuse components were also found by S. Bianchi (2008). This author presented a model of NGC 891 using his own code TRADING, which included an explicit treatment of clumps (both quiescent and active). Although his model was unable to fit the MIR emission of NGC 891, it is noteworthy that the best-fitting model in his study assigned the same 50 per cent of the total dust mass to clumps, as was found in this work. However, in our work, we use a very large number ($\sim 10^8$) of optically thin clouds, while the Bianchi model uses a relative small number of optically thick clouds, which may be the reason why we produce a better solution. We note that the large number of clouds used in our model would be very difficult to implement in an explicit calculations like that of S. Bianchi (2008), proving once again the efficacy of the pseudo-grain approach.

**Table 4.** NGC 891: wavelength-dependent model parameters for main disc and bulge, for the diffuse and clumpy models, denoted with 'd' and 'c', respectively. All length parameters are given in kpc.

| Parameter | Filter | | | | | | | |
|---|---|---|---|---|---|---|---|---|
| | $B$ | $V$ | $I$ | $J$ | $K_s$ | 3.6 µm | 4.5 µm | 5.8 µm |
| d | | | | | | | | |
| $h_s^{disc-m}$ | $10 \pm 1$ | $6.5 \pm 0.2$ | $4.6 \pm 0.1$ | $3.8 \pm 0.1$ | $3.8 \pm 0.1$ | $4.0 \pm 0.1$ | $4.0 \pm 0.1$ | $3.6 \pm 0.2$ |
| $z_s^{disc-m}$ | $0.42 \pm 0.02$ | $0.41 \pm 0.02$ | $0.40 \pm 0.02$ | $0.34 \pm 0.01$ | $0.31 \pm 0.01$ | $0.28 \pm 0.01$ | $0.28 \pm 0.01$ | $0.27 \pm 0.01$ |
| $R_e$ | $1.44 \pm 0.05$ | $1.44 \pm 0.05$ | $1.44 \pm 0.05$ | $1.44 \pm 0.05$ | $1.60 \pm 0.05$ | $1.60 \pm 0.05$ | $1.60 \pm 0.05$ | $1.44 \pm 0.05$ |
| $b/a$ | $0.43 \pm 0.02$ | $0.43 \pm 0.01$ | $0.43 \pm 0.01$ | $0.43 \pm 0.01$ | $0.43 \pm 0.01$ | $0.43 \pm 0.01$ | $0.43 \pm 0.01$ | $0.43 \pm 0.01$ |
| c | | | | | | | | |
| $h_s^{disc-m}$ | $10 \pm 1$ | $7.0 \pm 0.2$ | $5.3 \pm 0.1$ | $4.4 \pm 0.1$ | $4.0 \pm 0.1$ | $4.2 \pm 0.1$ | $4.2 \pm 0.1$ | $3.6 \pm 0.1$ |
| $z_s^{disc-m}$ | $0.50 \pm 0.02$ | $0.50 \pm 0.02$ | $0.46 \pm 0.01$ | $0.36 \pm 0.01$ | $0.31 \pm 0.01$ | $0.30 \pm 0.01$ | $0.31 \pm 0.02$ | $0.27 \pm 0.02$ |
| $R_e$ | $1.44 \pm 0.04$ | $1.44 \pm 0.04$ | $1.44 \pm 0.04$ | $1.44 \pm 0.04$ | $1.60 \pm 0.04$ | $1.60 \pm 0.04$ | $1.60 \pm 0.04$ | $1.44 \pm 0.04$ |
| $b/a$ | $0.47 \pm 0.01$ | $0.47 \pm 0.01$ | $0.47 \pm 0.01$ | $0.47 \pm 0.01$ | $0.47 \pm 0.01$ | $0.47 \pm 0.01$ | $0.47 \pm 0.01$ | $0.47 \pm 0.01$ |

Finally, we note that the clumpy model provides a solution with $\tau_B^f = 1.6$, which is much lower than that of the diffuse model ($\tau_B^f = 4.4$), and more in line with the solution obtained by E. M. Xilouris et al. (1998, 1999) from the optical fits of NGC 891.

### 4.4 Comparison between the diffuse and clumpy models of NGC 891

The key distinction between the diffuse and clumpy models is that, in the latter, a fixed fraction (50 per cent) of the total dust mass is assigned to the clumpy component. As a result, the most notable difference between the best-fitting models is the optical depth of the diffuse component. With 50 per cent of the dust mass in clumps ($f^c = 0.5$), and with the consequent change in the values of the geometrical parameters, the optical depth of the diffuse component in the clumpy model is only approximately half that of the purely diffuse model.

The effects of the reduced optical depth are a better fit to the data, in particularly at the critical NIR 2MASS $J$ and $K_s$ bands, essentially solving the energy balance problem in NGC 891. This is illustrated in Figs 11 and 12, which directly compare the diffuse and the clumpy model fits at these wavelengths, both in the surface-brightness profiles and in the maps.

#### *4.4.1 Geometric parameters*

Since the dust opacity varies between the diffuse and the clumpy models, the fitted geometrical parameters are also expected to differ between models. Indeed, inspection of the values of these parameters, as listed in Tables 2 and 4, shows the expected variation.

Both the diffuse and the clumpy models follow a similar stellar distribution to that of the MW (G. Natale et al. 2022), comprising a bulge and a disc that linearly declines in emissivity from the inner radius to the centre.

The radial distribution of stars in the main stellar disc is overall more extended in the clumpy model. This trend can be seen in Table 4, with the values starting to converge between models at longer NIR wavelengths. This difference can be attributed to attenuation effects. As the dust density declines exponentially with radius, its attenuation effect will also decrease with radius. In the outer regions of the galaxy, the disc becomes more optically thin, resulting in minimal attenuation of stellar emission. In contrast, the translucent regions at intermediate radii suppress stellar emission, making the surface-brightness profile appear flatter than it would in the absence of dust. Greater attenuation leads to stronger suppression of emission, causing the diffuse model to favour a shorter stellar scale length.

For the same reason, the scale height of the main stellar disc, $z_s^{m-disc}$, differs significantly between the clumpy and diffuse models, with $z_s^{m-disc}(B) = 0.50$ and 0.42 pc, respectively. Scale heights at all other wavelengths are listed in Table 4.

The radial distribution of stars (and consequently dust) in the main thin disc are found to be equivalent with a scale length of $h^{m-tdisc} = 5$ kpc derived in both the clumpy and diffuse models. Conversely, the scale length of the main dust disc is shorter in the clumpy model than in the diffuse model, with $h_d^{m-disc} = 10$ and 11 kpc, respectively; however, these values are consistent within the uncertainties.

The increased attenuation in the diffuse model can account for the difference in the bulge parameters. The diffuse model has a considerably flatter bulge with respect to the clumpy model, with its emission more strongly confined to the mid-plane. This helps to compensate for the higher attenuation in the mid-plane. With a lower optical depth, such confinement is not necessary in the clumpy model. Interestingly, unlike the scale lengths and scale heights, the values determined for $b/a$ for the two models do not seem to converge at longer wavelengths.

#### *4.4.2 Global properties*

*Dust optical depth and dust mass.* The maximum face-on $B$-band optical depths are $\max(\tau_B^f) = 4.4 \pm 0.3$ for the diffuse model and max $\left(\tau_B^{\mathrm{f}}\right) = 1.6 \pm 0.1$ for the clumpy model. In both cases, these maxima occur at the galaxy's centre. The lower central optical depth in the clumpy model arises directly from its defining assumption: 50 per cent of the total dust mass is placed in clumps, thereby reducing the opacity of the diffuse component.

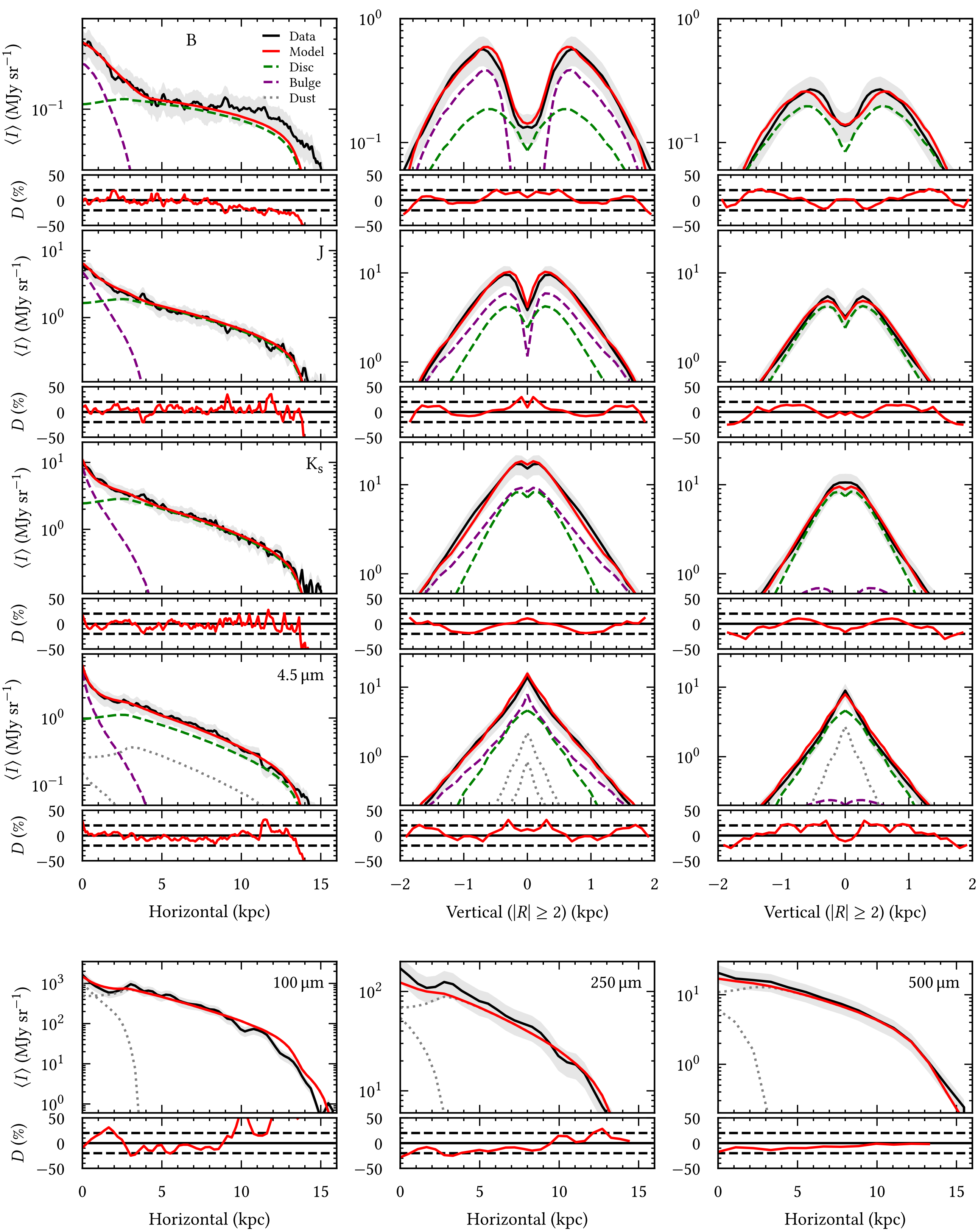

**Figure 8.** Same as Fig. 5, for the clumpy model of NGC 891.

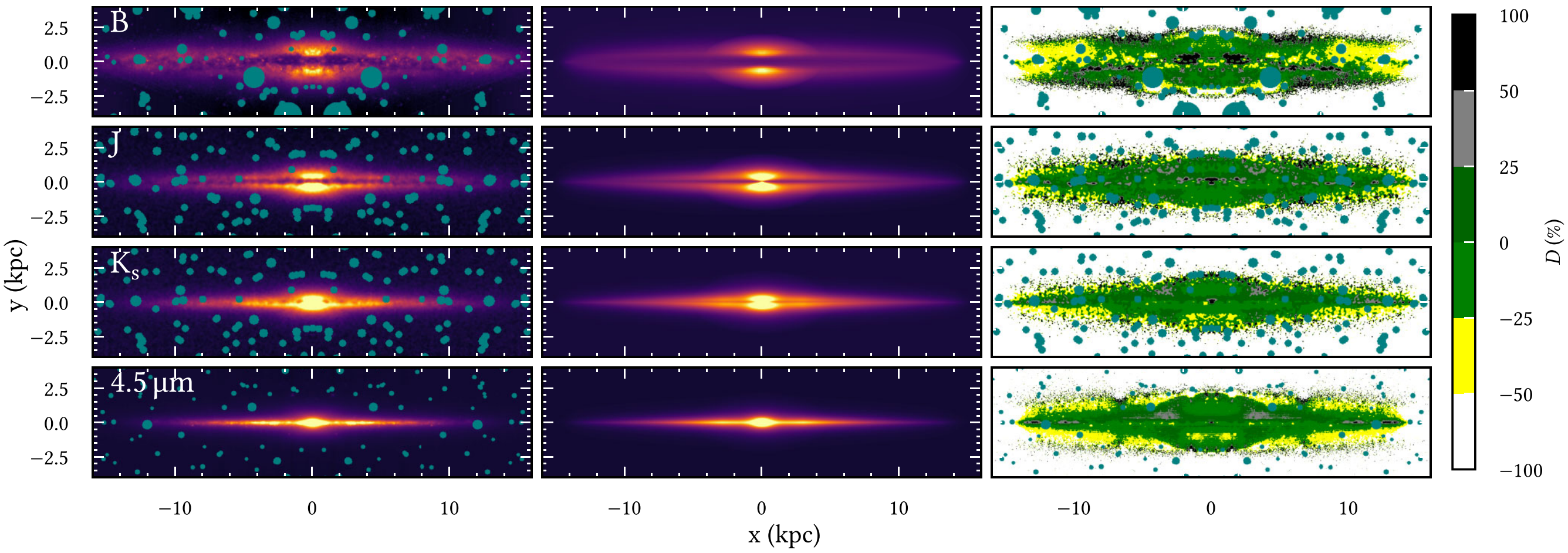

**Figure 9.** Same as Fig. 6, but for the clumpy model of NGC 891.

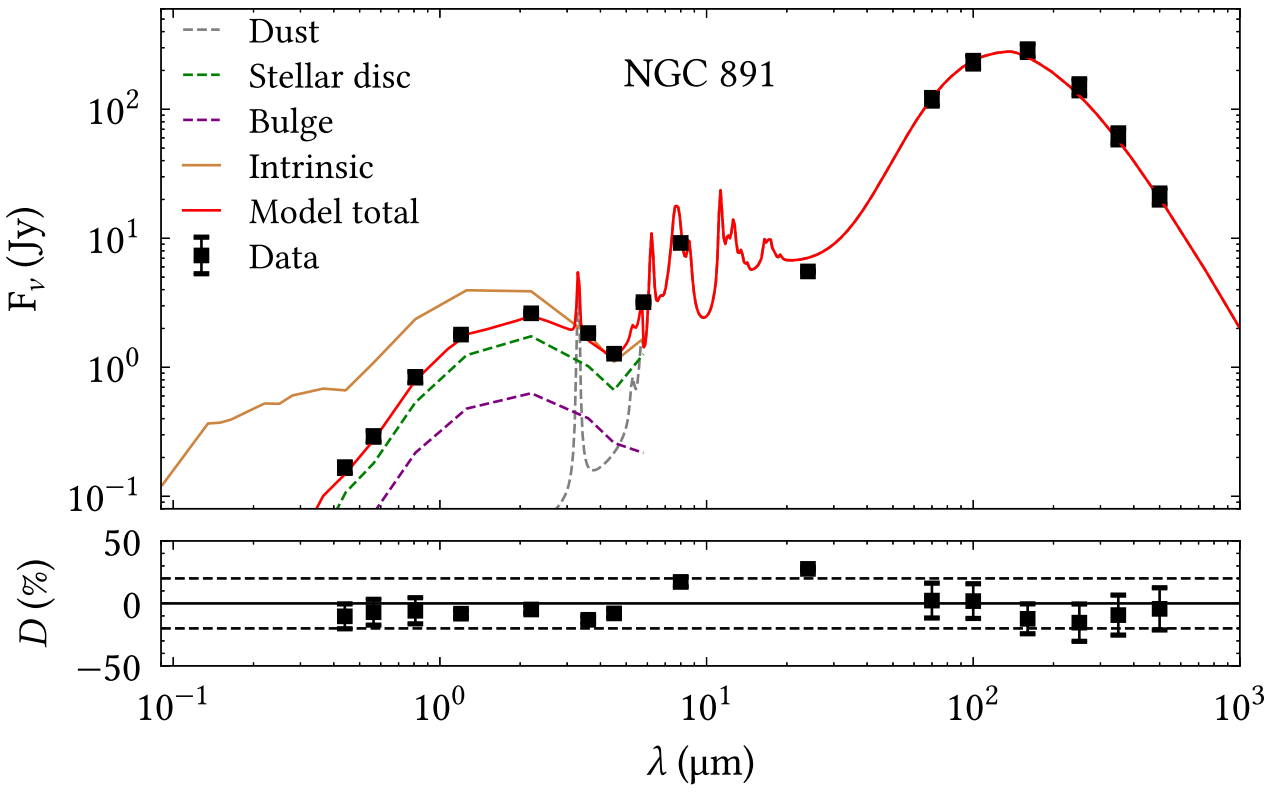

**Figure 10.** Same as Fig. 7, for the clumpy model of NGC 891.

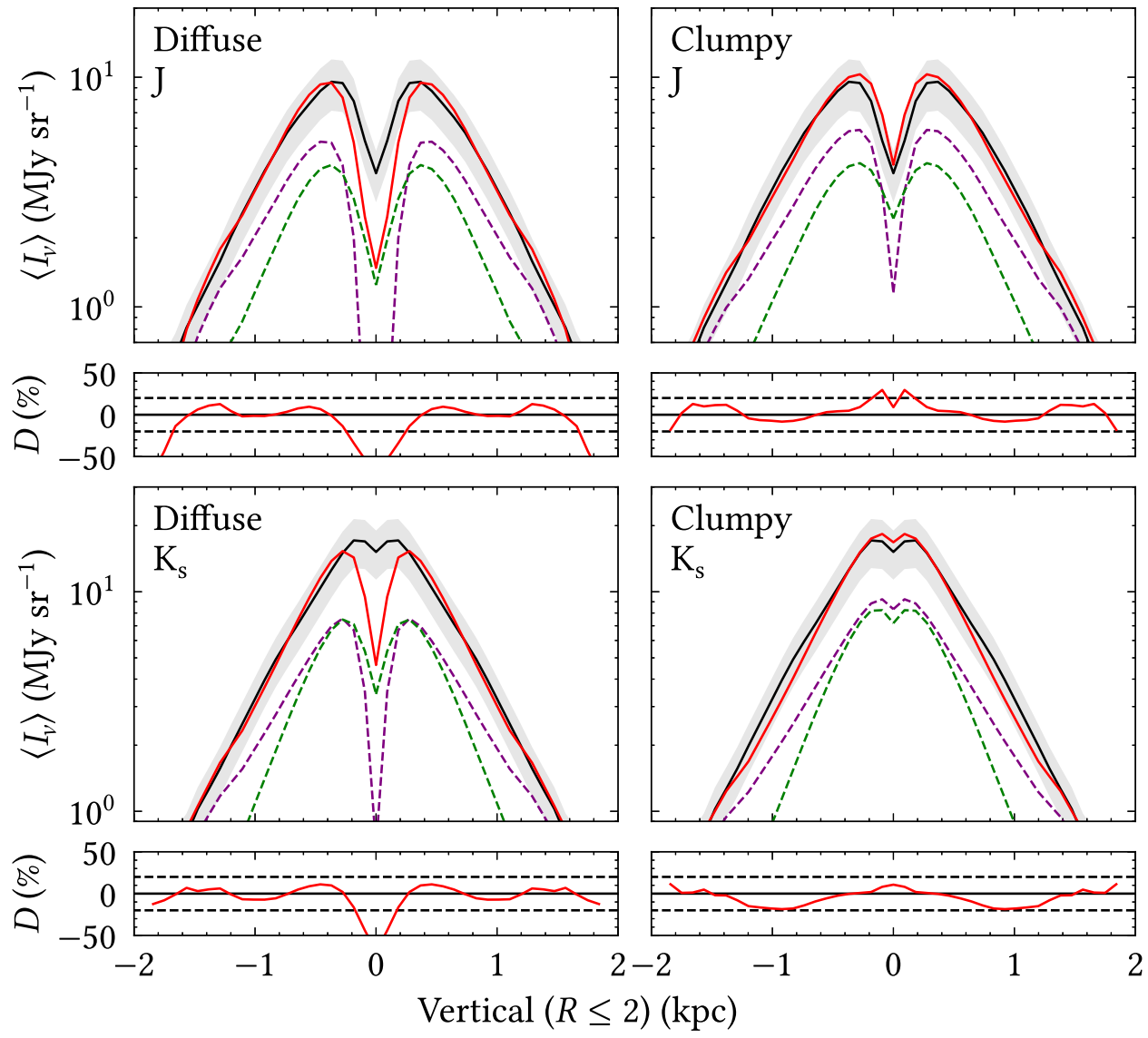

**Figure 11.** Comparison between the diffuse (left panels) and clumpy (right panels) models of NGC 891 in the 2MASS $J$ and $K_s$ bands. The plots follow the same format as Fig. 5.

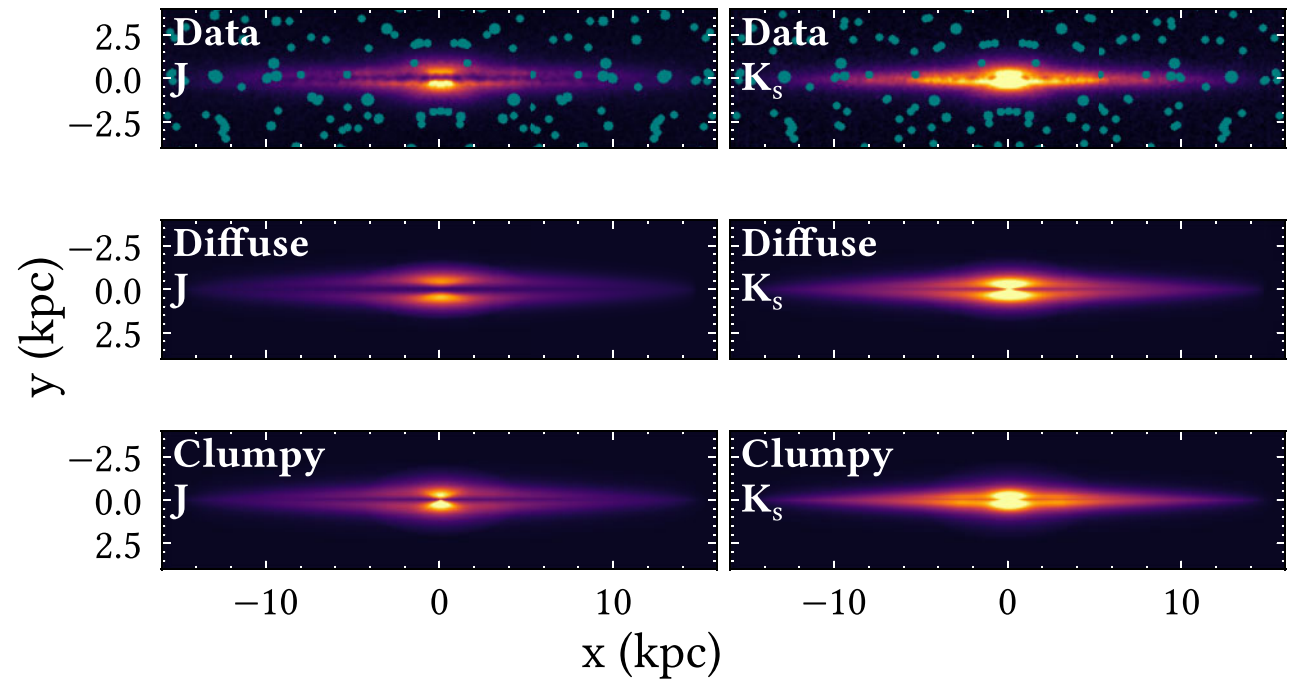

**Figure 12.** Surface-brightness maps of NGC 891 in the 2MASS $J$ and $K_s$ bands. The top panels show the observed data reflected about the vertical axis passing through the galaxy's centre, the middle panels show the diffuse model, and the bottom panels show the clumpy model.

When weighting the average dust opacity by the surface area, it was found that $\langle\tau_{B,\mathrm{area}}^{\mathrm{f}}\rangle = 1.0 \pm 0.1$ for the diffuse model and $0.4 \pm 0.1$ for the clumpy model. The global view of dust opacity across the galaxy thus reveals that in the diffuse model the galaxy is roughly optically thick on average, whereas the clumpy model suggests the galaxy is optically thin.

*Star formation rate.* The global SFR of NGC 891 is found to be $\mathrm{SFR} = 7.2 \pm 0.6$ and $4.6 \pm 0.4\,\mathrm{M}_\odot\,\mathrm{yr}^{-1}$ for the diffuse and clumpy models, respectively. Since the clumpy model has a lower dust opacity in the diffuse medium, it produces less dust attenuation, which in turn leads to a solution with a lower intrinsic UV luminosity, and consequently, a lower SFR. This explains the differences in the derived solutions for NGC 891.

In the previous model by C. C. Popescu et al. (2004), a much smaller value for the SFR was obtained, with $\mathrm{SFR} = 3.8\,\mathrm{M}_\odot\,\mathrm{yr}^{-1}$. This was solely due to the use of the ISOPHOT data in the 160 and

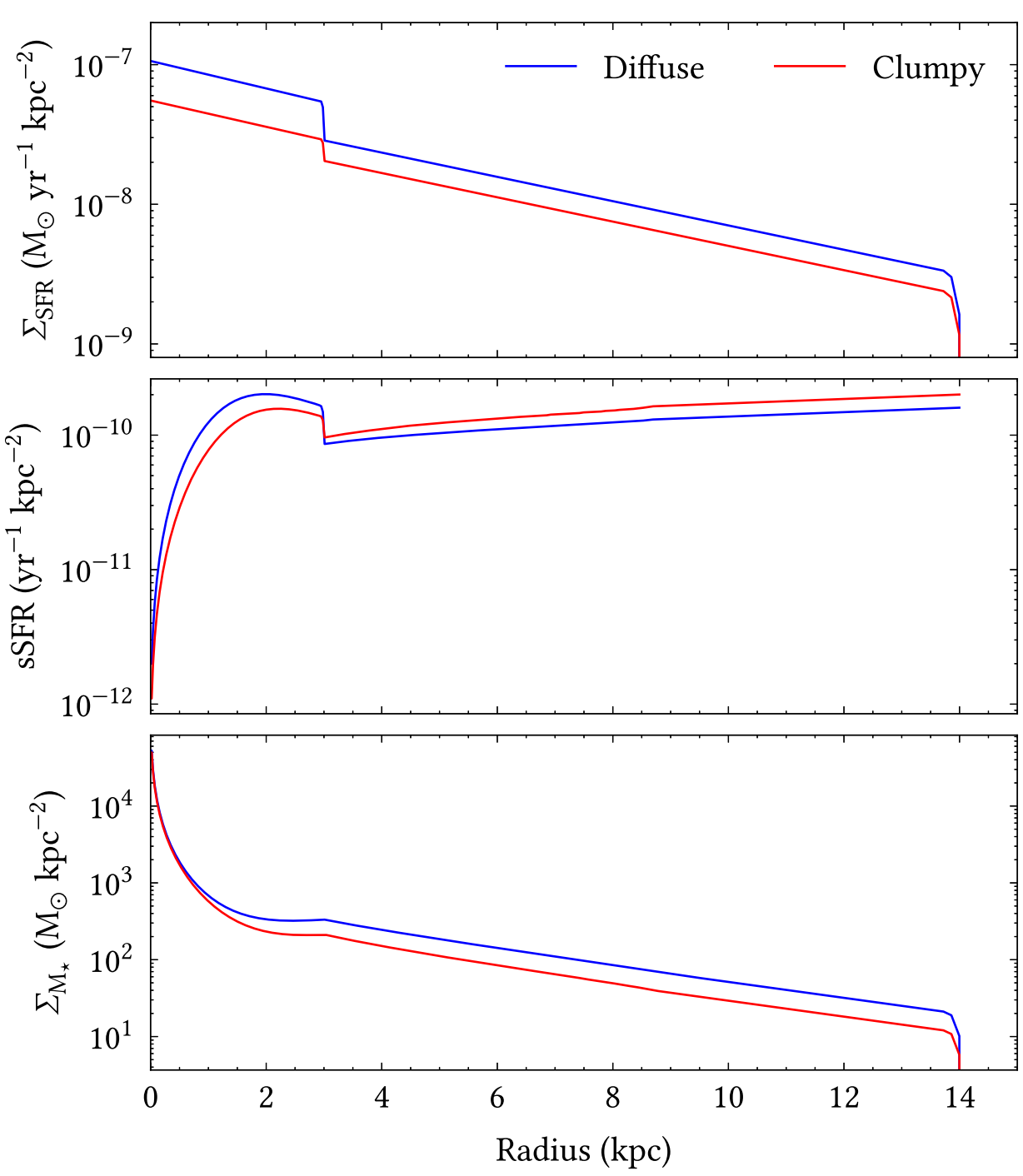

**Figure 13.** NGC 891 surface density of SFR ($\Sigma_{\rm SFR}$) (top), specific SFR (sSFR) (middle), and surface density of stellar mass ($\Sigma_{M_\star}$) (bottom). The diffuse and clumpy models are plotted with the blue, and red lines, respectively.

200 µm range, which, as discussed in the preamble of this section, provided flux densities that were approximately 1.5 times lower than those from the more recent *Herschel* observations, that were used in this work.

J.-H. Shinn & K.-I. Seon (2015) find a global star formation rate of SFR $= 2.4\,{\rm M}_\odot\,{\rm yr}^{-1}$ using the relation between SFR and IR luminosity (R. C. Kennicutt 1998), however it should be noted that H$\alpha$ is non-negligibly attenuated in NGC 891. M. Persic & Y. Rephaeli (2007) obtain SFR $= 3.9$ and $3.4\,{\rm M}_\odot\,{\rm yr}^{-1}$ when correcting for cirrus emission–UV photons scattered at high latitudes in the halo. J. H. Yoon et al. (2021) give an SFR $= 5.0\,{\rm M}_\odot\,{\rm yr}^{-1}$.

In the diffuse model, the main disc contributes 87.1 per cent to the total SFR, with the remaining 12.9 per cent provided by the inner disc. In the clumpy model however, the main disc contributes a higher percentage to the total, namely, 94.4 per cent ($4.37\,{\rm M}_\odot\,{\rm yr}^{-1}$). The difference in the relative contributions of the main and inner disc in to the total SFR can again be attributed to the higher dust attenuation of the diffuse model with respect to the clumpy one, with the attenuation having a larger effect in the inner regions of the galaxy.

#### 4.4.3 Radial trends

In addition to the global properties of NGC 891, their spatial variation within the galaxy is also examined. Fig. 13 shows the radial variation of surface density of SFR ($\Sigma_{\rm SFR}$), specific SFR (sSFR), and surface density of stellar mass ($\Sigma_{M_\star}$). The stellar mass is calculated from the NIR flux following the method of M. Eskew, D. Zaritsky & S. Meidt (2012). The radial trends in $\Sigma_{\rm SFR}$ are similar between the clumpy and diffuse models. Given the larger global SFR, it is not surprising that the diffuse model shows higher values of $\Sigma_{\rm SFR}$ across the extent of the galaxy, particularly in the inner disc. Conversely, the values of $\Sigma_{M_\star}$ for the two models converge near the centre. As a result, in the inner regions, the diffuse model has a higher sSFR until the interface between the morphological components is reached, beyond which the clumpy model consistently shows a higher sSFR.

**Table 5.** Basic properties of galaxies in this sample. The distances and inclination are those used in modelling.

| Galaxy | $D$ (Mpc) | $i$ (deg) | Type |
|---|---|---|---|
| M33 | 0.859 | 56.0 | SA(s)cd |
| M51 | 8.58 | 20.3 | SA(s)bc pec |
| M101 | 6.71 | 18.0 | SAB(rs)cd |
| NGC 628 | 9.50 | 15.0 | SA(s)c |
| NGC 891 | 9.85 | 89.7 | SA(s)b |
| NGC 3521 | 10.7 | 72.7 | SAB(rs)bc |
| NGC 3938 | 17.9 | 14.0 | SA(s)c |

## 5 APPLICATION OF THE CLUMPY MODEL TO A SMALL GALAXY SAMPLE

The success in the modelling of NGC 891 (Section 4) using the clumpy model and solving the energy balance problem raises the question of whether the clumpy model produces the right energy balance for other galaxies, and if it is generally applicable for galaxies of any inclination. To test this, a small sample of galaxies was built spanning the whole inclination range. The sample includes the galaxies that have previously been modelled with the pure diffuse model, and were described in Section 1: M33, TP20; NGC 628, M. T. Rushton et al. (2022); M51, C. J. Inman et al. (2023); and M101 and NGC 3938, D. Pricopi et al. (2025). In addition, NGC 3521 was also included in this sample to represent a galaxy with an intermediate inclination ($i = 73°$; F. Walter et al. 2008; L. S. Pilyugin, G. Tautvaišienė & M. A. Lara-López 2023). The fitting procedure used for the non edge-on galaxies is described in TP20 and Section 3.3.1. The panchromatic data used to constrain the clumpy models are largely the same as those used in the original diffuse models, with the exception of NGC 891 for which, as discussed in Section 4.1, newer data have become available.

The galaxies contained in this small sample considered for modelling are tabulated with some of their basic properties in Table 5. The geometrical parameters derived for the diffuse and clumpy models of each galaxy can be seen in Appendix. B.

The galaxies from the sample have been modelled with two or more morphological components (see Table 6 for an overview). The only galaxy containing only two morphological components, a disc and a bulge, is the face-on NGC 3938. This galaxy presents exceptionally smooth and simple profiles, with no signs for breaks or different slopes outside the bulge region. The profiles look even smoother than those of the edge-on galaxy from the sample, NGC 891. A bulge is also known to exist in all the galaxies from the sample, except for M33, and was thus included in the model galaxies. In addition, an inner disc was needed for modelling NGC 891, and the majority of non-edge-on galaxies from the sample. This inner disc is usually more prominent in the dust (dust disc) and young stellar population distribution (thin disc). Older stellar populations contained in stellar discs more often exhibit distributions that extend exponentially to the centre of the galaxy, with no evidence of an additional inner disc counterpart (NGC 628, NGC 891, and NGC 3521).

**Table 6.** The morphological components adopted in the modelling of our galaxy sample: n – nuclear disc, i – inner disc, m – main disc, o – outer disc, and b – bulge.

| Galaxy | | | | | |
|---|---|---|---|---|---|
| M33 | n | i | m | o | – |
| M51 | – | i | m | o | b |
| M101 | n | i | m | – | b |
| NGC 628 | – | i | m | – | b |
| NGC 891 | – | i | m | – | b |
| NGC 3521 | – | i | m | o | b |
| NGC 3938 | – | – | m | – | b |

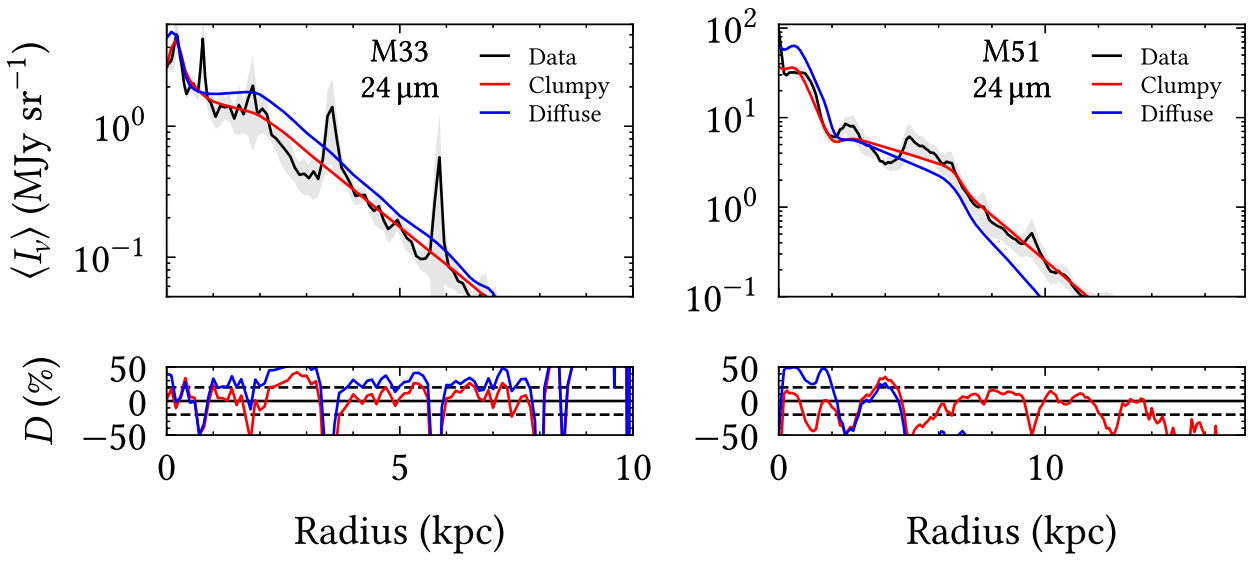

**Figure 14.** Comparison of the diffuse model (blue line) and clumpy model (red line) fits to 24 μm surface-brightness profiles (black line) for selected galaxies in the sample. The shaded region indicates the errors associated with the data. The per cent differences between the model totals and the observed profiles $D$ [ per cent] are plotted in the lower panels in the colour corresponding to the main plot.

Unlike the modelling of NGC 891, some of the face-on galaxies were modelled with two extra morphological component, a nuclear disc and an outer disc. The nuclear disc is only prominent in M33 and M101, and only appears in the distribution of young stars (the thin disc). An outer disc was needed for M33, M51, and NGC 3521.

### 5.1 Comparison between the predictions of the diffuse and clumpy models for the surface-brightness profiles

With the exception of the edge-on galaxy NGC 891, for which the diffuse model fails to account for the dust lane in the optical/NIR bands, the diffuse and clumpy models both provide reasonable fits for galaxies at non-edge-on inclinations. Nevertheless, the clumpy model tends to show slightly better fits than the diffuse model. In particular, the diffuse models tend to overpredict the data at 24 μm and underpredict at 500 μm, mainly in the central regions, as shown in Figs 14 and 15. For these specific cases, the clumpy model tends to perform better. Thus, at 24 μm, the clumpy models usually predict less emission (with respect to the diffuse component), due to some small suppression of the PAH and warm dust emission in the clumps. None the less, the differences between the predictions of the diffuse and clumpy models are rather small, and overall it can be concluded that both the diffuse and clumpy models provide a good fit for the spatial and SED of face-on galaxies. It shows that the RT solution in systems under this orientation is completely degenerate.

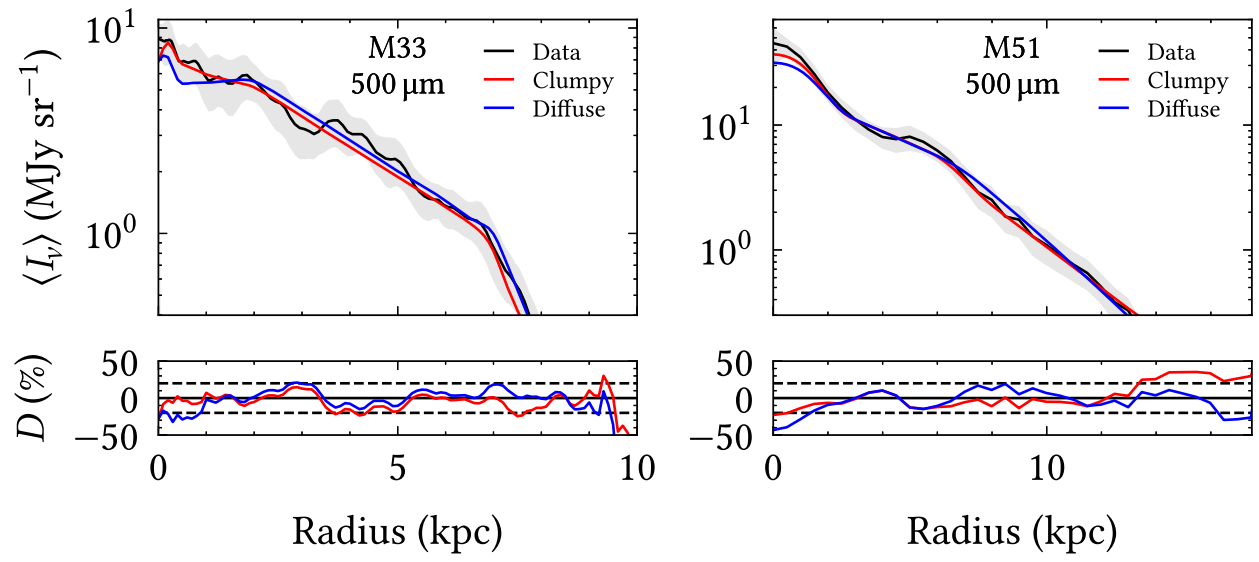

**Figure 15.** Same as Fig. 14 at 500 μm.

#### 5.1.1 Dust opacity

The inclusion of the clumpy component produced, as expected, a solution with less dust opacity than in the pure diffuse model. This can be seen in the values of the maximum dust opacity, max($\tau_B^{\rm f}$) listed in Table 7 for the diffuse and clumpy models, and also in Fig. 16, where max($\tau_B^{\rm f}$) – diffuse is plotted against max($\tau_B^{\rm f}$) – clumpy. The global $\tau_B^{\rm f}$ is calculated as the sum of $\tau_B^{\rm f}$ from the different morphological components at each radial position. Although at any radial position one morphological component dominates, there is still some partial overlap between components, and therefore the need to define a global $\tau_B^{\rm f}$. For this galaxy sample, max($\tau_B^{\rm f}$) is at the inner radius of the inner disc for M33, M51, and NGC 891 [see Table 8 for the max($\tau_B^{\rm f}$) of the individual morphological components]. For NGC 628 and NGC 3521, the maximum opacity occurs at the inner radius of the main disc, while for M101, the situation is more complex, with the radius being different in the diffuse and clumpy models.

The decrease in max($\tau_B^{\rm f}$) between the diffuse and clumpy models ranges by factors of 1.3 to 2.8 (see again Table 7 and Fig. 16). NGC 891 exhibits the largest difference, by a factor of 2.8, while M33 and M101 display the smallest differences, each by a factor of 1.3. Naively one would expect that these factors to follow the $\frac{1}{1-f^{\rm c}}$ ratio. For example, for solutions with $f^{\rm c} = 0.5$ $\left(\frac{1}{1-f^{\rm c}} = 2\right)$ the decrease in dust opacity between the diffuse and the clumpy model would then follow the $y = 2x$ line, as displayed in Fig. 16. However the values seem to fluctuate and are usually between $y = x$ and $2x$ line, due to the modulating effect of the luminosity sources heating the dust and the geometric effects. This can be explained with reference to the optimization routine, particularly that the dust opacity is constrained from the 500 μm data. Thus, in the clumpy model, the dust opacity needed to fit the submm data is lower, due to the additional clumpy component contributing to this emission. A lower dust opacity means that the dust attenuation in the UV will be lower, resulting in a solution with lower UV luminosity, and thus with lower SFR. In the iterative optimization routine, this will produce a slightly lower amplitude for the dust emission SED, including a small reduction in the 500 μm predicted emission. As a brief aside, it is noted that the 500 μm, while on the Rayleigh–Jeans side of the dust emission SED and thus tracing dust column density, is still not deep enough to have zero contribution from the heating sources. It is this very small contribution that is discussed here. To compensate for this small effect, the dust opacity would need to be

**Table 7.** The maximum and average face-on dust opacity, $\max(\tau_B^f)$ and $\langle\tau_{B,\,\mathrm{area}}^f\rangle$, of the galaxy sample derived using the diffuse (column under 'd') and clumpy (columns under 'c') models. $f^c$ is the fraction of total dust mass in the clumpy component.

| Galaxy | $f^c$ | | $\max(\tau_B^f)$ | | $\langle\tau_{B,\,\mathrm{area}}^f\rangle$ | |
|---|---|---|---|---|---|---|
| | d | c | d | c | d | c |
| M33 | 0 | 0.5 | 1.3 ± 0.1 | 1.0 ± 0.1 | 0.4 ± 0.1 | 0.2 ± 0.1 |
| M51 | 0 | 0.6 | 5.3 ± 0.3 [1] | 2.9 ± 0.2 | 0.5 ± 0.1 | 0.2 ± 0.1 |
| M101 | 0 | 0.4 | 2.4 ± 0.2 [2] | 1.9 ± 0.2 | 0.3 ± 0.1 | 0.2 ± 0.1 |
| NGC 628 | 0 | 0.5 | 1.7 ± 0.2 [3] | 0.8 ± 0.1 | 0.4 ± 0.1 | 0.3 ± 0.1 |
| NGC 891 | 0 | 0.5 | 4.4 ± 0.3 | 1.6 ± 0.1 | 1.0 ± 0.1 | 0.4 ± 0.1 |
| NGC 3521 | 0 | 0.5 | 2.8 ± 0.2 | 1.5 ± 0.1 | 0.5 ± 0.1 | 0.3 ± 0.1 |
| NGC 3938 | 0 | 0.5 | 1.6 ± 0.1 [2] | 0.9 ± 0.1 | 0.7 ± 0.1 | 0.3 ± 0.1 |

*Notes.* The values are calculated within the truncation radius of the respective galaxy. (1) C. J. Inman et al. (2023); (2) D. Pricopi et al. (2025); and (3) M. T. Rushton et al. (2022).

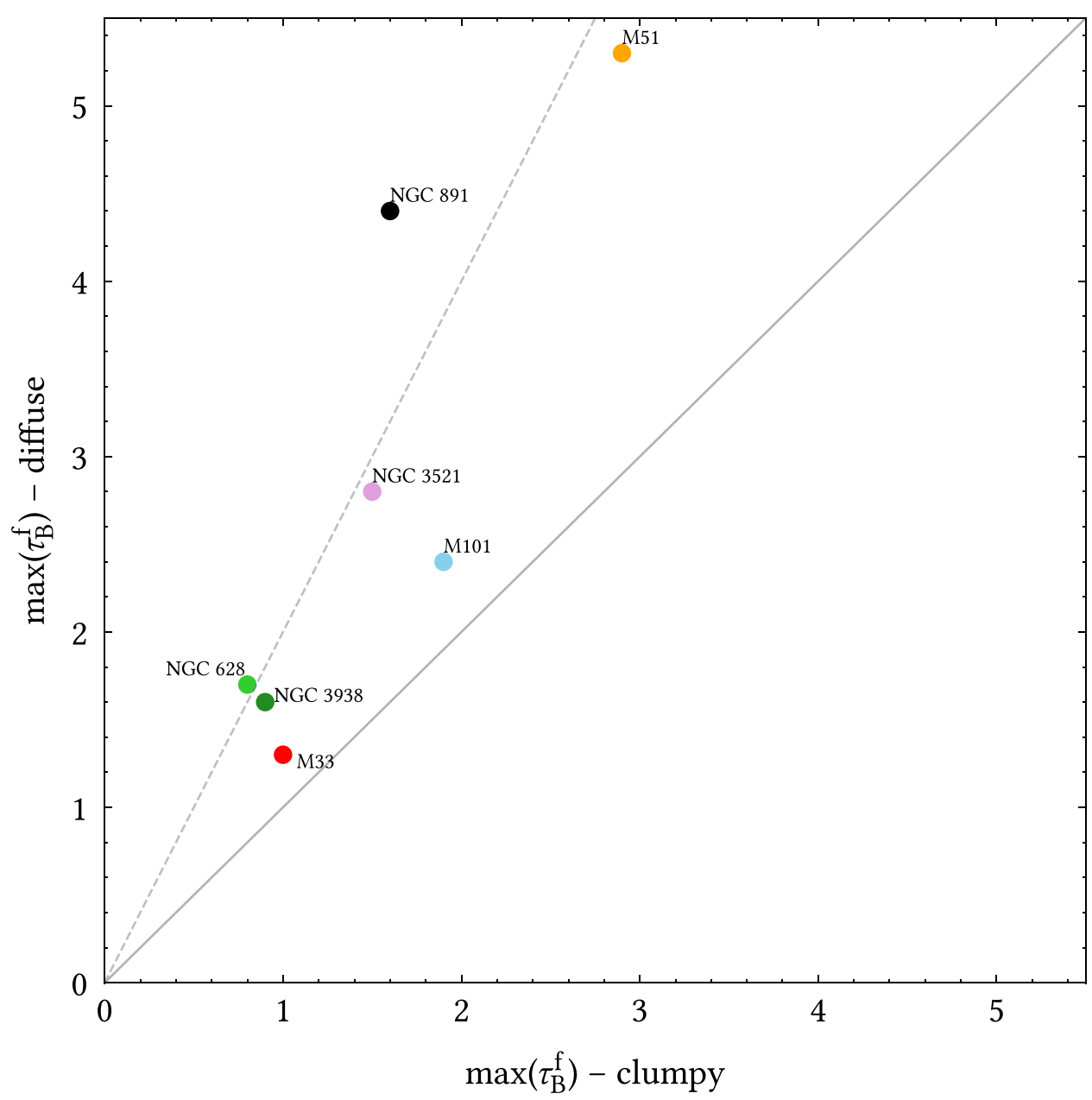

**Figure 16.** Comparison between the maximum optical depths, $\max(\tau_B^f)$ derived with the diffuse ($y$-axis) and clumpy ($x$-axis) models. The lines $y = x$ and $y = 2x$ are plotted with the grey-solid and grey-dashed lines, respectively.

increased to offset the reduction resulting from the lower SFR. So the overall effect is a decrease in the dust opacity, but by a smaller factor than the $\frac{1}{1-f^c}$ ratio. The geometric effects also play a role. A lower dust opacity may alter the derived scale length of the young stellar disc (see Table 13 and Fig. 17), which, in turn, will alter the derived SFR, producing further effects in the derived dust opacity. Furthermore, in some cases, the diffuse model underestimated the submm emission, yielding poorer fits than the clumpy model (see Fig. 15).

The clumpy solution indicates that, with respect to the maximum dust opacity (occurring in the central regions at the inner radius of the inner disc or main disc), NGC 3938, NGC 628, and M33 are optically thin, with $\tau_B^f < 1$, NGC 3521, and NGC 891 are slightly optically thick, while M101 and M51 are optically thick. By contrast, the diffuse model predicted that all the galaxies in this sample are optically thick in their centre. This makes a big difference in our understanding of dust opacity in nearby galaxies.

Another way to compare the opacity predicted by the two types of models is to inspect the average surface density weighted dust opacity, $\langle\tau_{B,\,\mathrm{area}}^f\rangle$ (see again Table 7). As with the $\max(\tau_B^f)$, the clumpy models have lower $\langle\tau_{B,\,\mathrm{area}}^f\rangle$ than the diffuse models, with factors ranging between 2.5 (in M51 and NGC 891) and 1.3 (in NGC 628). Interestingly, the galaxy with the lowest difference in average dust opacity between the diffuse and clumpy model, NGC 628, is the galaxy with the second highest difference in the maximum dust opacity. This is mainly due to the morphological/geometrical structure of NGC 628. On the one hand, the galaxy has a very small inner disc, with rapidly varying surface brightness at the position of the maximum opacity, with geometric parameters rather difficult to derive in the models (diffuse/clumpy). On the other hand, on large scales, the galaxy exhibits a relatively smooth variation in azimuthally averaged surface brightness, which would make the overall average opacity less dependent on fluctuations in the derived geometric parameters.

The clumpy model predicts that on average, disc galaxies are optically thin when seen face-on. Based on $\langle\tau_{B,\,\mathrm{area}}^f\rangle$, NGC 891 has the highest opacity $\langle\tau_{B,\,\mathrm{area}}^f\rangle = 0.4$, followed by NGC 3938 with $\langle\tau_{B,\,\mathrm{area}}^f\rangle = 0.3$. However, M51 has the highest maximum opacity in the inner region $\max(\tau_B^f) = 2.9$, followed by M101 with $\max(\tau_B^f) = 1.6$. Thus, galaxies with the highest opacity in their inner regions do not necessarily have the largest average opacity. This can be partially explained by the extent of their outer discs and the truncation radius. Both M101 and M51 possess extended, faint outer discs that may bias the area-weighted average optical depth toward lower values. M101, the largest galaxy in this sample, has a truncation radius of $R_t = 30$ kpc, and therefore has the largest surface area. M51 is relatively centrally dense, with a faint outer disc extending out to 20 kpc. When only the area within the truncation radius of the main disc of M51 ($R_t = 7$ kpc) is considered, the area-weighted average optical depth increases to $\langle\tau_{B,\,\mathrm{area}}^f\rangle = 0.8$.

The various morphological components of the galaxies in the sample show the same decrease in dust opacity between the diffuse and the clumpy models, both in $\max(\tau_B^f)$ (Table 8) and in $\langle\tau_{B,\,\mathrm{area}}^f\rangle$ (Table 9), but with different degrees of variation.

**Table 8.** Maximum optical depth as seen from face-on in the *B* band for each morphological component in the galaxy sample for the diffuse and clumpy models. The generic morphological components: inner, main, and outer discs are denoted by the subscripts 'i', 'm', and 'o' respectively. Values under 'd' and 'c' correspond to the diffuse and clumpy models, respectively.

| Galaxy | $\max\left(\tau^{\mathrm{f}}_{B,\,\mathrm{i}}\right)$ | | $\max\left(\tau^{\mathrm{f}}_{B,\,\mathrm{m}}\right)$ | | $\max\left(\tau^{\mathrm{f}}_{B,\,\mathrm{o}}\right)$ | |
|---|---|---|---|---|---|---|
| | d | c | d | c | d | c |
| M33 | 1.3 ± 0.1 | 1.0 ± 0.1 | 0.9 ± 0.1 | 0.5 ± 0.1 | 0.4 ± 0.1 | 0.1 ± 0.1 |
| M51 | 5.3 ± 0.3 | 2.9 ± 0.2 | 2.6 ± 0.1 | 1.0 ± 0.1 | 1.3 ± 0.1 | 0.3 ± 0.1 |
| M101 | 1.7 ± 0.2 | 1.6 ± 0.2 | 1.9 ± 0.1 | 1.0 ± 0.1 | – | – |
| NGC 628 | 0.8 ± 0.1 | 0.3 ± 0.1 | 1.5 ± 0.1 | 0.8 ± 0.1 | – | – |
| NGC 891 | 4.4 ± 0.3 | 1.6 ± 0.1 | 3.3 ± 0.2 | 1.3 ± 0.1 | – | – |
| NGC 3521 | – | – | 2.8 ± 0.2 | 1.5 ± 0.1 | 0.8 ± 0.1 | 0.4 ± 0.1 |
| NGC 3938 | – | – | 1.6 ± 0.1 | 0.9 ± 0.1 | – | – |

With respect to the average dust opacity, it is noted that in the clumpy models the outer discs have, as expected, very low opacity, of around $\langle\tau^{\mathrm{f}}_{B,\,\mathrm{area}}\rangle \sim 0.1$.

### 5.1.2 *SFR*

As already inferred in Section 5.1.1, the clumpy model predicts smaller SFRs than the diffuse model. Inspection of the tabulated values from Table 10 shows a reduction in SFR by factors ranging between 1.2 to 2.0. The decrease directly follows from the lower UV attenuation predicted by the clumpy models. The decreasing factors do not seem to correlate with inclination, which is reassuring, as any such correlation could potentially indicate a systematic bias resulting from the line-of-sight optical depth. The decreasing factors also do not correlate with the face-on dust opacity of the galaxy. This means that the variation in SFR derived from the diffuse and clumpy models only depends on the detailed structure and morphology of the galaxy under study. The largest variation in SFR was found for NGC 3521 (a factor of 2.0) and the minimum variation for NGC 3938 (a factor of 1.2).

The morphological components of the galaxy sample overall follow the same trend of decreasing SFR between the diffuse and the clumpy model, as seen in Tables 11 and 12. As expected, the closest variation to that shown in the global SFR is in the main disc (see Table 12), which usually dominates the bolometric output of a galaxy. The largest departure from the global properties is in the inner disc (see Table 11), which exhibits a much larger spread in the decreasing factors, from 1.2 in M33 to 4.8 in M101. It is clear that the inner discs, with their smaller sizes and rapid variation in surface brightness, are more prone to large differences in the fitting parameters. In addition, the inner discs lie in the regions with either maximum dust opacity, or in regions with a linear decrease in dust opacity. This produces complex dust attenuation in the inner regions, with values strongly dependent on the type of model (diffuse or clumpy).

### 5.1.3 *sSFR*

The specific star formation rate, sSFR, derived with the clumpy models are typically smaller by 1.01 to 1.89 times than those from the diffuse models (see Table 10). This trend arises naturally from the definition $\mathrm{sSFR} = \mathrm{SFR}/M_\star$, where $M_\star$ is the stellar mass. As previously discussed, the clumpy model produces lower SFRs due to the relatively lower intrinsic UV luminosities. Stellar masses ($M_\star$) are derived from the intrinsic fluxes at 3.6 and 4.5 μm using the calibration of M. Eskew et al. (2012). Because the NIR wavelengths are less affected by dust attenuation, the reduction in $M_\star$ between models is relatively modest compared to the decrease in the UV-derived SFR, leading to an overall decrease in sSFR. Although smaller, the sSFRs derived with the clumpy models still suggest these galaxies are moderately star-forming.

### 5.1.4 *The fraction of stellar light absorbed and re-radiated by dust*

The fraction of the stellar light absorbed and re-radiated by dust, $f^{\mathrm{abs}}$ is listed in Table 10. This metric provides insight into the efficiency with which dust processes stellar radiation, and is influenced by the dust's microscopic properties, composition, and spatial distribution relative to the distribution of stars. $f^{\mathrm{abs}}$ derived from the clumpy models tends to be higher by factors of 1.06–1.37, compared to the diffuse models. NGC 3521 is exceptional to this trend with $f^{\mathrm{abs}}$ decreasing to 0.76 times that of the diffuse model.

### 5.1.5 *The contribution of different stellar populations to dust heating*

The contribution of different stellar populations to dust heating is a very important property of star-forming galaxies, although still a matter of debate in the literature. This section examines the fraction of the energy absorbed by dust coming from young stars, denoted as $F^{\mathrm{dust}}_{\mathrm{young}}$, under the assumption that the contribution of the old stellar populations represent the $1 - F^{\mathrm{dust}}_{\mathrm{young}}$. These values are presented in columns 8 and 9 of Table 10. Across the sample, the clumpy models consistently yield lower values of $F^{\mathrm{dust}}_{\mathrm{young}}$, indicating a reduced contribution from the young stars to the total dust heating. This reduction arises from the lower efficiency with which UV photons are absorbed in the clumpy dust configuration. The clumpy models are characterized by a flatter extinction curve (Fig. 3) whereby UV radiation, primarily emitted by young stars, is less attenuated than in the diffuse models. While the reduction in attenuation is significant in the UV, it becomes negligible in the optical and NIR, hence the contribution of older stars to dust heating remains largely unchanged. However, this is not true for NGC 891 where $F^{\mathrm{dust}}_{\mathrm{young}}$ is larger in the clumpy model. This is attributed to the failure of the diffuse model to fit the optical data, which limits the validity of a direct comparison of these values.

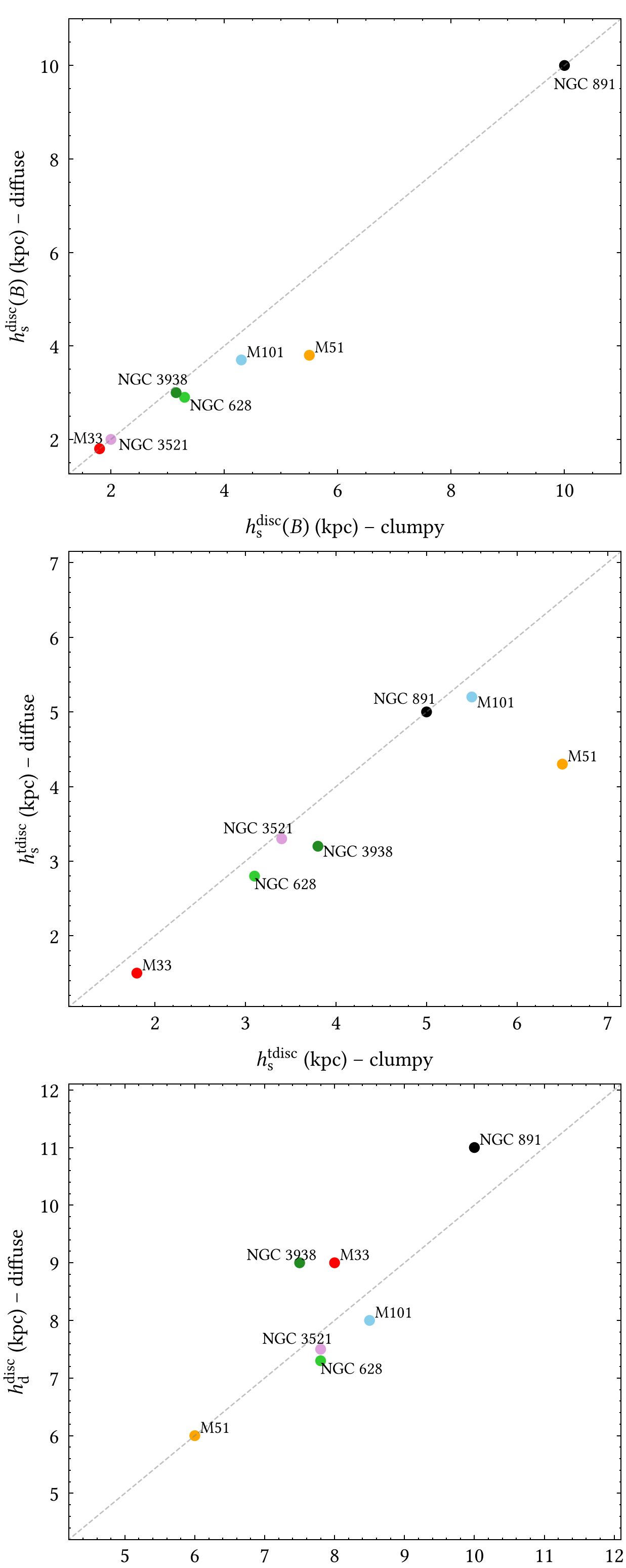

**Figure 17.** Comparison between the scale lengths of the stellar disc (in the $B$ band) $h_s^{disc}(B)$, thin stellar disc $h_s^{tdisc}$, and dust disc $h_d^{disc}$ derived with the diffuse and clumpy models. In all cases, the scale lengths refer to the main disc morphological component. The $x$- and $y$-coordinates display the clumpy and diffuse derived values, respectively. The line $y = x$ is plotted with the grey-dashed line to help visualize any trends.

### 5.1.6 *Scale lengths*

As already discussed and explained in Section 4.4.1 for NGC 891, the scale length of the stellar discs are usually longer in the clumpy model than in the diffuse model. This can be seen in the top and middle panel of Fig. 17, where the $x$- and $y$-axes display the clumpy and diffuse derived values for the main disc morphological component. It is clear that the points all lie either on the $y = x$ line (a few) or below it (most of them), with various degree of departure from it.

For the scale length of the dust, there is no trend (see bottom panel of Fig. 17), with galaxies both above and below the equality line. Table 13 lists the scale lengths of all the disc components of the galaxy sample: inner, main, and outer discs, derived with the diffuse and clumpy models.

## 5.2 Spatial variation of key properties in the clumpy model

As shown in this section, the clumpy model solves the long standing energy balance problem in edge-on galaxies and provides consistent solutions to galaxies of any orientation. It is therefore adopted as the standard model. Because of this it is important to analyse the statistical trends of this galaxy sample using the new model.

Since the defining characteristic of the clumpy model is lower dust opacity, it is insightful to study the radial trends in dust opacity. Fig. 18 shows the $\tau_B^f$ versus radial distance (in units of their inner radius) for all the galaxies in the sample. The plot shows that overall most of the galaxies are optically thin when viewed face-on, throughout their discs. The inner regions of M51 and NGC 3521 are definitively optically thick. The figure also shows rather transparent extended outer discs.

Fig. 19 shows the radial variation of SFR surface density ($\Sigma_{SFR}$), sSFR, and the stellar mass surface density ($\Sigma_{M_\star}$). To enable better comparison between galaxies, each plot is normalized to the inner radius of its main dust disc, and the radius is given in units of the inner radius. The SFR surface densities show an overall exponential decline over their main discs, with NGC 3521 exhibiting the longest scale length, and NGC 628 the shortest. $\Sigma_{M_\star}$ is relatively consistent within this galaxy sample: a sharp decline from the centre until the limit of the bulge is reached, after which $\Sigma_{M_\star}$ declines exponentially. Again, NGC 628 shows the shortest scale length, and NGC 3521 the longest.

The sSFR across the sample shows similar behaviour beyond the inner radius; a gradual exponential increase with radius. This is expected for late-type spiral galaxies under the inside–out growth scenario, where the main and outer discs are actively forming stars. M101 and NGC 891 show the greatest rate of increase with radius. While the sSFR for NGC 891 is increasing only modestly, and appears to begin to plateau at larger radii, M101 exhibits a consistently increasing sSFR. This enhanced SFR could be explained by interactions with its environment, particularly with NGC 5474 (J. C. Mihos et al. 2013; S. T. Linden & J. C. Mihos 2022). The outer discs of M51 and M33 exhibit a marked decline in sSFR. TP20 note that the outer disc of M33 may be influenced by environmental effects being a satellite in the Local Group. Gas could be removed from the outer disc via ram-pressure stripping, thereby quenching star formation. Similarly, M51 is a heavily interacting with its companion galaxy, M51b, which significantly influences the star formation in its outer disc.

**Table 9.** The average surface density weighted dust opacity, $\langle \tau^{f}_{B,\,\mathrm{area}} \rangle$ for each morphological component in the galaxy sample for the diffuse (column under 'd') and clumpy (column under 'c') models. The generic morphological components: inner, main, and outer discs are denoted by the subscripts 'i', 'm', and 'o' respectively.

| Galaxy | $\langle \tau^{f}_{B,\,\mathrm{area,\,i}} \rangle$ | | $\langle \tau^{f}_{B,\,\mathrm{area,\,m}} \rangle$ | | $\langle \tau^{f}_{B,\,\mathrm{area,\,o}} \rangle$ | |
|---|---|---|---|---|---|---|
| | d | c | d | c | d | c |
| M33 | 0.2 ± 0.1 | 0.2 ± 0.1 | 0.6 ± 0.1 | 0.3 ± 0.1 | 0.1 ± 0.1 | 0.1 ± 0.1 |
| M51 | 4.5 ± 0.3 | 2.5 ± 0.1 | 1.8 ± 0.1 | 0.7 ± 0.1 | 0.3 ± 0.1 | 0.1 ± 0.1 |
| M101 | 0.3 ± 0.1 | 0.2 ± 0.1 | 0.3 ± 0.1 | 0.2 ± 0.1 | – | – |
| NGC 628 | 0.1 ± 0.1 | 0.03 ± 0.1 | 0.4 ± 0.1 | 0.3 ± 0.1 | – | – |
| NGC 891 | 2.2 ± 0.1 | 0.8 ± 0.1 | 0.9 ± 0.1 | 0.4 ± 0.1 | – | – |
| NGC 3521 | – | – | 1.6 ± 0.1 | 0.9 ± 0.1 | 0.4 ± 0.1 | 0.2 ± 0.1 |
| NGC 3938 | – | – | 0.7 ± 0.1 | 0.3 ± 0.1 | – | – |

**Table 10.** Comparison of global properties of the diffuse ('d') and clumpy ('c') models of the galaxy sample, including SFR, sSFR, fraction of stellar light reradiated by dust $f^{\mathrm{abs}}$, and the percentage of the dust heating powered by the young stellar disc, $F^{\mathrm{dust}}_{\mathrm{young}}$.

| Galaxy | SFR ($\mathrm{M_\odot\ yr^{-1}}$) | | sSFR ($\times 10^{-11}\ \mathrm{yr^{-1}}$) | | $f^{\mathrm{abs}}$ | | $F^{\mathrm{dust}}_{\mathrm{young}}$ (per cent) | |
|---|---|---|---|---|---|---|---|---|
| | d | c | d | c | d | c | d | c |
| M33 | 0.48 ± 0.04 | 0.30 ± 0.02 | 7.35 ± 0.66 | 7.08 ± 0.64 | 0.33 | 0.35 | 80 | 69 |
| M51 | 4.1 ± 0.4 | 2.9 ± 0.2 | 4.36 ± 0.38 | 3.30 ± 0.18 | 0.46 | 0.63 | 61 | 45 |
| M101 | 3.2 ± 0.2 | 2.5 ± 0.2 | 6.23 ± 0.55 | 4.54 ± 0.40 | 0.33 | 0.39 | 71 | 62 |
| NGC 628 | 2.0 ± 0.2 | 1.7 ± 0.1 | 9.52 ± 0.85 | 8.77 ± 0.78 | 0.38 | 0.51 | 67 | 62 |
| NGC 891 | 7.2 ± 0.6 | 4.6 ± 0.4 | 6.06 ± 0.53 | 6.04 ± 0.54 | 0.37 | 0.48 | 66 | 72 |
| NGC 3521 | 3.3 ± 0.2 | 1.7 ± 0.1 | 3.49 ± 0.31 | 1.99 ± 0.18 | 0.33 | 0.25 | 58 | 37 |
| NGC 3938 | 2.2 ± 0.1 | 1.8 ± 0.1 | 5.13 ± 0.46 | 3.84 ± 0.34 | 0.34 | 0.40 | 64 | 56 |

**Table 11.** SFR for the nuclear and inner morphological components in the galaxy sample. These are denoted by the subscript 'n' and 'i' respectively. Values under 'd' and 'c' correspond to the diffuse and clumpy models respectively.

| Galaxy | $\mathrm{SFR_n}$ ($\mathrm{M_\odot\ yr^{-1}}$) | | $\mathrm{SFR_i}$ ($\mathrm{M_\odot\ yr^{-1}}$) | |
|---|---|---|---|---|
| | d | c | d | c |
| M33 | $(1.0 \pm 0.1) \times 10^{-3}$ | $(9.1 \pm 0.2) \times 10^{-4}$ | $(1.70 \pm 0.06) \times 10^{-2}$ | $(1.40 \pm 0.08) \times 10^{-2}$ |
| M51 | – | – | 1.4 ± 0.2 | 0.48 ± 0.07 |
| M101 | $(6.0 \pm 0.3) \times 10^{-2}$ | $(3.2 \pm 0.2) \times 10^{-2}$ | $(2.9 \pm 0.2) \times 10^{-2}$ | $(6.0 \pm 0.5) \times 10^{-3}$ |
| NGC 628 | – | – | $(4.0 \pm 0.4) \times 10^{-2}$ | $(3.0 \pm 0.3) \times 10^{-2}$ |
| NGC 891 | – | – | 0.94 ± 0.08 | 0.26 ± 0.03 |
| NGC 3521 | – | – | $(2.7 \pm 0.2) \times 10^{-2}$ | $(8.0 \pm 0.8) \times 10^{-3}$ |
| NGC 3938 | – | – | – | – |

**Table 12.** Continuation of Table 11. SFR for the main and outer morphological components in the galaxy sample. These are denoted by the subscript 'm' and 'o' respectively. Values under 'd' and 'c' correspond to the diffuse and clumpy models, respectively.

| Galaxy | $\mathrm{SFR_m}$ ($\mathrm{M_\odot\ yr^{-1}}$) | | $\mathrm{SFR_o}$ ($\mathrm{M_\odot\ yr^{-1}}$) | |
|---|---|---|---|---|
| | d | c | d | c |
| M33 | 0.45 ± 0.04 | 0.27 ± 0.02 | $(1.8 \pm 0.2) \times 10^{-2}$ | $(1.2 \pm 0.1) \times 10^{-2}$ |
| M51 | 2.73 ± 0.20 | 1.75 ± 0.13 | 0.57 ± 0.05 | 0.64 ± 0.06 |
| M101 | 3.09 ± 0.23 | 2.43 ± 0.18 | – | – |
| NGC 628 | 1.96 ± 0.15 | 1.63 ± 0.12 | – | – |
| NGC 891 | 6.31 ± 0.47 | 4.37 ± 0.33 | – | – |
| NGC 3521 | 2.49 ± 0.19 | 1.30 ± 0.10 | 0.82 ± 0.07 | 0.37 ± 0.03 |
| NGC 3938 | 2.20 ± 0.16 | 1.73 ± 0.13 | – | – |

**Table 13.** Scale lengths of all disc components of the galaxy sample. Where applicable, the nuclear, inner, main, and outer discs are denoted with 'n', 'i', 'm', and 'o' respectively. Values under 'd' and 'c' correspond to the diffuse and clumpy models respectively. All values are given in kpc.

| Galaxy | Morph comp | Parameter (kpc) $h_s^{tdisc}$ d | c | $h_s^{disc}(B)$ d | c | $h_d^{disc}$ d | c |
|---|---|---|---|---|---|---|---|
| M33 | n | 0.02 | 0.02 | – | – | – | – |
| | i | 0.10 | 0.13 | 0.05 | 0.05 | 0.15 | 0.15 |
| | m | 1.5 | 1.8 | 1.8 | 1.8 | 9.0 | 8.0 |
| | o | 0.6 | 0.65 | 1.0 | 1.0 | 1.0 | 1.3 |
| M51 | i | 0.6 | 0.55 | 0.74 | 0.72 | 5 | 4.7 |
| | m | 4.3 | 6.5 | 3.8 | 5.5 | 6.0 | 6.0 |
| | o | 1.45 | 1.80 | 2.0 | 2.4 | 3.6 | 4.5 |
| M101 | n | 0.06 | 0.08 | – | – | – | – |
| | i | 0.4 | 0.5 | 0.4 | 0.5 | 0.90 | 0.75 |
| | m | 5.2 | 5.5 | 3.7 | 4.3 | 8.0 | 8.5 |
| NGC 628 | i | 0.2 | 0.15 | – | – | – | – |
| | m | 2.8 | 3.1 | 2.9 | 3.3 | 7.3 | 7.8 |
| NGC 891 | i | 3.0 | 3.0 | – | – | 3.0 | 3.0 |
| | m | 5.0 | 5.0 | 10 | 10 | 10 | 10 |
| NGC 3521 | i | 0.5 | 0.5 | – | – | – | – |
| | m | 3.3 | 3.4 | 2.0 | 2.0 | 7.5 | 7.8 |
| | o | 2.3 | 2.8 | 4.1 | 4.1 | 7.5 | 7.0 |
| NGC 3938 | m | 3.2 | 3.8 | 3.00 | 3.15 | 9.0 | 7.5 |

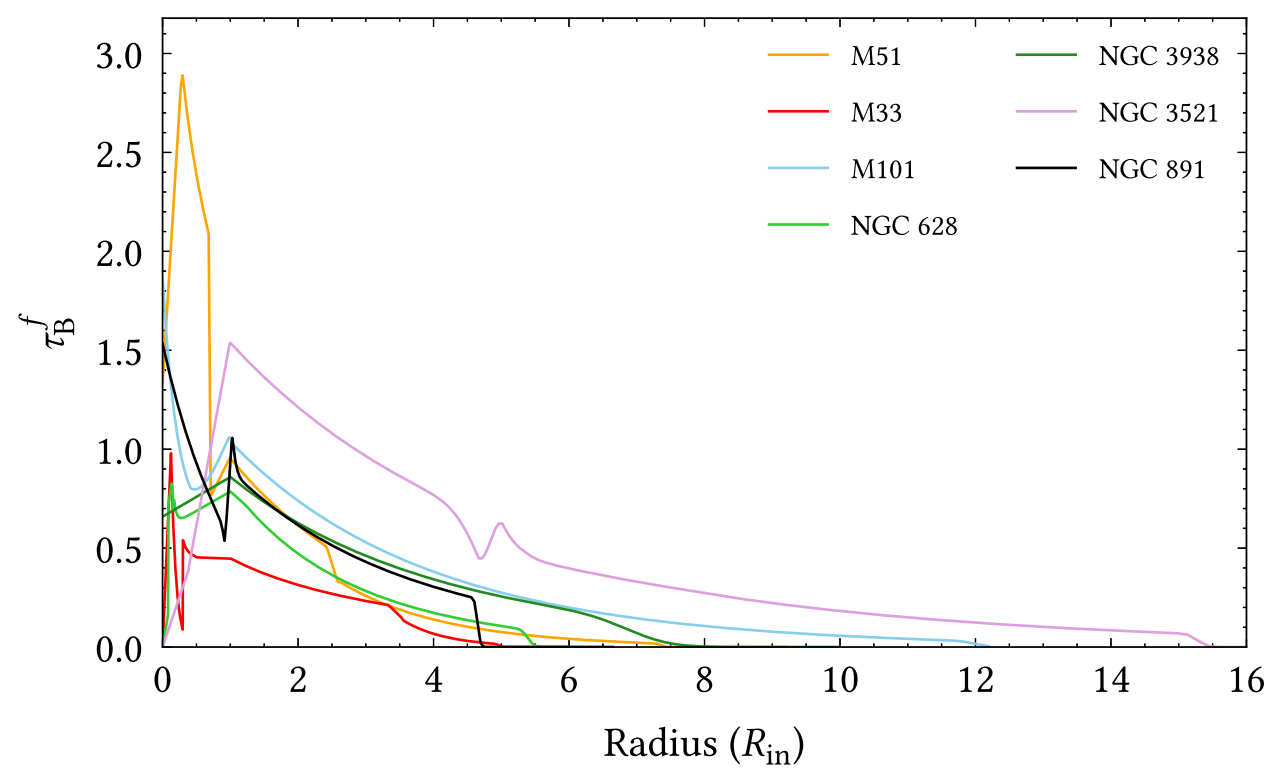

**Figure 18.** The radial variation of $\tau_B^f$ for each galaxy in the sample as a function of radius in units of the inner radius of the main dust disc.

Fig. 20 shows the radial variation of $F_{young}^{dust}$ and $F_{old}^{dust}$ for the clumpy models across the galaxy sample. In all cases, the energy absorbed near the galaxy centres is dominated by the bulge. The only exception is M33, which does not have a bulge but instead hosts a nuclear thin disc comprised of young stars. A consistent trend is observed: the contribution from young stars increases with radius until the influence of the bulge diminishes, after which $F_{young}^{dust}$ either plateaus or continues to gradually increase before declining near the edge of the galaxy. M51 largely follows this behaviour within $R \approx 7$ kpc, beyond which it abruptly reverses trend exhibiting a decreasing $F_{young}^{dust}$. As discussed by C. J. Inman et al. (2023), the spiral structure of M51 is well preserved up to this radius, beyond which interactions with its lenticular companion, M51b, become significant. These interactions compromise the assumption of axisymmetry of the model beyond this point.

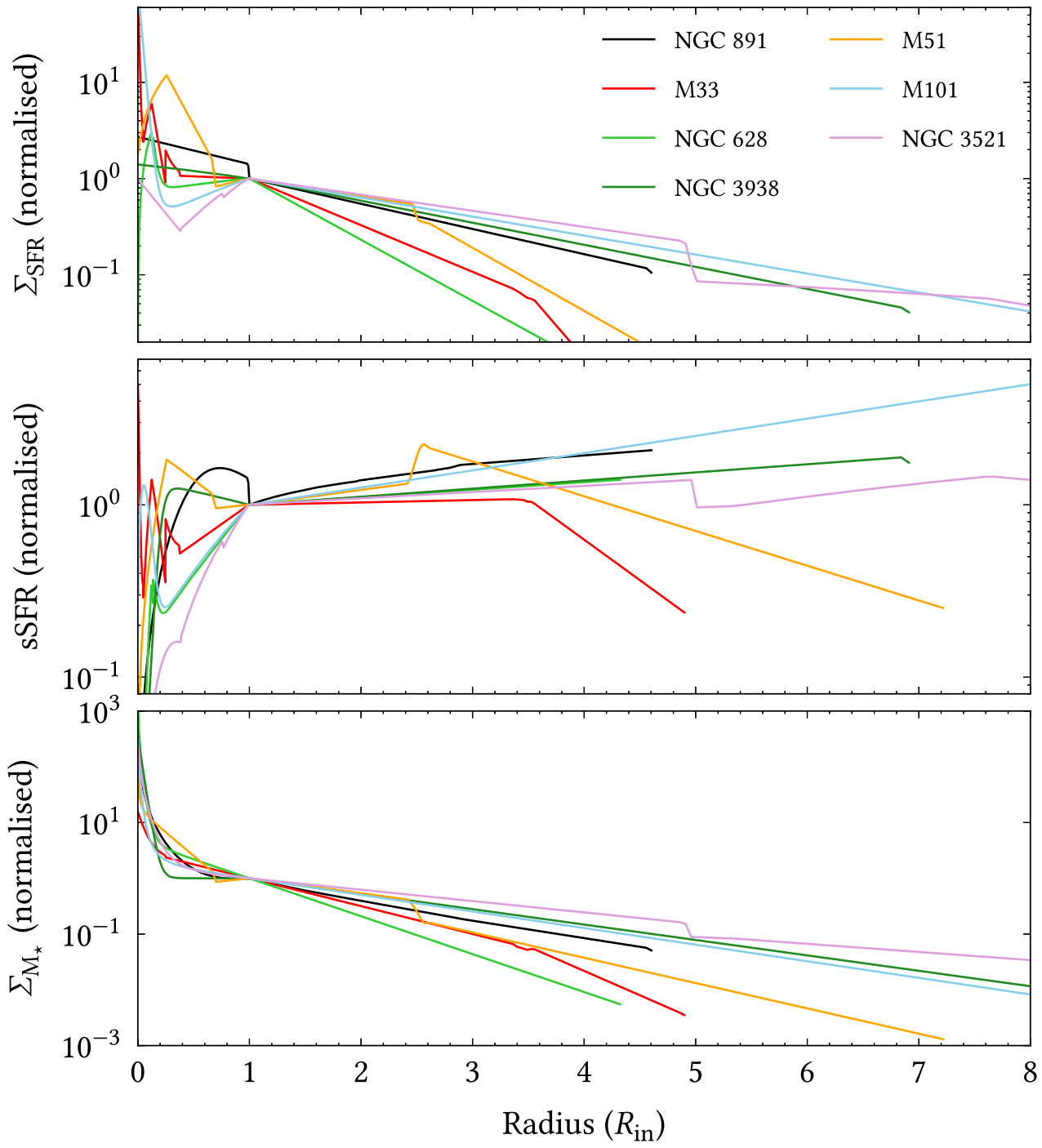

**Figure 19.** Radial profiles of surface densities (face-on) for key stellar parameters derived for the galaxy sample. The top panel shows the surface density of star formation $\Sigma_{SFR}$. The middle panel shows the sSFR. The bottom panel shows the surface density of stellar mass $\Sigma_{M_\star}$. Each quantity for each galaxy has been normalized at the inner radius of the main dust disc.

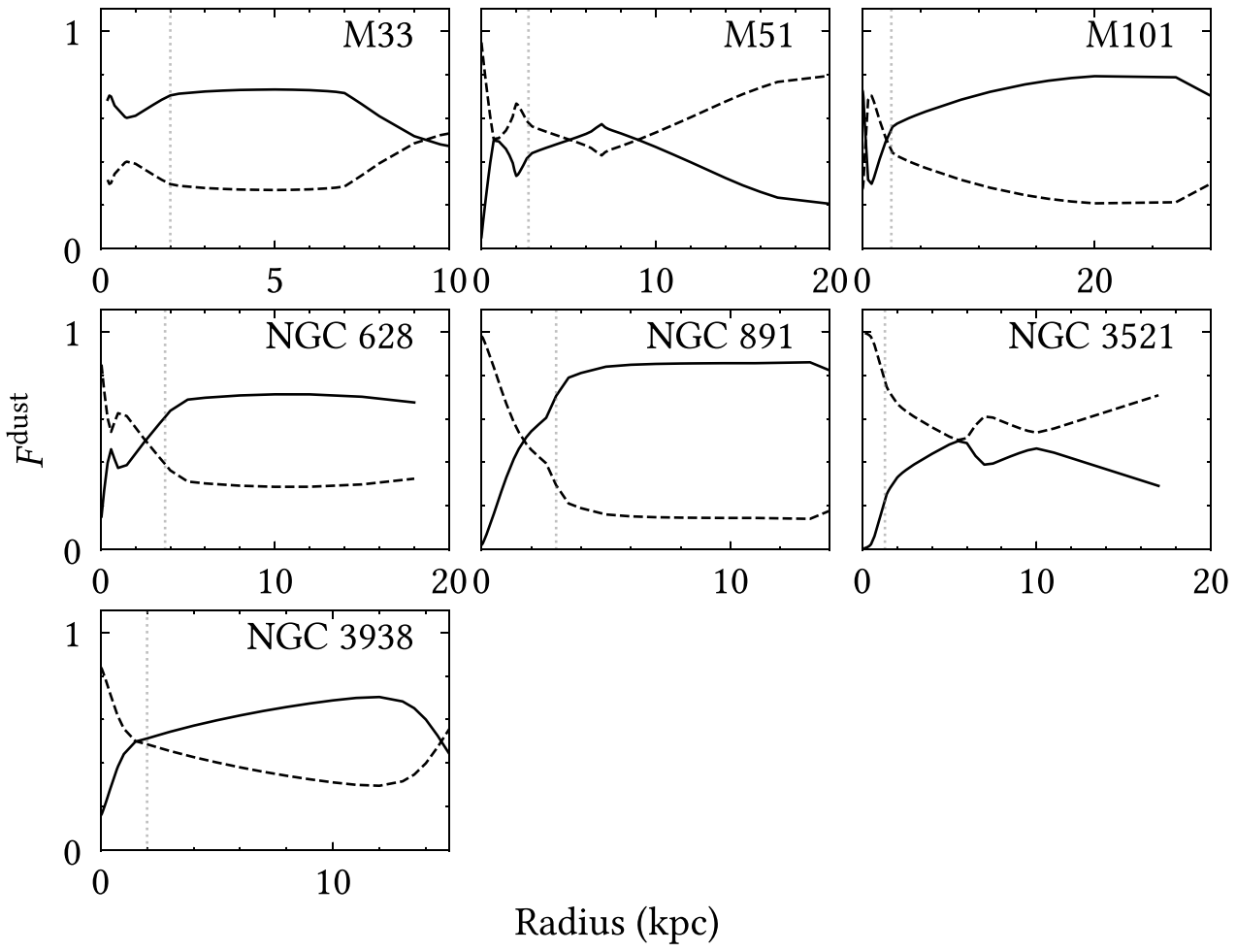

**Figure 20.** Radial profiles of the total energy absorbed by the young ($F_{young}^{dust}$, solid line) and old ($F_{old}^{dust}$, dashed line) stellar populations. The inner radius of the main dust disc of each galaxy is plotted with the grey dotted lines.

### 5.2.1 *Trends and variability in the galaxy sample*

A total of seven galaxies have now been modelled using the RT codes of PT11 updated with the clumpy formalism. From this small sample, trends in various properties begin to emerge. For example, each galaxy was successfully modelled with an MW-type dust whose properties are described in J. C. Weingartner & B. T. Draine (2001), and B. T. Draine & A. Li (2007), without needing to invoke different grain properties. However, in previous work using the diffuse model (C. J. Inman et al. 2023) it was found that in the case of M51, allowing for varying dust properties provided better fits in the PAH region and a more realistic UV intrinsic stellar SED. In that work, the derived attenuation curve suggested the need for a dust model with a reduced 2200 Å bump in the inner disc, which was modelled with a Large-Magellanic-Cloud-type dust (J. C. Weingartner & B. T. Draine 2001).

In this work, it was found that the inner disc of M51 was much better fit using the clumpy model with an MW-type dust, demonstrating possible degeneracies between the microscopic and macroscopic dust properties. Since the clumpy model provides a consistent approach for all the galaxies in the sample, it is concluded that an MW-type dust also provides a consistent solution for all modelled galaxies within the framework of the clumpy model.

In addition, no evidence is found of a 'submm excess' (C. Bot et al. 2010; A. Rémy-Ruyer et al. 2013; I. Hermelo et al. 2016; T. G. Williams et al. 2019), as the submm emission is well fitted across the galaxy sample, without need of modifying the dust grain emissivity. This can be attributed to the fitting of the detailed geometry, informed by the panchromatic data set.

It is found that on average a $f^{c} = 0.5 \pm 0.1$ is able to adequately reproduce the panchromatic data. Six of the eight galaxies in this sample were best fitted with an $f^{c} = 0.5$, while M101 and M51 were best fitted with an $f^{c} = 0.4$ and 0.6, respectively.

## 6 SUMMARY

The goal of this work was to solve the so-called energy balance problem between direct stellar light and dust emission (C. C. Popescu et al. 2000; S. Bianchi et al. 2000; K. M. Dasyra et al. 2005; S. Bianchi 2007, 2008; M. Baes et al. 2010; C. C. Popescu et al. 2011; W. Saftly et al. 2015; G. De Geyter et al. 2015; A. V. Mosenkov et al. 2018). This problem was encountered when modelling the UV/optical/FIR/submm SEDs of edge-on spiral galaxies with RT models, and was attributed to the inability of these models to account for both the appearance of the dust lanes in edge-on galaxies as well as for the observed submm dust emission. In general, it was found that the solution needed to fit the optical/NIR images produced an underestimation of the submm emission by factors of $3 - 5$, while solutions constrained to fit the submm emission predicted too prominent dust lanes. On the other hand, good agreement between model and observations was found when modelling face-on galaxies (I. De Looze et al. 2014; S. Viaene et al. 2017; S. Verstocken et al. 2020; A. Nersesian et al. 2020a, b; C. J. Inman et al. 2023; D. Pricopi et al. 2025).

In order to solve this long-standing problem and find a consistent solution for both face-on and edge-on galaxies, this work uses quiescent clumps in the RT modelling. The approach is to use quiescent dust clouds (with no internal stellar sources) heated by the diffuse interstellar RFs as 'pseudo-dust grains' with equivalent optical constants given by the attenuation properties of the clouds. This introduces a subgrid approach in the RT models, allowing the parsec and subparsec scales to be incorporated into the large-scale calculations of the galaxy discs (kiloparsec scales).

RT calculations of a single dust cloud illuminated externally by stellar RFs were done to quantify the amount of absorption and scattering produced by the clump. From these, the absorption and scattering efficiencies of the cloud, $Q_{\rm abs}$ and $Q_{\rm sca}$, as well as the corresponding cross-sections, $C_{\rm abs}$ and $C_{\rm sca}$, were calculated. The temperature of the dust within the cloud was also calculated, enabling the derivation of the dust emission SEDs of the clouds. Since the calculation of the dust emission SED depends on the intensity and colour of the RFs heating the dust in which the clump is embedded, SEDs were calculated for clumps exposed to RFs of different strength and colour, and from this, a corresponding library of dust emission SEDs for the clump was produced. In all these calculations, the same RT codes and techniques were utilized as for the large-scale galaxy calculations, but for the geometry of the clump. Specifically, the RT codes from PT11 were used. The results were also checked using the RT code DART-Ray (G. Natale et al. 2014, 2015, 2017). Both codes employ a ray-tracing algorithm accounting for both the absorption and anisotropic scattering of stellar light by dust grains of various sizes and chemical composition. The dust model used is comprised of astronomical silicates and carbonaceous grains in form of graphites and PAH molecules, with a grain size distribution from J. C. Weingartner & B. T. Draine (2001), and optical constants from B. T. Draine & H. M. Lee (1984) and B. T. Draine & A. Li (2007).

In the present model, the clumps have a volume emissivity described by spherically symmetric I. King (1962) profile, a core radius $r_{\rm c} = 1$ pc, and a truncation radius $r_{\rm t} = 1$ pc, with the dust opacity through the clump (from the centre to the truncation radius) of $\tau_B^{\rm c,s} = 1$. These clumps describe the cirrus-like clouds observed in the diffuse ISM, being optically thin in the optical range, but becoming progressively optically thick in the UV. Clumps that are optically thick in the optical range have also been calculated, but were not found to provide adequate solutions to the galaxy models, within the framework of the existing dust grain properties.

The derivation of the extinction and emission properties of the clump enabled the virtualization of the macroscopic dust cloud into a microscopic pseudo-grain, that was supplemented into the existing dust grain mixture of graphite, silicate grains, and PAH molecules. This modified dust model was found to be characterized by a flatter extinction curve. Thus, a new parameter was added to the RT models of galaxies: the fraction of total dust mass in clumps $f^{c}$.

The new formalism was tested and applied to the edge-on galaxy NGC 891, for which the energy balance problem was first detected (C. C. Popescu et al. 2000). The clumpy model with $f^{c} = 0.5$ achieves better fits to the data than the pure diffuse model, and most critically reproduces the dust lane seen in the $J$ and $K_{\rm s}$ bands, thus solving the energy balance problem for this galaxy.

The new formalism was then applied to a small sample of galaxies of various inclinations, to test the general applicability of the model. The sample was chosen to include the face-on galaxies already modelled by us with the pure diffuse model, M33 TP20, NGC 628 (M. T. Rushton et al. 2022), M51 (C. J. Inman et al. 2023), and M101 and NGC 3938 (D. Pricopi et al. 2025). In addition, the sample includes the intermediate-inclination galaxy NGC 3521. It was found that the clumpy model successfully accounted for the panchromatic images of all galaxies in the sample.

The results of this analysis indicate that the clumpy models are characterized by a reduction in attenuation with respect to the pure diffuse models, particularly at shorter wavelengths in the UV. As a result the clumpy models predict a 21–64 per cent reduction in maximum face-on optical depth in the $B$ band ($\tau_B^{\rm f}$) when compared to their purely diffuse counterparts. This is primarily due to the allocation of a portion of the dust mass away from the diffuse component, but also in part due to the differences in modelling that arises from using a modified dust model. The reduced optical depth is significant as only two of the seven galaxies modelled with the clumpy model yield solutions with an optically thick centre (M101 and M51), with a further two galaxies (NGC 891 and NGC 3521) being moderately optically thick. The remaining three galaxies are optically thin with $\tau_B^f \leq 1$. In contrast, the diffuse models predict that all the galaxies are optically thick, or moderately optically thick in their centres. As a consequence to the reduced attenuation, the clumpy models predict values of SFR 15–51 per cent lower than their purely diffuse counterparts. Because dust attenuation does not strongly affect the NIR bands, the stellar masses determined using the 3.6 and 4.5 μm fluxes described by M. Eskew et al. (2012) are relatively unchanged between the clumpy and diffuse models. This, in conjunction with a reduced SFR, implies that the clumpy model systematically derives smaller sSFR when compared to the pure diffuse counterpart.

To conclude, in addition to solving the energy balance problem, this study also gives an answer to the long-standing debate of whether spiral galaxies are optically thin or thick. In their fundamental paper, 'Are spiral galaxies optically thin or thick?' E. M. Xilouris et al. (1999) gave the following answer to the question:

'The face-on central optical depth is less than one in all optical bands, indicating that typical spiral galaxies like the ones that we have modelled would be completely transparent if they were to be seen face-on.' The conclusion was reached from the RT analysis of the optical/NIR images of spiral galaxies. The inclusion of the dust emission in the overall RT modelling complicated the situation and changed the answer in favour of galaxies being optically thick in their centre (C. C. Popescu et al. 2000, 2011). However, the energy balance problem in edge-on galaxies indicated that the definitive solution was yet to be found. The incorporation of the quiescent clumps into the model returns the answer to the original one provided by E. M. Xilouris et al. (1999), albeit with the backup of the panchromatic dimension. Indeed, in this work, it was found that spiral galaxies are overall optically thin when viewed face-on, although some are still optically thick in their centre.

## ACKNOWLEDGEMENTS

The authors would like to thank an anonymous referee for their expert comments, that very much helped improve the manuscript. CJI acknowledges support from a Science and Technology Facilities Council studentship grant (grant no. ST/W507386/1). CJI would like to thank for making use of the computing facilities of the MPIK.

CCP would like to acknowledge discussions with Dr Richard Tuffs from MPIK regarding the pseudo-grain concept. This concept was born from joint discussions that go back many years ago, as well as from more recent ones regarding the practicalities of implementation. CCP would also like to thank for the visiting positions offered in the past at the Max-Planck-Institut für Kernphysik while collaborating with Dr Tuffs.

This work is based in part on observations made with the NASA Galaxy Evolution Explorer. *GALEX* is operated for NASA by the California Institute of Technology under NASA contract NAS5-98034. This research has made use of the NASA/IPAC Infrared Science Archive, which is operated by the Jet Propulsion Laboratory, California Institute of Technology, under contract with the National Aeronautics and Space Administration.

This work is also based on the Sloan Digital Sky Survey (SDSS) data. Funding for the SDSS IV has been provided by the Alfred P. Sloan Foundation, the U.S. Department of Energy Office of Science, and the Participating Institutions. SDSS acknowledges support and resources from the Center for High-Performance Computing at the University of Utah. The SDSS web site is www.sdss4.org. SDSS is managed by the Astrophysical Research Consortium for the Participating Institutions of the SDSS Collaboration including the Brazilian Participation Group, the Carnegie Institution for Science, Carnegie Mellon University, Center for Astrophysics | Harvard & Smithsonian (CfA), the Chilean Participation Group, the French Participation Group, Instituto de Astrofísica de Canarias, The Johns Hopkins University, Kavli Institute for the Physics and Mathematics of the Universe (IPMU)/University of Tokyo, the Korean Participation Group, Lawrence Berkeley National Laboratory, Leibniz Institut für Astrophysik Potsdam (AIP), Max-Planck-Institut für Astronomie (MPIA Heidelberg), Max-Planck-Institut für Astrophysik (MPA Garching), Max-Planck-Institut für Extraterrestrische Physik (MPE), National Astronomical Observatories of China, New Mexico State University, New York University, University of Notre Dame, Observatório Nacional/MCTI, The Ohio State University, Pennsylvania State University, Shanghai Astronomical Observatory, United Kingdom Participation Group, Universidad Nacional Autónoma de México, University of Arizona, University of Colorado Boulder, University of Oxford, University of Portsmouth, University of Utah, University of Virginia, University of Washington, University of Wisconsin, Vanderbilt University, and Yale University.

This work has also made use of data products from the Two Micron All Sky Survey, which is a joint project of the University of Massachusetts and the Infrared Processing and Analysis Center/California Institute of Technology, funded by the National Aeronautics and Space Administration and the National Science Foundation. This work is based in part on observations made with the *Spitzer Space Telescope*, which is operated by the Jet Propulsion Laboratory, California Institute of Technology under a contract with NASA. We also utilize observations performed with the ESA *Herschel Space Observatory* (G. L. Pilbratt et al. 2010), in particular to do photometry using the PACS (A. Poglitsch et al. 2010) and SPIRE (M. J. Griffin et al. 2010) instruments.

## DATA AVAILABILITY

The data underlying this article will be shared on reasonable request to the corresponding author.

## APPENDIX A: NGC 891 FITS TO THE SURFACE-BRIGHTNESS PROFILES AND MAPS

In this appendix, the fits to the surface-brightness profiles and maps of NGC 891 are presented for all wavelengths where data were used.

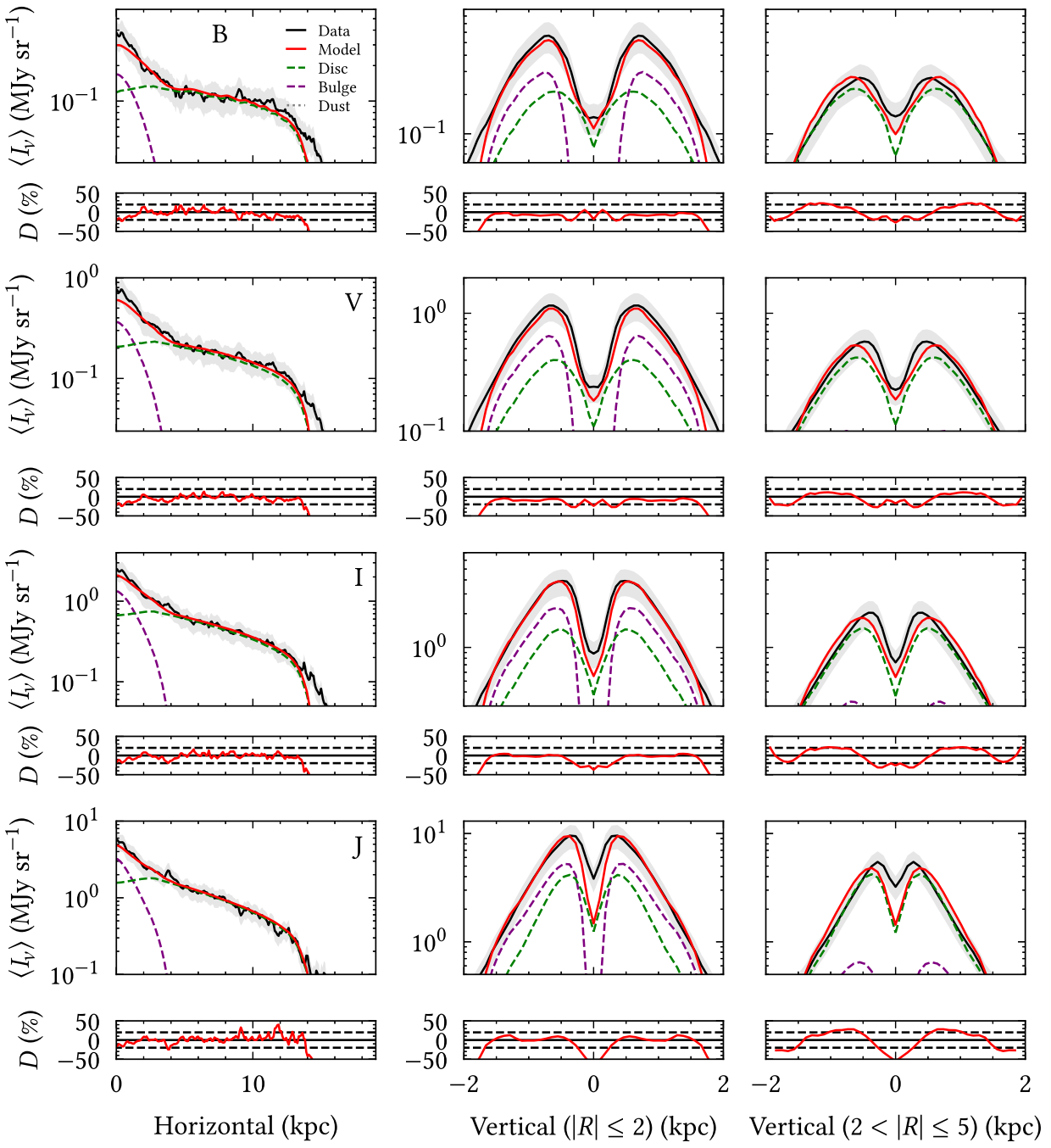

**Figure A1.** NGC 891: averaged horizontal SB profiles (mirrored about the central vertical axis, LH panel) and averaged vertical SB profiles (middle and RH panel) of the pure diffuse model at the optical wavelengths. The observed SB profiles are plotted with solid black line with the shaded banding indicating the uncertainty. The solid red line indicates the model total, with the individual component contributions plotted with the dashed and dotted lines. The percent differences between the model total and the observed profiles, $D$[ per cent], are plotted in the panels below each profile, with the dashed horizontal lines indicating $\pm 20$ per cent deviation.

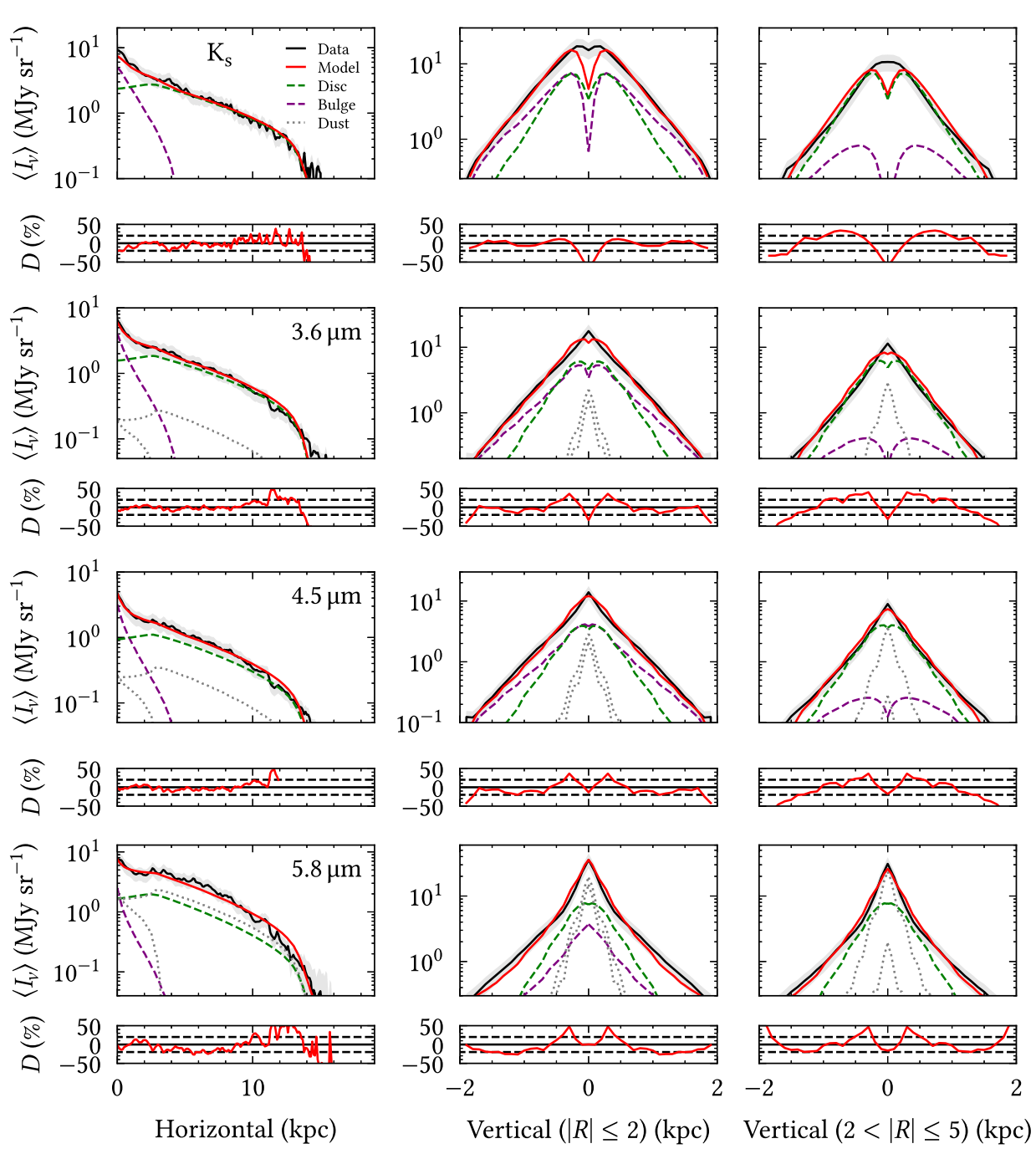

**Figure A2.** NGC 891: same as Fig. A1 for the diffuse model at NIR wavelengths.

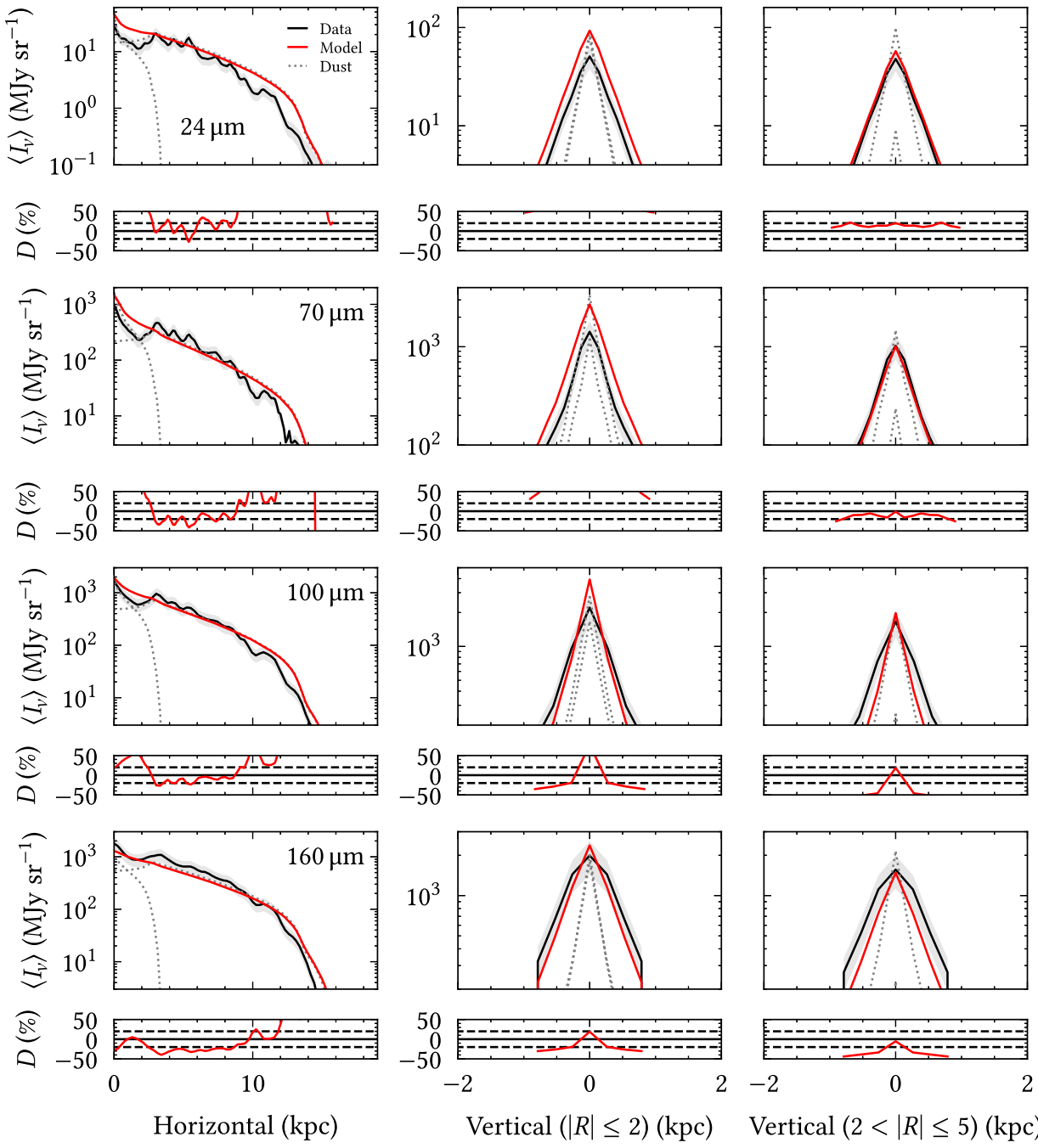

**Figure A3.** NGC 891: same as Fig. A1 for the diffuse model at MIR/FIR wavelengths.

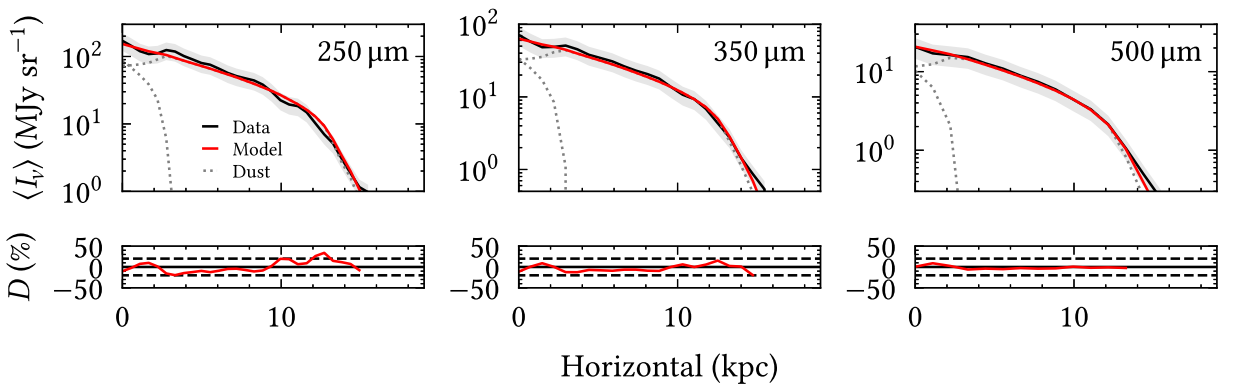

**Figure A4.** NGC 891: averaged horizontal SB profiles (mirrored about the central vertical axis) of the pure diffuse model at wavelengths in the FIR/submm. The observed SB profiles are plotted with the solid black line with the shaded banding indicating the uncertainty. The solid red line indicates the model total, with the individual component contributions plotted with the dashed and dotted lines. The percent differences between the model total and the observed profiles, $D$[ per cent], are plotted in the panels below each profile, with the dashed horizontal lines indicating $\pm 20$ per cent deviation.

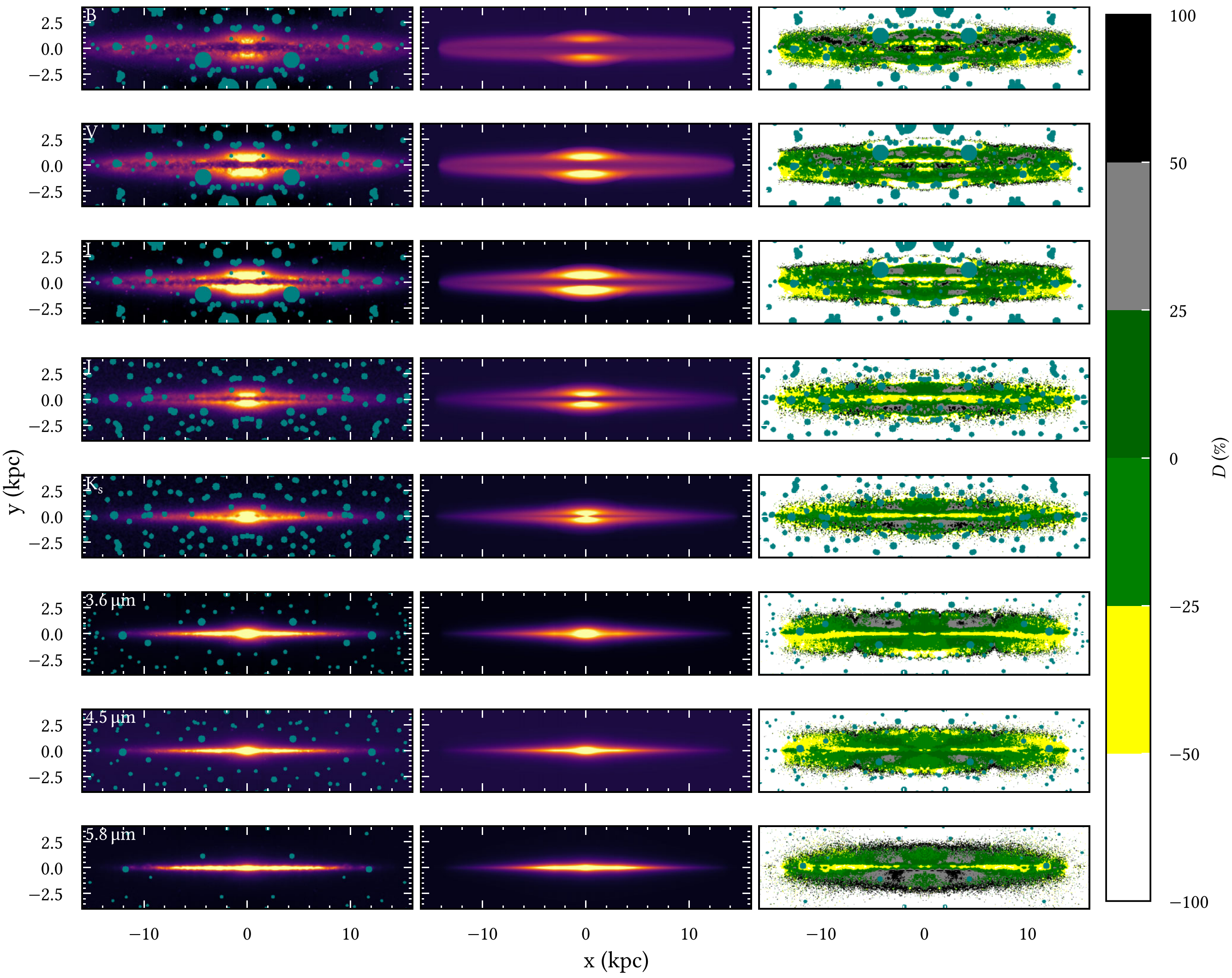

**Figure A5.** NGC 891: fits of the pure diffuse model to the surface-brightness maps. Left: surface-brightness maps of NGC 891, mirrored about their vertical central axis. Middle: surface-brightness maps of the corresponding diffuse model images. Right: residuals between the data and model images calculated as $D = (M - O)/O$. Masked foreground stars are marked in teal.

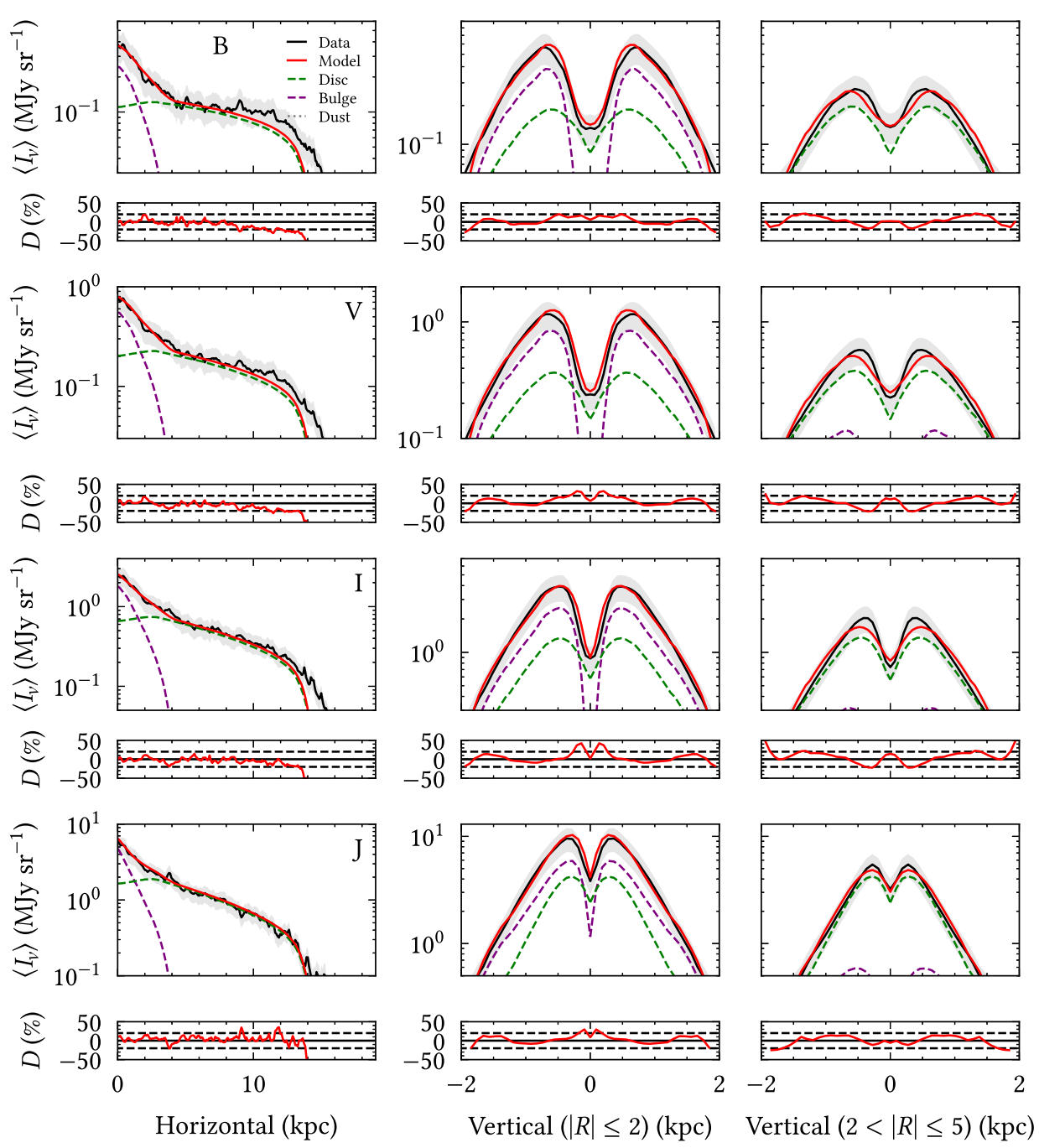

**Figure A6.** NGC 891: averaged horizontal SB profiles (mirrored about the central vertical axis, LH panel) and averaged vertical SB profiles (middle and RH panels) of the clumpy model at the optical wavelengths. The observed SB profiles are plotted with solid black line with the shaded banding indicating the uncertainty. The solid red line indicates the model total, with the individual component contributions plotted with the dashed and dotted lines. The percent differences between the model total and the observed profiles, $D$[ per cent], are plotted in the panels below each profile, with the dashed horizontal lines indicating $\pm 20$ per cent deviation.

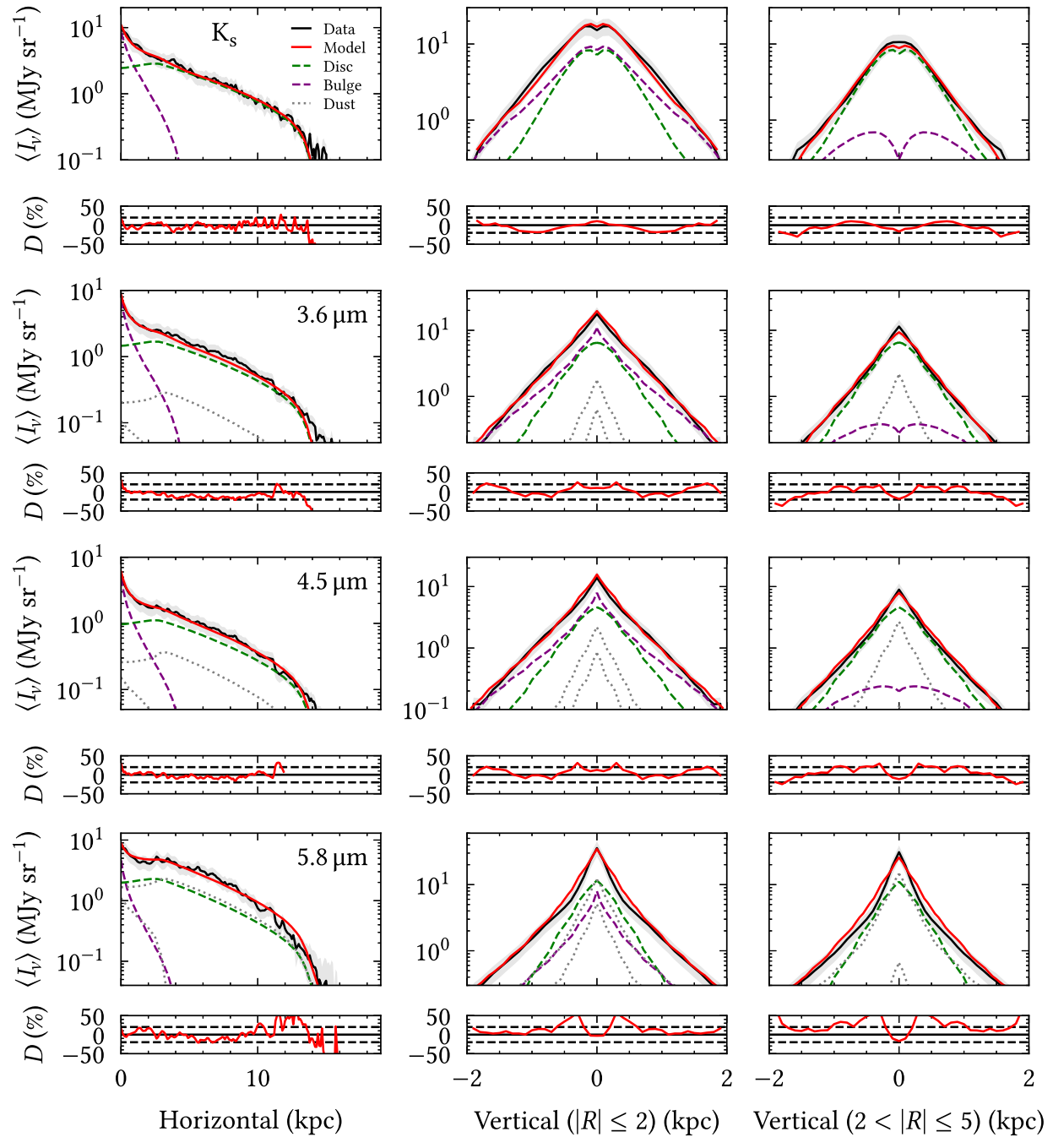

**Figure A7.** NGC 891: same as Fig. A6 for the clumpy model at NIR wavelengths.

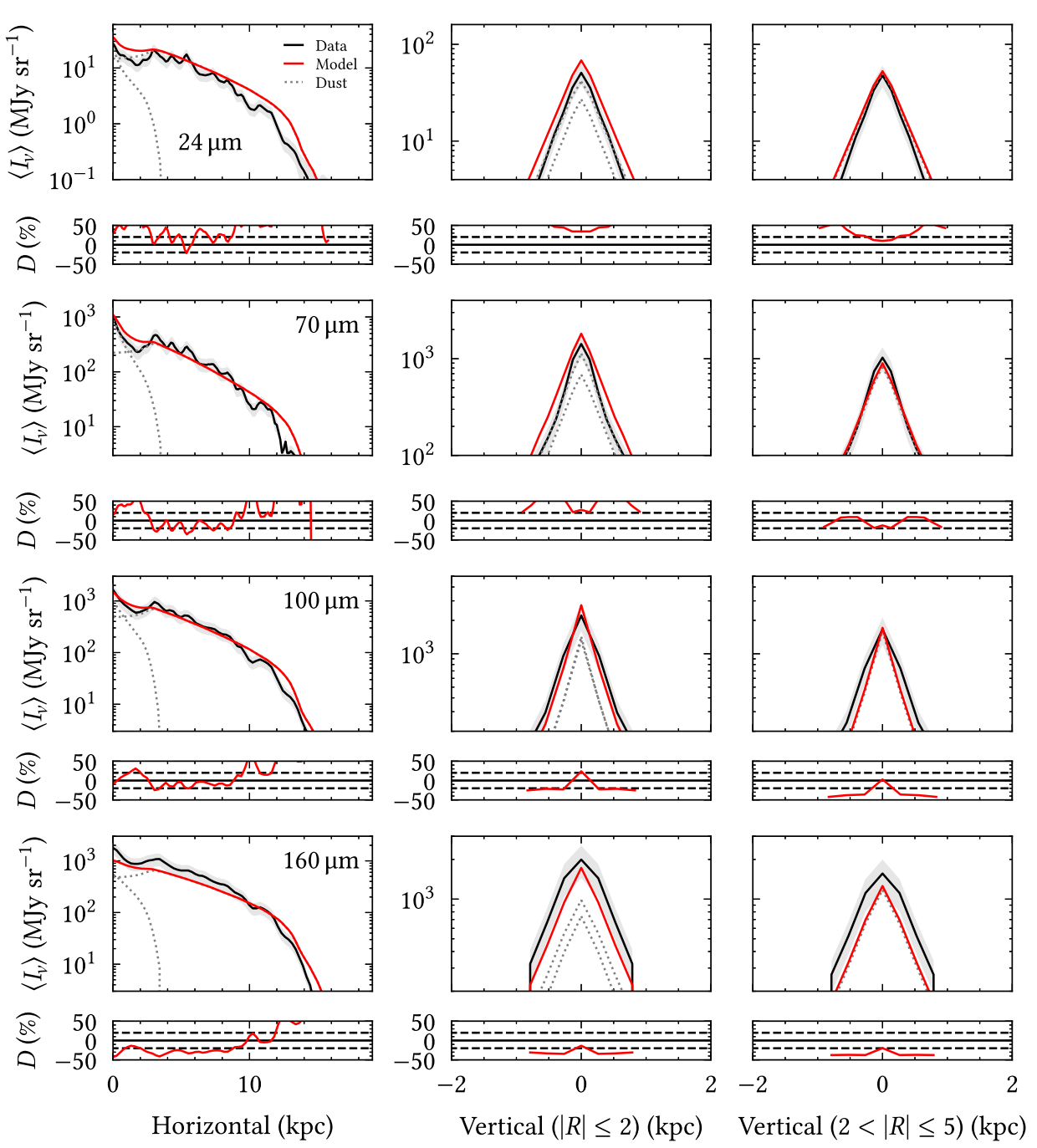

**Figure A8.** NGC 891: same as Fig. A6 for the clumpy model at MIR/FIR wavelengths.

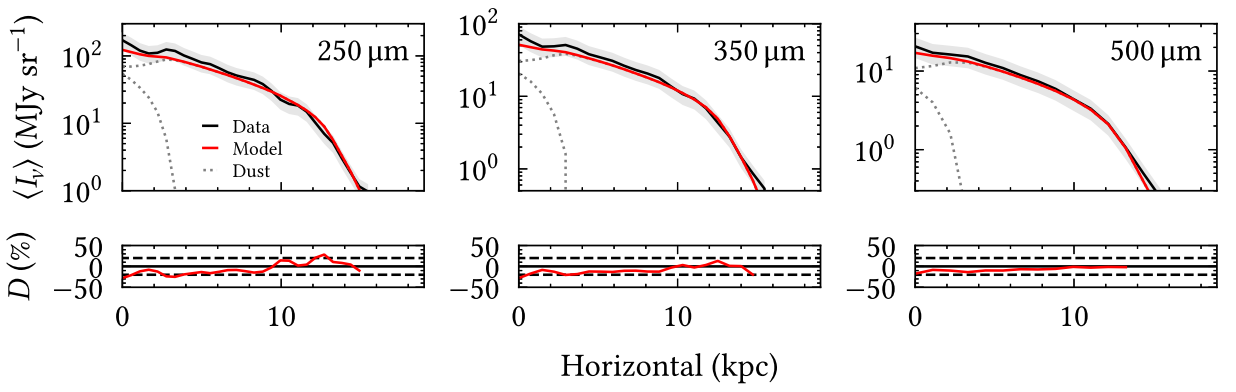

**Figure A9.** NGC 891: averaged horizontal SB profiles (mirrored about the vertical axis) of the clumpy model at wavelengths in the FIR/submm. The observed SB profiles are plotted with the solid black line with the shaded banding indicating the uncertainty. The solid red line indicates the model total, with the individual component contributions plotted with the dashed and dotted lines. The percent differences between the model total and the observed profiles, $D$[ per cent], are plotted in the panels below each profile, with the dashed horizontal lines indicating $\pm 20$ per cent deviation.

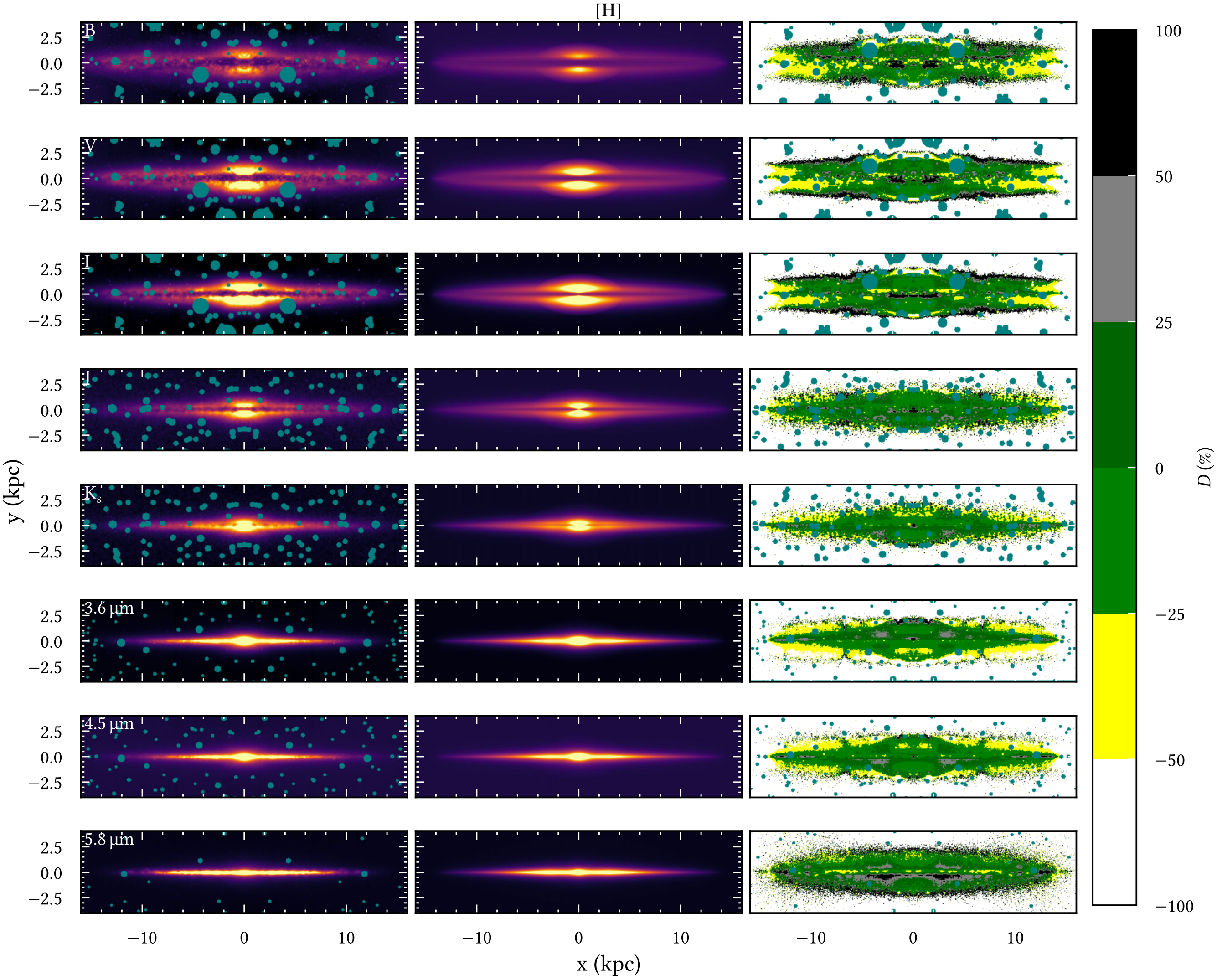

**Figure A10.** NGC 891: fits of the clumpy model to the surface-brightness maps. Left: surface-brightness maps of NGC 891, mirrored about their vertical central axis. Middle: surface-brightness maps of the corresponding diffuse model images. Right: residuals between the data and model images calculated as $D = (M - O)/O$. Masked foreground stars are marked in teal.

## APPENDIX B: GEOMETRICAL PARAMETERS DERIVED FOR THE DIFFUSE AND CLUMPY MODELS

This appendix contains tables listing the geometrical parameters constrained for diffuse and clumpy models for the non-edge-on galaxies in the galaxy sample. The appendices contain the derived fitted and fixed parameters for each galaxy: Tables B1-B2 (M33), Tables B3-B4 (M51), Tables B5-B6 (M101), Tables B7-B8 (NGC 628), Tables B9-B10 (NGC 3521), Tables B11-B12 (NGC 3938). All wavelengths dependent parameters of the clumpy model for each galaxy is given in Table B13.

**Table B1.** The free geometrical parameters of the best-fitting diffuse (middle column), and clumpy models (right column) of M33, denoted with 'd' and 'c', respectively. The superscripts 'n', 'i', 'm', and 'o' found in the notation for the different parameters denote the nuclear, inner, main, and outer discs, respectively.

| Parameter | d | c |
|---|---|---|
| $h_{\rm s}^{\rm tdisc-(n,\,i,\,m,\,o)}$ | (0.02, 0.10, 1.50, 0.60) | (0.02, 0.13, 1.80, 0.65) |
| $h_{\rm s}^{\rm disc-(i,\,m,\,o)}(B)$ | (0.05, 1.80, 1.00) | (0.05, 1.80, 1.00) |
| $h_{\rm d}^{\rm disc-(i,\,m,\,o)}$ | (0.15, 9.00, 1.00) | (0.15, 8.00, 1.30) |
| $\chi_{\rm s}^{\rm tdisc-(n,\,i,\,m,\,o)}$ | (1.00, 0.00, 0.75, −20.00) | (1.00, −0.80, 1.10, −20.00) |
| $\chi_{\rm s}^{\rm disc-(i,\,m,\,o)}$ | (1.00, 1.00, −20.00) | (1.00, 1.00, −20.00) |
| $\chi_{\rm d}^{\rm disc-(i,\,m,\,o)}$ | (0.00, 0.75, −20.00) | (−1.10, 1.00, −20.00) |

**Table B2.** The geometrical parameters fixed from data and theoretical considerations of M33. The values are the same between the diffuse and clumpy models. The superscripts 'n', 'i', 'm', and 'o' found in the notation for the different parameters denote the nuclear, inner, main, and outer discs, respectively.

| Parameter | Value |
|---|---|
| $R_{\rm in,s}^{\rm tdisc-(n,\,i,\,m,\,o)}$ | (0.00, 0.25, 2.00, 7.10) |
| $R_{\rm in,s}^{\rm disc-(i,\,m,\,o)}$ | (0.00, 0.00, 7.10) |
| $R_{\rm in,d}^{\rm disc-(i,\,m,\,o)}$ | (0.25, 2.00, 7.10) |
| $R_{\rm tin,s}^{\rm tdisc-(n,\,i,\,m,\,o)}$ | (0.00, 0.00, 0.60, 6.76) |
| $R_{\rm tin,s}^{\rm disc-(i,\,m,\,o)}$ | (0.00, 0.00, 6.76) |
| $R_{\rm tin,d}^{\rm disc-(i,\,m,\,o)}$ | (0.00, 0.60, 6.76) |
| $R_{\rm t,s}^{\rm tdisc-(n,\,i,\,m,\,o)}$ | (0.10, 0.75, 7.00, 10.00) |
| $R_{\rm t,s}^{\rm disc-(i,\,m,\,o)}$ | (0.50, 7.00, 10.00) |
| $R_{\rm t,d}^{\rm disc-(i,\,m,\,o)}$ | (1.00, 7.00, 10.00) |
| $z_{\rm s}^{\rm tdisc-(n,\,i,\,m,\,o)}$ | (0.09, 0.09, 0.09, 0.09) |
| $z_{\rm s}^{\rm disc-(i,\,m,\,o)}$ | (0.19, 0.19, 0.19) |
| $z_{\rm d}^{\rm disc-(i,\,m,\,o)}$ | (0.16, 0.16, 0.16) |

**Table B3.** Same as Table B1 for M51.

| Parameter | d | c |
|---|---|---|
| $h_{\rm s}^{\rm tdisc-(i,\,m,\,o)}$ | (0.60, 4.30, 1.45) | (0.55, 6.50, 1.80) |
| $h_{\rm s}^{\rm disc-(i,\,m,\,o)}(B)$ | (0.74, 3.80, 2.00) | (0.72, 5.50, 2.40) |
| $h_{\rm d}^{\rm disc-(i,\,m,\,o)}$ | (5.00, 6.00, 3.60) | (4.70, 6.00, 4.50) |
| $\chi_{\rm s}^{\rm tdisc-(i,\,m,\,o)}$ | (0.10, −0.70, 2.00) | (0.15, −0.60, 2.00) |
| $\chi_{\rm s}^{\rm disc-(i,\,m,\,o)}$ | (0.00, 1.50, −5.00) | (0.00, 1.50, −5.00) |
| $\chi_{\rm d}^{\rm disc-(i,\,m,\,o)}$ | (0.45, 1.00, −1.00) | (0.50, 0.75, −1.00) |
| $R_{\rm e}$ | 0.35 | 0.40 |
| $\frac{b}{a}$ | 1.00 | 1.00 |

**Table B4.** Same as Table B2 for M51.

| Parameter | value |
|---|---|
| $R_{\rm in,s}^{\rm tdisc-(i,\,m,\,o)}$ | (0.70, 2.70, 7.10) |
| $R_{\rm in,s}^{\rm disc-(i,\,m,\,o)}$ | (0.00, 2.70, 7.10) |
| $R_{\rm in,d}^{\rm disc-(i,\,m,\,o)}$ | (0.80, 2.70, 7.10) |
| $R_{\rm tin,s}^{\rm tdisc-(i,\,m,\,o)}$ | (0.00, 1.86, 6.70) |
| $R_{\rm tin,s}^{\rm disc-(i,\,m,\,o)}$ | (0.00, 1.86, 6.76) |
| $R_{\rm tin,d}^{\rm disc-(i,\,m,\,o)}$ | (0.00, 1.86, 6.70) |
| $R_{\rm t,s}^{\rm tdisc-(i,\,m,\,o)}$ | (1.90, 6.80, 20.00) |
| $R_{\rm t,s}^{\rm disc-(i,\,m,\,o)}$ | (1.90, 6.80, 20.00) |
| $R_{\rm t,d}^{\rm disc-(i,\,m,\,o)}$ | (1.90, 6.80, 20.00) |
| $z_{\rm s}^{\rm tdisc-(i,\,m,\,o)}$ | (0.09, 0.09, 0.09) |
| $z_{\rm s}^{\rm disc-(i,\,m,\,o)}$ | (0.19, 0.19, 0.19) |
| $z_{\rm d}^{\rm disc-(i,\,m,\,o)}$ | (0.16, 0.16, 0.16) |
| $n_{\rm s}$ | 4.0 |

**Table B5.** Same as Table B1 for M101.

| Parameter | d | c |
|---|---|---|
| $h_{\rm s}^{\rm tdisc-(n,\,i,\,m)}$ | (0.06, 0.40, 5.20) | (0.08, 0.50, 5.50) |
| $h_{\rm s}^{\rm disc-(i,\,m)}(B)$ | (0.40, 3.70) | (0.50, 4.30) |
| $h_{\rm d}^{\rm disc-(i,\,m)}$ | (0.90, 8.00) | (0.75, 8.50) |
| $\chi_{\rm s}^{\rm tdisc-(n,\,i,\,m)}$ | (0.00, 1.60, 0.10) | (0.00, 1.60, 0.10) |
| $\chi_{\rm s}^{\rm disc-(i,\,m)}$ | (0.00, 0.10) | (0.00, 0.10) |
| $\chi_{\rm d}^{\rm disc-(i,\,m)}$ | (0.00, 0.40) | (0.00, 0.40) |
| $R_{\rm e}$ | 0.46 | 0.46 |
| $\frac{b}{a}$ | 0.60 | 0.60 |

**Table B6.** Same as Table B2 for M101.

| Parameter | Value |
|---|---|
| $R_{\rm in,s}^{\rm tdisc-(n,\,i,\,m)}$ | (0.00, 0.00, 2.50) |
| $R_{\rm in,s}^{\rm disc-(i,\,m)}$ | (0.00, 0.00) |
| $R_{\rm in,d}^{\rm disc-(i,\,m)}$ | (0.00, 2.50) |
| $R_{\rm tin,s}^{\rm tdisc-(n,\,i,\,m)}$ | (0.00, 0.00, 0.00) |
| $R_{\rm tin,s}^{\rm disc-(i,\,m)}$ | (0.00, 0.00) |
| $R_{\rm tin,d}^{\rm disc-(i,\,m)}$ | (0.00, 0.00) |
| $R_{\rm t,s}^{\rm tdisc-(n,\,i,\,m)}$ | (1.00, 2.50, 30.00) |
| $R_{\rm t,s}^{\rm disc-(i,\,m)}$ | (2.50, 30.00) |
| $R_{\rm t,d}^{\rm disc-(i,\,m)}$ | (2.50, 30.00) |
| $z_{\rm s}^{\rm tdisc-(n,\,i,\,m)}$ | (0.09, 0.09, 0.09) |
| $z_{\rm s}^{\rm disc-(i,\,m)}$ | (0.40, 0.40) |
| $z_{\rm d}^{\rm disc-(i,\,m)}$ | (0.27, 0.27) |
| $n_{\rm s}$ | 2.0 |

**Table B7.** Same as Table B1 for NGC 628.

| Parameter | d | c |
|---|---|---|
| $h_{\rm s}^{\rm tdisc-(i,\,m)}$ | (0.20, 2.80) | (0.15, 3.10) |
| $h_{\rm s}^{\rm disc-m}(B)$ | 2.90 | 3.30 |
| $h_{\rm d}^{\rm disc-m}$ | 7.30 | 7.80 |
| $\chi_{\rm s}^{\rm tdisc-(i,\,m)}$ | (0.00, 0.35) | (0.00, 0.70) |
| $\chi_{\rm s}^{\rm disc-m}$ | 1.00 | 1.00 |
| $\chi_{\rm d}^{\rm disc-m}$ | 0.80 | 0.75 |
| $R_{\rm e}$ | 0.82 | 0.82 |
| $\frac{b}{a}$ | 0.60 | 0.60 |

**Table B8.** Same as Table B2 for NGC 628.

| Parameter | Value |
|---|---|
| $R_{\rm in,s}^{\rm tdisc-(i,\,m)}$ | (0.55, 4.50) |
| $R_{\rm in,s}^{\rm disc-m}$ | 0 |
| $R_{\rm in,d}^{\rm disc-m}$ | 3.70 |
| $R_{\rm tin,s}^{\rm tdisc-(i,\,m)}$ | (0, 0.62) |
| $R_{\rm tin,s}^{\rm disc-m}$ | 0 |
| $R_{\rm tin,d}^{\rm disc-m}$ | 0.30 |
| $R_{\rm t,s}^{\rm tdisc-(i,\,m)}$ | (20.0, 20.0) |
| $R_{\rm t,s}^{\rm disc-m}$ | 20.0 |
| $R_{\rm t,d}^{\rm disc-m}$ | 20.0 |
| $z_{\rm s}^{\rm tdisc-(i,\,m)}$ | (0.09, 0.09) |
| $z_{\rm s}^{\rm disc-m}$ | 0.22 |
| $z_{\rm d}^{\rm disc-m}$ | 0.14 |
| $n_{\rm s}$ | 2 |

**Table B9.** Same as Table B1 for NGC 3521.

| Parameter | d | c |
|---|---|---|
| $h_{\rm s}^{\rm tdisc-(i,\,m,\,o)}$ | (0.50, 3.30, 2.30) | (0.50, 3.40, 2.80) |
| $h_{\rm s}^{\rm disc-(m,\,o)}$ | (2.00, 4.10) | (2.00, 4.10) |
| $h_{\rm d}^{\rm disc-(m,\,o)}$ | (7.50, 7.50) | (7.80, 7.00) |
| $\chi_{\rm s}^{\rm tdisc-(i,\,m,\,o)}$ | (0.10, −0.50, 2.50) | (0.10, −0.60, 2.50) |
| $\chi_{\rm s}^{\rm disc-(m,\,o)}$ | (1.00, 2.00) | (1.00, 2.00) |
| $\chi_{\rm d}^{\rm disc-(m,\,o)}$ | (−0.10, 1.00) | (0.00, 1.00) |
| $R_{\rm e}$ | 0.55 | 0.60 |
| $\frac{b}{a}$ | 0.50 | 0.50 |

**Table B10.** Same as Table B2 for NGC 3521.

| Parameter | d |
|---|---|
| $R_{\rm in,s}^{\rm tdisc-(i,\,m,\,o)}$ | (0.00, 1.30, 10.00) |
| $R_{\rm in,s}^{\rm disc-(m,\,o)}$ | (0.00, 7.00) |
| $R_{\rm in,d}^{\rm disc-(m,\,o)}$ | (1.30, 7.00) |
| $R_{\rm tin,s}^{\rm tdisc-(i,\,m,\,o)}$ | (0.00, 0.50, 6.50) |
| $R_{\rm tin,s}^{\rm disc-(m,\,o)}$ | (0.00, 6.50) |
| $R_{\rm tin,d}^{\rm disc-(m,\,o)}$ | (0.00, 6.50) |
| $R_{\rm t,s}^{\rm tdisc-(i,\,m,\,o)}$ | (1.00, 6.30, 20.00) |
| $R_{\rm t,s}^{\rm disc-(m,\,o)}$ | (6.50, 20.00) |
| $R_{\rm t,d}^{\rm disc-(m,\,o)}$ | (6.30, 20.00) |
| $z_{\rm s}^{\rm tdisc-(i,\,m,\,o)}$ | (0.09, 0.09, 0.09) |
| $z_{\rm s}^{\rm disc-(m,\,o)}$ | (0.19, 0.19) |
| $z_{\rm d}^{\rm disc-(m,\,o)}$ | (0.16, 0.16) |
| $n_{\rm s}$ | 2.0 |

**Table B11.** Same as Table B1 for NGC 3938.

| Parameter | d | c |
|---|---|---|
| $h_{\rm s}^{\rm tdisc}$ | 3.20 | 3.80 |
| $h_{\rm s}^{\rm disc}(B)$ | 3.00 | 3.15 |
| $h_{\rm d}^{\rm disc}$ | 9.00 | 7.50 |
| $\chi_{\rm s}^{\rm tdisc}$ | 1.10 | 1.40 |
| $\chi_{\rm s}^{\rm disc}$ | 1.00 | 1.00 |
| $\chi_{\rm d}^{\rm disc}$ | 1.00 | 0.60 |
| $R_{\rm e}$ | 0.53 | 0.53 |
| $\frac{b}{a}$ | 0.94 | 0.94 |

**Table B12.** Same as Table B2 for NGC 3938.

| Parameter | Value |
|---|---|
| $R_{\mathrm{in,s}}^{\mathrm{tdisc}}$ | 2 |
| $R_{\mathrm{in,s}}^{\mathrm{disc}}$ | 2 |
| $R_{\mathrm{in,d}}^{\mathrm{disc}}$ | 2 |
| $R_{\mathrm{tin,s}}^{\mathrm{tdisc}}$ | 0 |
| $R_{\mathrm{tin,s}}^{\mathrm{disc}}$ | 0 |
| $R_{\mathrm{tin,d}}^{\mathrm{disc}}$ | 0 |
| $R_{\mathrm{t,s}}^{\mathrm{tdisc}}$ | 14 |
| $R_{\mathrm{t,s}}^{\mathrm{disc}}$ | 17 |
| $R_{\mathrm{t,d}}^{\mathrm{disc}}$ | 14 |
| $z_{\mathrm{s}}^{\mathrm{tdisc}}$ | 0.09 |
| $z_{\mathrm{s}}^{\mathrm{disc}}$ | 0.19 |
| $z_{\mathrm{d}}^{\mathrm{disc}}$ | 0.16 |
| $n_{\mathrm{s}}$ | 1 |

**Table B13.** Values for the wavelength-dependent parameters of the clumpy model for each galaxy.

| Galaxy | Parameter | $B$ | $V$ | $I$ | $J$ | $K_s$ | 3.6 μm | 4.5 μm | 5.8 μm |
|---|---|---|---|---|---|---|---|---|---|
| M33 | $h_s^{disc-m}$ | 1.80 | 1.75 | 1.65 | 1.05 | 1.15 | 1.75 | 1.80 | 1.40 |
| | $h_s^{disc-i}$ | 0.05 | 0.07 | 0.12 | 0.05 | 0.10 | 0.12 | 0.13 | 0.09 |
| | $h_s^{disc-o}$ | 1.00 | 1.00 | 1.00 | 1.00 | 1.00 | 1.00 | 1.00 | 1.00 |
| M51 | $h_s^{disc-m}$ | 5.50 | 4.60 | 4.18 | 3.70 | 3.80 | 4.40 | 4.60 | 7.00 |
| | $h_s^{disc-i}$ | 0.72 | 0.68 | 0.70 | 0.67 | 0.66 | 0.72 | 0.77 | 1.10 |
| | $h_s^{disc-o}$ | 2.40 | 2.30 | 2.30 | 1.90 | 2.00 | 2.60 | 2.60 | 2.00 |
| M101 | $h_s^{disc-m}$ | 4.30 | 4.30 | 4.10 | 3.50 | 3.60 | 3.65 | 3.60 | 3.70 |
| | $h_s^{disc-i}$ | 0.50 | 0.50 | 0.50 | 0.40 | 0.40 | 0.30 | 0.20 | 0.10 |
| NGC 628 | $h_s^{disc-m}$ | 3.30 | 3.20 | 2.90 | 2.68 | 2.64 | 2.90 | 2.90 | 2.80 |
| NGC 891 | $h_s^{disc-m}$ | 10.00 | 7.00 | 5.30 | 4.40 | 4.00 | 4.20 | 4.20 | 3.60 |
| | $z_s$ | 0.50 | 0.50 | 0.46 | 0.36 | 0.31 | 0.30 | 0.31 | 0.27 |
| | $\frac{b}{a}$ | 0.47 | 0.47 | 0.47 | 0.47 | 0.47 | 0.47 | 0.47 | 0.47 |
| | $R_e$ | 1.44 | 1.44 | 1.44 | 1.44 | 1.60 | 1.60 | 1.60 | 1.44 |
| NGC 3521 | $h_s^{disc-m}$ | 2.00 | 2.00 | 2.20 | 2.60 | 2.80 | 2.80 | 2.90 | 4.30 |
| | $h_s^{disc-o}$ | 4.10 | 4.10 | 4.30 | 4.10 | 4.15 | 3.90 | 3.80 | 3.70 |
| NGC 3938 | $h_s^{disc-m}$ | 3.15 | 3.00 | 3.00 | 2.65 | 2.50 | 2.80 | 2.80 | 2.80 |
| NGC 5907 | $h_s^{disc-m}$ | 8.20 | 6.70 | 6.30 | 5.00 | 5.00 | 4.80 | 4.70 | 4.50 |
| | $z_s$ | 0.52 | 0.50 | 0.52 | 0.45 | 0.43 | 0.44 | 0.43 | 0.42 |
| | $\frac{b}{a}$ | 0.48 | 0.48 | 0.44 | 0.36 | 0.50 | 0.38 | 0.38 | 0.38 |
| | $R_e$ | 1.00 | 1.00 | 1.00 | 1.00 | 0.80 | 1.00 | 1.00 | 1.00 |

This paper has been typeset from a TEX/LATEX file prepared by the author.